\documentclass[twocolumn,superscriptaddress,aps,preprintnumbers,amsmath,amssymb,prl,nofootinbib]{revtex4-1}

\usepackage[titletoc,toc,title]{appendix}
\usepackage{titletoc}
\usepackage{titlesec}

\usepackage{textgreek}
\usepackage{amsmath}
\usepackage{amssymb}
\usepackage{amsfonts}
\usepackage{graphicx}
\usepackage{xcolor}
\usepackage{xfrac}
\usepackage{footmisc}
\usepackage{comment}
\usepackage{pifont}
\usepackage{times}
\usepackage[hidelinks]{hyperref}
\usepackage{,bm}
\usepackage{enumitem}
\usepackage{centernot}
\usepackage{cancel}
\usepackage{slashed}
\usepackage{relsize}
\usepackage{amsmath,amssymb,mathrsfs}
\usepackage{mdframed}
\usepackage{tcolorbox}
\usepackage{bm}
\usepackage{times}
\usepackage{epsfig}
\usepackage{verbatim}
\usepackage{bm}
\usepackage[utf8]{inputenc}
\usepackage{graphics}
\usepackage{graphicx,epsfig,amssymb,amsmath,color,cancel}
\usepackage{subfigure}
\usepackage[english]{babel}
\usepackage{adjustbox}
\usepackage{soul}
\usepackage[normalem]{ulem}

\definecolor{zima_blue}{HTML}{1393C1}
\hypersetup{setpagesize=false,bookmarksnumbered=true,bookmarksopen=true,%
colorlinks=true,linkcolor=zima_blue,urlcolor=zima_blue,citecolor=zima_blue,linktocpage=false}

\usepackage{orcidlink}

\usepackage[
    starfontserif 
    ]{starfont}
\usepackage{amsmath}

\DeclareSymbolFont{starfontsym}{OT1}{sts}{m}{n}
\DeclareMathSymbol{\mathSun}{\mathord}{starfontsym}{115}
\DeclareMathSymbol{\mathMercury}{\mathord}{starfontsym}{102}
\DeclareMathSymbol{\mathVenus}{\mathord}{starfontsym}{103}
\DeclareMathSymbol{\mathTerra}{\mathord}{starfontsym}{76}
\DeclareMathSymbol{\mathvarTerra}{\mathord}{starfontsym}{108}
\DeclareMathSymbol{\mathMoon}{\mathord}{starfontsym}{100}
\DeclareMathSymbol{\mathvarMoon}{\mathord}{starfontsym}{97}
\DeclareMathSymbol{\mathMars}{\mathord}{starfontsym}{104}
\DeclareMathSymbol{\mathJupiter}{\mathord}{starfontsym}{106}
\DeclareMathSymbol{\mathSaturn}{\mathord}{starfontsym}{83}
\DeclareMathSymbol{\mathUranus}{\mathord}{starfontsym}{70}
\DeclareMathSymbol{\mathvarUranus}{\mathord}{starfontsym}{65}
\DeclareMathSymbol{\mathNeptune}{\mathord}{starfontsym}{71}
\DeclareMathSymbol{\mathPluto}{\mathord}{starfontsym}{74}
\DeclareMathSymbol{\mathvarPluto}{\mathord}{starfontsym}{72}

\newcommand{\td}{{\rm d}}

\begin{document}




\author{Yann Gouttenoire~\orcidlink{0000-0003-2225-6704}}
\email{yann.gouttenoire@gmail.com}
\affiliation{School of Physics and Astronomy, Tel-Aviv University, Tel-Aviv 69978, Israel}
\affiliation{PRISMA+ Cluster of Excellence $\&$ MITP, Johannes Gutenberg University, 55099 Mainz, Germany}

\title{WIMPs and new physics interpretations of the PTA signal are incompatible}
\begin{abstract}

In order to explain the large amplitude of the nano-Hertz stochastic gravitational wave background observed in pulsar timing arrays (PTA), primordial sources must be particularly energetic.  This is correlated to the generation of large density fluctuations, later collapsing into ultra-compact mini-halo (UCMHs).  We demonstrate that if dark matter is made of WIMPs, then photon and neutrino fluxes from UCMHs produced by curvature peaks, first-order phase transition and domain wall interpretations of the PTA signal, exceed current bounds.

\end{abstract}

\maketitle


\textit{\textbf{Introduction.}} In June 2023, several pulsar timing array (PTA) collaborations~\cite{NANOGrav:2023gor,Antoniadis:2023rey,Reardon:2023gzh,Xu:2023wog,InternationalPulsarTimingArray:2023mzf} reported a stochastic gravitational wave background at nano-Hertz frequencies with an amplitude of about $\Omega_{\rm GW}\sim 10^{-9}$. If this signal originates from the early universe, it must be produced by extremely energetic processes~\cite{Giblin:2014gra}, see~\cite{Ratzinger:2020koh,Madge:2023dxc,NANOGrav:2023hvm,EPTA:2023xxk,Bian:2023dnv,Figueroa:2023zhu,Ellis:2023oxs} for reviews. Various works have showed that early universe interpretations of the PTA signal -- whether they are curvature peaks \cite{Nakama:2016gzw,Chen:2019xse,Inomata:2020xad,Vaskonen:2020lbd,DeLuca:2020agl,Franciolini:2023pbf}, first-order phase transitions (FOPTs) \cite{Gouttenoire:2023bqy,Ellis:2023oxs,Lewicki:2024ghw}, or domain wall networks (DWs)~\cite{Ferreira:2022zzo,Kitajima:2023cek,Gouttenoire:2023ftk} -- are correlated with large density fluctuations, associated with the production of solar-mass primordial black holes (PBHs), testable at LIGO. However, forming PBHs requires a high density threshold $(\delta \gtrsim 0.5)$ which limits their abundance and weakens the correlation between the PTA signal and PBH production. In contrast, density fluctuations as low as $\delta \gtrsim 10^{-4}$ results in the formation of ultra-compact dark matter mini-halos (UCMHs)~\cite{Ricotti:2009bs}, which supplement the distribution of smoother DM halos predicted in standard
$\Lambda$CDM~\cite{Diemand:2005vz,Zhao:2005py,Moore:2005uu,Green:2005fa,Loeb:2005pm,Delos:2022yhn}. 
Under the hypothesis that dark matter is composed of weakly interacting massive particles (WIMPs)~\cite{Cirelli:2024ssz}, such enhancements in density result in significantly increased photon and neutrino signals~\cite{Scott:2009tu,Josan:2010vn,Bringmann:2011ut,Nakama:2017qac,Delos:2018ueo,Blanco:2019eij}. 

In this \textit{letter}, we characterize the UCMH population arising from the curvature peak, FOPT and DW interpretations of the PTA signal. Under the minimal hypothesis that DM is made of WIMPs, we calculate the photon and neutrino fluxes at earth, including both galactic and extra-galactic emission. Using Fermi-LAT and atmospheric neutrino measurement, we demonstrate that these three new physics interpretations of the PTA signal are incompatible with multi-TeV WIMP DM.


\textit{\textbf{Ultra-Compact Mini-Halos.}}
Consider a primordial comoving curvature perturbation with initial amplitude $\mathcal{R}(k)$. For modes entering the horizon before matter-radiation equality ($k\gg k_{\rm eq}$), the comoving density contrast evolves as  
$\delta(k,t) \simeq 6.4\, \mathcal{R}(k)\,\ln\left[0.66\,ka/(k_{\rm eq}a_{\rm eq})\right]$ 
during radiation era, and as
\begin{equation}
\label{eq:delta_R}
\delta(k,t) \simeq 9.6\, \mathcal{R}(k)\,\frac{a}{a_{\rm eq}}\,\ln\left[0.13\,\frac{k}{k_{\rm eq}}\right],
\end{equation}
during matter era~\cite{Hu:1995en}, with $k_{\rm eq}\simeq (98~\rm Mpc)^{-1}$ and $a^{-1}_{\rm eq}\simeq 3401$~\cite{ParticleDataGroup:2020ssz}. Effects from baryons and dark energy are neglected.
Regions of size $R=k^{-1}$ collapse~\cite{1967ApJ...147..859P} when the linear density contrast reaches $\delta_c\simeq 1.69$ in the matter era~\cite{Dodelson:2003ft} and $\delta_c\simeq 3$ in the radiation era~\cite{Blanco:2019eij,StenDelos:2022jld}. The minimal curvature perturbation required for collapse at a scale factor $a_c$ during the matter era $(a_c\gg a_{\rm eq})$ is
\begin{align}
\label{eq:R_min_R_zc}
\mathcal{R}^{\rm min}(R,a_c)\simeq \frac{9}{4}\,\delta^{\rm min}
\simeq 3.2 \times 10^{-4} \frac{a_c^{-1}}{100}\frac{\mathcal{L}(\rm pc)}{\mathcal{L}(R)},
\end{align}
with $\mathcal{L}(R)\equiv \ln\left(0.13/k_{\rm eq}R\right)$. Thus, a curvature power spectrum with variance $\left<\mathcal{R}^2\right>\gtrsim 10^{-7}$ leads to collapse at a much larger redshift than typical DM halos $a_c^{-1}\gg 30$~\cite{Delos:2022yhn}.
Due to the low velocity dispersion, these early-collapsing regions initialy develop steep profiles $\rho(r)\propto r^{-\alpha}$, with $\alpha=9/4$ for scale-invariant curvature power spectrum $\mathcal{P}_{\mathcal{R}}(k)$~\cite{Fillmore:1984wk,Bertschinger:1985pd} and $\alpha=3/2$ (Moore profile) for peaked $\mathcal{P}_{\mathcal{R}}(k)$~\cite{Ishiyama:2010es,Ishiyama:2014uoa,Anderhalden:2013wd,Angulo:2016qof,Ogiya:2017hbr,Delos:2017thv,Delos:2019mxl}. In case of frequent mergers, the slope relaxes to $\alpha=1$ (NFW profile) \cite{Ogiya:2016hyo,Delos:2017thv,Delos:2019mxl}. In this work, we consider UCMHs density profiles with $\alpha=1$ and $\alpha=3/2$, parameterized as
\begin{equation}
\label{eq:UCMH_rho_density}
\rho_{\rm UCMH}(r)=\min\left[\frac{\rho_h}{(r/r_h)^\alpha(1+r/r_h)^{3-\alpha}},\,\rho_{\rm max}\right].
\end{equation}
For collapse during matter era, N-body simulations suggest $\rho_h\simeq30\,\overline{\rho}_{\rm DM,0}\,a_c^{-3}$ and $r_h\simeq0.68\,a_c\,k_\star^{-1}$ where $k_\star$ is the comoving peak momentum of the curvature power spectrum and $\overline{\rho}_{\rm DM,0}\simeq (1.8\,{\rm meV})^4$~\cite{Delos:2017thv}. Instead, for collapse during radiation era, simulations indicate $\rho_h\simeq g\,\overline{\rho}_{\rm rad,0}\,a_f^{-4}$ with $g\in[1,10^3]$ and $\overline{\rho}_{\rm rad,0}\simeq (0.24\,{\rm meV})^4$~\cite{Delos:2023fpm}. Using $\rho_h\,r_h^3\simeq\overline{\rho}_{\rm DM,0}\,k_\star^{-3}$ one obtains $r_h \simeq 14(\overline{\rho}_{\rm DM,0}a_f^{4}/g \overline{\rho}_{\rm rad,0})^{1/3}k_{\star}^{-1}$. Here, $a_f\simeq e^2a_{\rm eq}$ is the scale factor at which the patch becomes locally matter-dominated, with a typical ellipticity of $e\simeq0.15$~\cite{StenDelos:2022jld,Delos:2023fpm}. We conservatively set $g=1$. DM annihilation flattens the profile in Eq.~\eqref{eq:UCMH_rho_density} above the value~\cite{Scott:2009tu}
\begin{equation}
\rho_{\rm max} = \frac{M_{\rm DM}}{\langle \sigma_{\rm ann} v \rangle \Delta t},
\end{equation}
where $\Delta t\simeq 13.8~\rm Gyrs$ is the time between halo formation and today.
The UCMH mass is approximately the DM mass within the horizon at mode entry
\begin{equation}
\label{eq:M_i_R}
M_{\rm UCMH} \simeq \overline{\rho}_{\rm DM,0} \frac{4\pi k_\star^{-3}}{3} \simeq  3.6 M_{\mathMoon}  \left( \frac{10^6~\rm Mpc^{-1}}{k_{\star}} \right)^3,
\end{equation}
with $M_{\mathMoon}\simeq 3.7\times 10^{-8}~M_{\mathSun}$ the Moon mass.
It has been claimed that the UCMH mass continues to grow through accretion of DM and baryonic matter during the matter era as  $M_{\rm UCMH}\propto a^{-1}$~\cite{Bertschinger:1985pd,Ricotti:2009bs,Bringmann:2011ut,Nakama:2017qac}. Such growth has not been observed in recent N-body simulations~\cite{Delos:2017thv}, hence we make the conservative choice of a constant UCMH mass.

We calculate the number density of UCMHs following the predictions from peak theory~\cite{Bardeen:1985tr}
\begin{equation}
\label{eq:n_peak_theory_2}
    \frac{dn(R,a_c)}{da_c} = \frac{1}{(2\pi)^2R^3}\left(\frac{\sigma_2}{\sqrt{3}\sigma_1} \right)^3e^{-\nu_c^2/2} G(\gamma,\gamma \nu_c),
\end{equation}
where the function $G(x,y)$ is given in \cite[Eq.~A19]{Bardeen:1985tr}, $\nu_c(R,a_c) \equiv \delta^{\rm min}(R,a_c)/\sigma_{0}(R)$ and $\gamma \equiv \sigma_1^2/\sigma_0\sigma_2$. The $\sigma_j$ are spectral moments of the curvature power spectrum $\mathcal{P}_{\mathcal{R}}(k)$ convolved with window and transfer functions, e.g.~\cite{Ferrante:2022mui}
\begin{equation}
\label{eq:sigma_j_moments}
    \sigma_j^2(R) = \frac{16}{81}\int \frac{dk}{k}(k R)^4\tilde{W}^2(k,R)T^2(k,R)\mathcal{P}_{\mathcal{R}}(k) k^{2j},
\end{equation}
 We set $\tilde{W}(k,R)=3j_1(kR)/kR$ for a real-space top hat window function~\cite{Young:2019osy}, and $T(k,R)=3j_1(kR/\sqrt{3})/(kR/\sqrt{3})$ in the comoving gauge~\cite{Josan:2009qn}. After freeze-out, DM continue to scatter elastically with SM until the temperature $T_{\rm KD}$ of kinetic decoupling. This leads to collisional damping of DM perturbations through an exponential factor suppressing the transfer function~\cite{Hofmann:2001bi,Loeb:2005pm,Green:2005fa}
\begin{equation}
    T(k,R) \rightarrow T(k,R) \times \exp\left[-\frac{1}{2}\left(\frac{k}{k_D}\right)^2\right],
\end{equation}
where $k_{\rm D}^{-1}\sim v_{\rm KD}/\mathcal{H}(T_{\rm KD})$ with $v_{\rm KD}$ the DM velocity at $T_{\rm KD}$~\cite{Hofmann:2001bi,Loeb:2005pm,Green:2005fa}. Using numerical factors in~\cite{Green:2005fa}, we get
\begin{align}
    k_{\rm D} &\simeq \frac{1.7 \times 10^{8}}{\rm Mpc} \left( \frac{m_{\rm DM}}{1~\rm TeV}\right)^{1/2}\left( \frac{T_{\rm KD}}{30~\rm MeV} \right)^{1/2}.
\end{align}
Depending on the SM-WIMP interactions, $T_{\rm KD}$ ranges from MeV to GeV~\cite{Profumo:2006bv,Bringmann:2009vf,Cornell:2012tb,Ando:2019rgx}. We conservatively set $T_{\rm KD} = 1$ MeV.

\textit{\textbf{Gamma-Ray and Neutrino Fluxes.}}
The number of photons or neutrinos $X=\gamma,\nu$ radiated per unit energy per ultracompact minihalo (UCMH) is given by
\begin{equation}
\label{eq:dLX_dE_0}
    \frac{dL}{dE} = \sum_{f} \langle \sigma_f v \rangle \frac{dN_{f\to X}}{dE} \! \int_0^{R} \!\!\!4\pi r^2\,dr \, \frac{1}{2} \left( \frac{\rho_{\rm UCMH}(r)}{m_{\rm DM}} \right)^{\!2},
\end{equation}
where $R$ is the UCMH radius, the factor $1/2$ assumes that DM is self-conjugate, the sum over $ f $ accounts for all possible final states from DM annihilation, each weighted by its thermally averaged annihilation cross-section $ \langle \sigma_f v \rangle $ and corresponding differential photon and neutrino yield $dN_{f\to X}/dE$, which we calculate with ${\tt PPPC4DMID}$~\cite{Cirelli:2010xx}. 
Integrating this luminosity along the line of sight, we get the diffuse flux at angle $\theta$ to the Galactic center, per unit of solid angle and energy
\begin{equation}
\label{eq:flux_photon_DM}
    \begin{aligned}
  \!      \frac{\td^2\mathcal{F}_X(\theta)}{\td\Omega \td E} \!&=\!\!\! \int_0^{\infty}\!\! \frac{\td s}{4\pi}\frac{\rho_{\rm DM}(s,\theta)}{\overline{\rho}_{\rm DM,0}}\!\! \int_0^{\infty}\!\!\!\!\!\!\td a_c\!\!\int_0^{\infty}\!\!\frac{\td R}{R} \frac{dn}{da_c} \frac{\td L[(1+z_e)E]}{\td E} .
    \end{aligned}
\end{equation}
where the DM density at a distance $s$ from the Earth at angle $\theta$ with the galactic center is 
\begin{equation}
   \! \rho_{\rm DM}(s,\theta) \!=\! \rho_{\rm MW}\!\!\left[\!\sqrt{\!s^2\!+\!r_{\odot}^2\!-\!2r_{\odot}s\cos{\theta}}\right]\!+\!\frac{e^{-\tau(E,z_e)}\!\!}{(1\!+\!z_e)^3}\overline{\rho}_{\rm DM,0}.
\end{equation}
The first piece is the DM halo density profile of the Milky-Way, assumed to follow NFW law $\rho_{\rm MW}(r)=\rho_s(r/r_s)^{-1}(1+r/r_s)^{-2}$ with $\rho_s\simeq 0.53~\rm GeV \,cm^{-3}$ and $r_s\simeq 16~\rm kpc$~\cite{Nesti:2013uwa}.
The second piece account for extra-galactic contribution, with the the transparency factor $e^{-\tau}$ accounting for the absorption over cosmological distances. We calculate the optical depth $\tau(E,z)$ with ${\tt PPPC4DMID}$~\cite{Cirelli:2010xx}. The redshift at emission $z_e$ is related to the comoving distance $s$ by $s = \int_0^{z_e} dz/H(z)$.
We consider two classes of constraints.
First, we impose that the photon flux $ d^2\mathcal{F}_\gamma/d\Omega dE $ must not exceed the value measured by Fermi–LAT at the galactic north pole for $\theta \in [60^\circ,~90^\circ] $ and $ E_\gamma \in [0.1,~10^2]~\rm GeV $~\cite[Fig.~13]{Fermi-LAT:2012edv}.
Second, we impose that the neutrino flux $ d^2\mathcal{F}_\nu/d\Omega dE $ must not exceed the atmospheric value measured by Frejus~\cite{Frejus:1994brq}, Amanda-II~\cite{IceCube:2010jrf}, Antares~\cite{ANTARES:2013iuz}, Super-Kamiokande~\cite{Super-Kamiokande:2015qek}, Icecube~\cite{IceCube:2016umi,IceCube:2023hou} and given in \cite[Fig.~14]{Super-Kamiokande:2015qek}. We conservatively fix $\theta=90^\circ$ to minimize contamination from astrophysical sources like pulsars and supernova remnants.

\begin{figure}[th!]
\centering
\raisebox{0cm}{\makebox{\includegraphics[width=0.5\textwidth, scale=1]{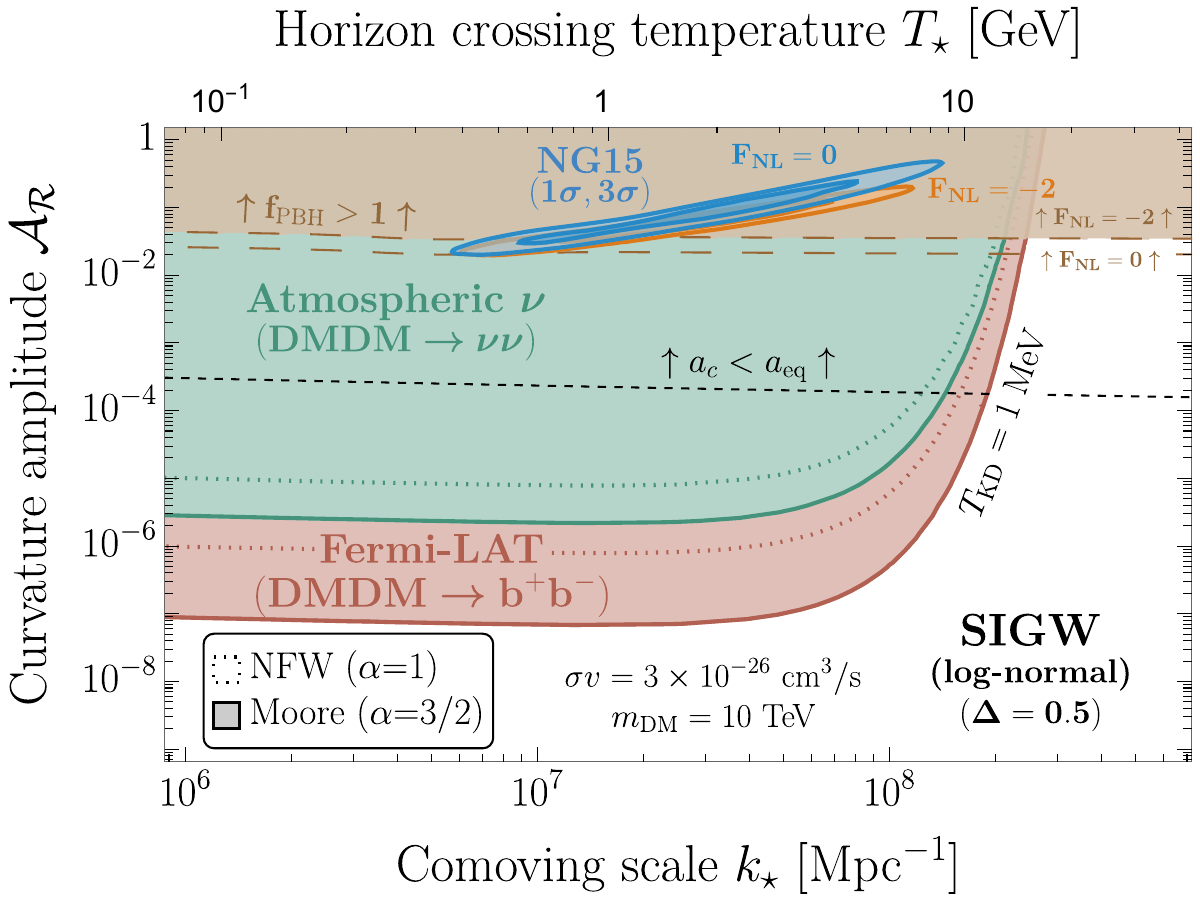}}}
\caption{Small amount of non-gaussianities (NGs), $F_{\rm NL}=\mathcal{O}(1)$, can relieve the tension between the SIGW interpretation of the PTA signal (blue and orange) and the PBH overclosure bound (brown). However, they are not enough to evade the cosmic-ray constraints (red and green) if DM is made of WIMPs. The horizontal dashed black line indicates when collapse occurs during the radiation era $a_c<a_{\rm eq}$. } 
\label{fig:SIGW}
\end{figure}

\textit{\textbf{Scalar-Induced Gravitational Waves.}}  
The curvature peaks, $\mathcal{R}(k)$, which are responsible for UCMH formation, can also source scalar-induced gravitational waves (SIGWs)~\cite{Domenech:2021ztg,Espinosa:2018eve,Kohri:2018awv}. In a generic scenario, the curvature perturbation can be parametrized by the power-series~\cite{Byrnes:2007tm}:
\begin{equation}
\mathcal{R} = \mathcal{R}_{\rm G} + F_{\rm NL}\Bigl(\mathcal{R}_{\rm G}^2 - \langle \mathcal{R}_{\rm G}^2 \rangle\Bigr)+\cdots,
\end{equation}
where $\mathcal{R}_{\rm G}$ obeys Gaussian statistics, which we describe with a log-normal power spectrum
\begin{equation}
\label{eq:PR_LN}
\mathcal{P}_{\mathcal{R}}^{\rm G}(k) = \frac{\mathcal{A}_\mathcal{R}}{\sqrt{2\pi}\Delta}\,\exp\!\left[-\frac{\log^2(k/k_\star)}{2\Delta^2}\right].
\end{equation}
Here, $\mathcal{A}_\mathcal{R}$ denotes the amplitude, $k_\star$ is the peak scale, and $\Delta$ characterizes the width of the peak. The parameter $F_{\rm NL}$ measure the level of Non-Gaussianities (NG). At zeroth order in $F_{\rm NL}$, the present-day SIGW energy density power spectrum is given by
\begin{equation}
\begin{aligned}
    \Omega_{\rm SIGW} = \int_0^\infty \!\!\td v \!\int_{1+v}^{|1-v|} \!\!\!\!\td u\, \,\overline{\mathcal{I}_{u,s}^2}  ~\mathcal{P}_{\mathcal{R}_{\rm G}}\left(ku\right) \mathcal{P}_{\mathcal{R}_{\rm G}}\left(kv\right) ,
\end{aligned}
\end{equation}
where the kernel $\overline{\mathcal{I}_{u,s}^2}$ accounts for the evolution of curvature and tensor modes up to the present day~\cite{Domenech:2021ztg,Espinosa:2018eve,Kohri:2018awv}. NG corrections involve terms of order $F_{\rm NL}^2\mathcal{P}_{\mathcal{R}_{\rm G}}^3$ and $F_{\rm NL}^4\mathcal{P}_{\mathcal{R}_{\rm G}}^4$ which can be found in Refs.~\cite{Unal:2018yaa,Cai:2018dig,Adshead:2021hnm}. Using \texttt{PTArcade} ~\cite{Mitridate:2023oar} and the relevant expression for $\Omega_{\rm SIGW}$, we calculate in Fig.~\ref{fig:SIGW} the 1$\sigma$ and 3$\sigma$  posterior contours for the parameters $\mathcal{A}_\mathcal{R}$ and $k_\star$  required to explain the NG15 signal~\cite{NANOGrav:2023gor}.
 Following the methodology of Refs.~\cite{Musco:2020jjb,Escriva:2019phb,Ferrante:2022mui,Frosina:2023nxu,Iovino:2024tyg}, we compute the PBH abundance from curvature peaks normalised to DM, $f_{\rm PBH}$. We show in Fig.~\ref{fig:SIGW} the overclosure bound, $f_{\rm PBH}<1$, excludes most of the PTA posterior for $F_{\rm NL}=0$, but that the tension can be relieved for negative NG, $F_{\rm NL}=-2$, as found in~\cite{Franciolini:2023pbf}.

 Plugging Eq.~\eqref{eq:PR_LN} into Eq.~\eqref{eq:sigma_j_moments}, we calculate the abundance UCMHs produced by curvature peaks, neglecting the impact of NGs.
 We assume that DM is a WIMP with of mass of $m_{\rm DM}=10~\rm TeV$, annihilating with a canonical cross-section ($\sigma v\sim 3 \times 10^{-26}~\rm cm^3/s$) into either $b\overline{b}$ or $\nu\overline{\nu}$. We derive the novel constraints resulting from Fermi-LAT (red) and atmospheric neutrinos (green), respectively.
We find that the curvature peak interpretation of the PTA signal, with small level of NGs $f_{\rm NG}=\mathcal{O}(1)$, is largely incompatible with DM being made of WIMPs.

\begin{figure}[th!]
\centering
\raisebox{0cm}{\makebox{\includegraphics[width=0.5\textwidth, scale=1]{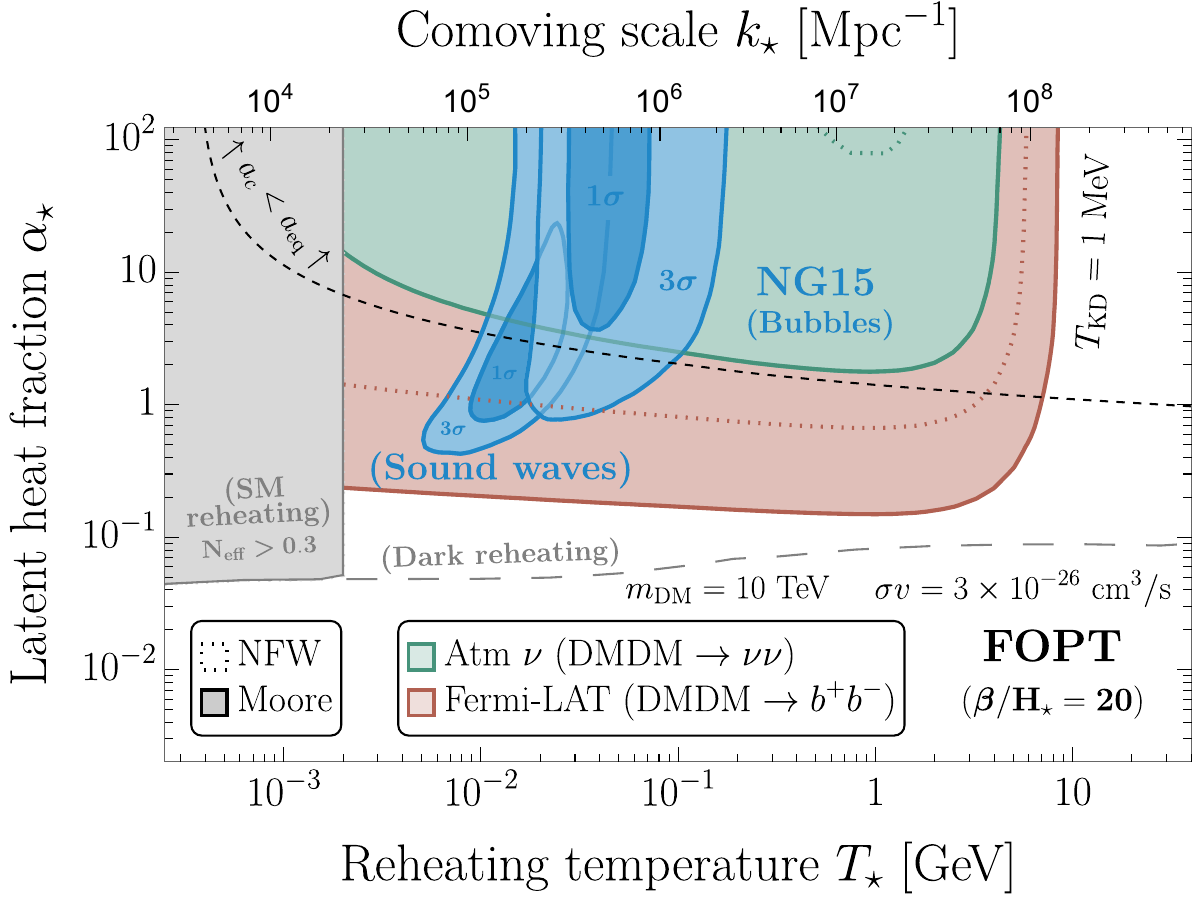}}}
{\makebox{\includegraphics[width=0.5\textwidth, scale=1]{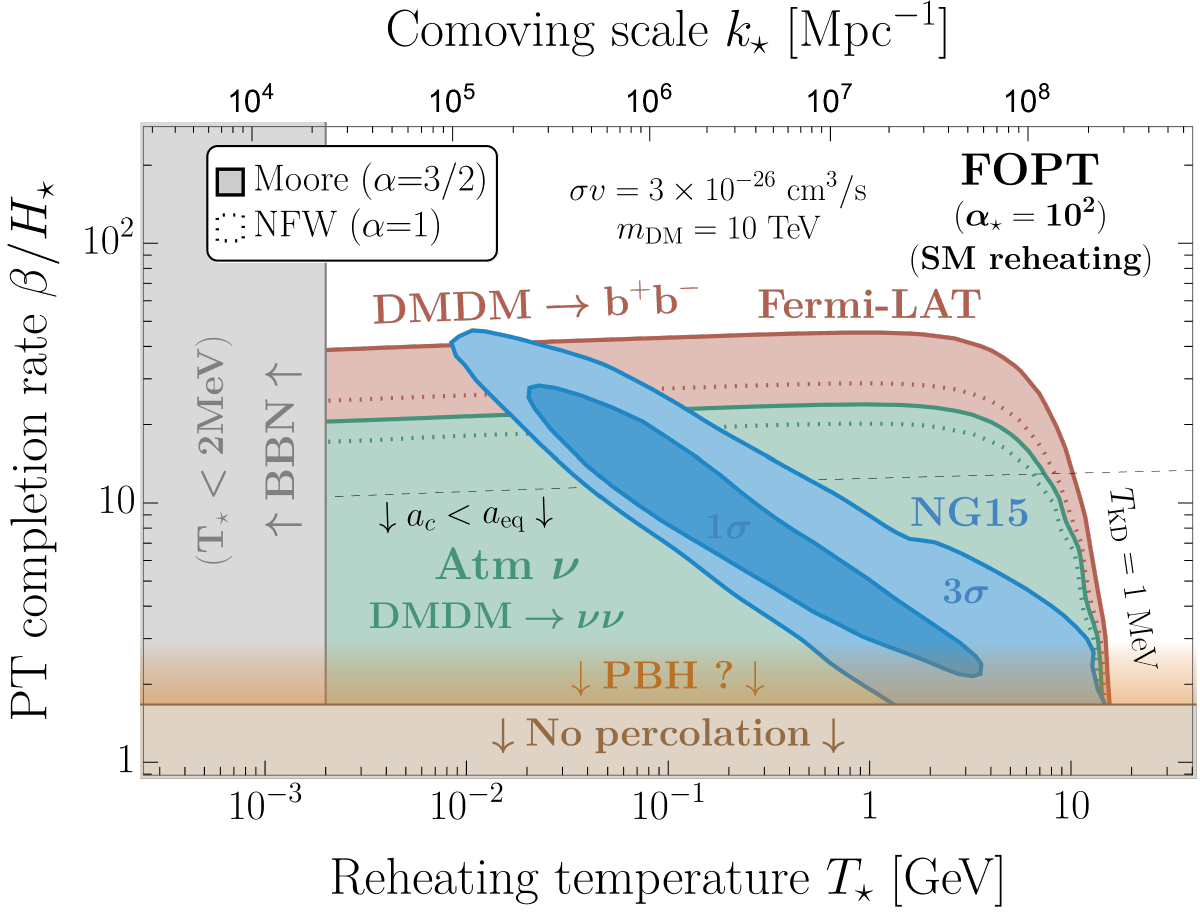}}}
\caption{The FOPT interpretation of the PTA signal shown in blue appears in tension with multi-TeV WIMP DM, when looking at the Fermi-LAT constraints on the photon flux (red). The atmospheric neutrino constraints exclude a large parameter space but are evaded for $\beta/H\gtrsim 20$ if UCMHs are modeled with a NFW density profile.  } 
\label{fig:FOPT}
\end{figure}

\textit{\textbf{First-order Phase Transitions.}} 
A FOPT involves a transition between vacuum states via bubble nucleation and subsequent expansion. The collisions of these bubbles, along with the generation of sound waves and turbulence in the plasma, can source GWs~\cite{Caprini:2015zlo,Caprini:2019egz,Gouttenoire:2022gwi}, providing a viable interpretation of the PTA signal, see dedicated studies~\cite{NANOGrav:2021flc,Xue:2021gyq,RoperPol:2022iel,Bringmann:2023opz} and reviews~\cite{NANOGrav:2023hvm,EPTA:2023xxk,Ellis:2023oxs,Madge:2023dxc}.  We reproduce the bayesian analysis of the FOPT intepretation of the PTA signal performed in~\cite{NANOGrav:2023hvm}.
We use the models ${\tt PT-BUBBLE}$ and ${\tt PT-SOUND}$ built-in in $\tt PTArcade$~\cite{Mitridate:2023oar}, associated to GW from bubble collision and sound waves respectively.  Those models assume the bubble wall velocity to be maximal, neglect the contribution from turbulence, and consider a wide range of spectral shapes as priors. 
The GW signal depends on three parameters: the percolation temperature $T_\star$, the latent heat fraction $\alpha_\star=\Delta V/\rho_{\rm rad}(T_\star)$ where $\Delta V$ is the vacuum energy difference, and the completion rate of the PT at $T_\star$, $\beta =\td\log(\Gamma)/\td t$,  where $\Gamma(t)$ is the bubble nucleation rate. Fig.~\ref{fig:FOPT} displays the NG15 posteriors in the $(T_\star,\alpha_\star)$ and $(T_\star,\beta/H_\star)$ planes. The PTA signal favor FOPTs which are slow ($\beta/H_\star\lesssim20$) and strong ($\alpha_\star\gtrsim 0.5$), as the ones arising from nearly flat scalar potentials~\cite{Baldes:2018emh,Lewicki:2024sfw,Goncalves:2025uwh,Balan:2025uke,Costa:2025csj}.

The stochastic nature of bubble nucleation during a FOPT also generates superhorizon curvature perturbations~\cite{Sasaki:1982fi,Liu:2022lvz,Giombi:2023jqq,Elor:2023xbz,Lewicki:2024ghw,Buckley:2024nen,Cai:2024nln,Jinno:2024nwb,Franciolini:2025ztf}. The curvature power spectrum has been computed analytically in~\cite{Jinno:2024nwb} and numerically in~\cite{Franciolini:2025ztf}. It can be parameterized as~\cite{Jinno:2024nwb}
\begin{equation}
\label{eq:PR_FOPT}
\mathcal{P}_\mathcal{R}(k)=\left(\frac{\alpha_\star}{1+\alpha_\star}\right)^{2}\!\left(\frac{H_\star}{\beta}\right)^{2}\frac{a_1(k_p/\beta)^3}{\Bigl(a_2+(k_p/\beta)^2\Bigr)^3}\Theta(H_\star-k_p),
\end{equation}
where $k_p=k/a_\star$, $a_1\simeq0.90$, $a_2\simeq0.24$. The Heaviside function $\Theta(H_\star-k_p)$ excludes sub-horizon modes whose treatment goes beyond the regime of validity of~\cite{Jinno:2024nwb,Franciolini:2025ztf}. Inserting Eq.~\eqref{eq:PR_FOPT} into Eq.~\eqref{eq:sigma_j_moments}, we compute the UCMH population and the corresponding photon and neutrino fluxes in Eq.~\eqref{eq:flux_photon_DM} under the WIMP hypothesis. Fig.~\ref{fig:FOPT} shows that the double assumption that the PTA signal originates from a FOPT, and DM is made of WIMPs, appears to be excluded by Fermi-LAT. The atmospheric neutrino constraints are evaded for $\beta/H\gtrsim 20$ and a NFW density profile.

The brown-shaded region in Fig.~\ref{fig:FOPT} marks where FOPTs do not complete. Percolation fails when the false vacuum fraction $F$, scaled by the universe volume expansion, $V_{\rm false} = a^3 F$, keeps expanding instead of decreasing at the percolation time -- when $63\%$ of the false vacuum has transitioned. Assuming a nucleation rate $\Gamma = H^4 e^{\beta (t - t_*)}$, we find that the transition does not complete if $\beta/H_\star < 1.67$. The potential for FOPTs near this no-completion boundary to produce primordial black holes through \textit{late-blooming} remains under active investigation~\cite{Kodama:1982sf,Liu:2021svg,Hashino:2021qoq,Kawana:2022olo,Lewicki:2023ioy,Gouttenoire:2023naa,Baldes:2023rqv,Gouttenoire:2023bqy,Gouttenoire:2023pxh,Jinno:2023vnr,Flores:2024lng,Lewicki:2024ghw,Ai:2024cka,Hashino:2025fse,Murai:2025hse,Zou:2025sow,Franciolini:2025ztf}.

\begin{figure}[th!]
\centering
\raisebox{0cm}{\makebox{\includegraphics[width=0.5\textwidth, scale=1]{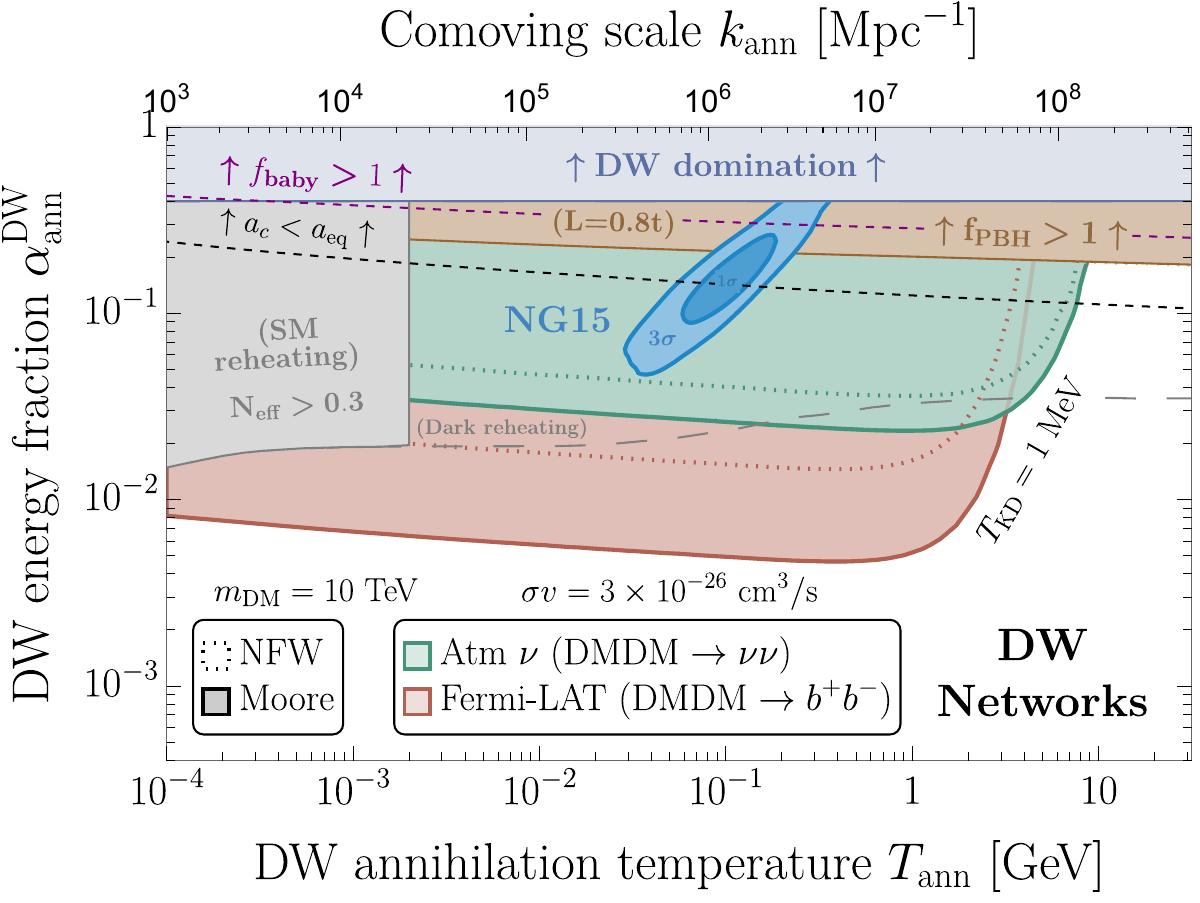}}}
\caption{The DW interpretation of the PTA signal (blue) is largely incompatible with multi-TeV WIMP DM (red+green), even if DM annihilate only into neutrinos with UCMHs modeled as a NFW density profile (dotted green). Close to the DW-domination boundary, $\alpha_{\rm ann}^{\rm DW}+ \alpha^{\rm V}_{\rm ann} >1$, DW networks can produce PBHs above the DM relic density (brown) or a baby-universe in our past-light cone (dashed purple)~\cite{Gouttenoire:2023gbn,Gouttenoire:2025ofv}.  The gray region is excluded by the $N_{\rm eff}$ constraints~\cite{Gouttenoire:2025ofv}. } 
\label{fig:DW}
\end{figure}

\textit{\textbf{Domain Wall Networks.}}
Domain walls are topological defects formed during a cosmological phase transition when the universe transitions to degenerate vacua. They are described by their surface tension $ \sigma $ and correlation length $L\simeq t/\mathcal{A}$ where $\mathcal{A} = 0.8 \pm 0.1$ \cite{Hiramatsu:2013qaa} for the $\mathbb{Z}_2$-symmetric potential. Behaving as ``2-dimensional dark energy''~\cite{Ipser:1983db}, their energy density redshift slower than radiation, $\rho_{\rm DW}=\sigma/L$ with $L \simeq t/\mathcal{A}$. In order to prevent DW to dominate the energy density of the universe, it is necessary to lift the degeneracy between the vacua by introducing an energy density difference $V_{\rm bias}$~\cite{Zeldovich:1974uw}. This generates a pressure $V_{\rm bias}$, which when larger than the pressure $C_d\, \sigma / L$ arising from the surface tension, quickly drive the DW network to annihilate after a time
\begin{equation}
    \label{eq:t_ann}
    t_{\rm ann} \simeq  C_d \,\mathcal{A}\,\frac{\sigma}{V_{\rm bias}} \;,
\end{equation}
where $C_d\sim 3$ according to lattice simulations \cite{Kawasaki:2014sqa}. The energy fraction stored in the DW surface at annihilation is 
\begin{align}
    \label{eq:alpha_DW}
     \alpha_{\rm ann}^{\rm DW} \equiv \frac{\rho_{\rm DW}}{\rho_{\rm rad}} \Big|_{t=t_{\rm ann}}
     \simeq C_d^{-1}\left( \frac{t_{\rm ann}}{t_{\rm dom }} \right)^2 \; ,
\end{align}
where $t_{\rm dom} \simeq\sqrt{3}M_{\rm pl}/2\sqrt{V_{\rm bias}}$ with $ M_{\rm pl} \simeq 2.44 \times 10^{18}~\rm GeV$, is the time when the bias energy would dominate the universe in the absence of annihilation. For $N_{\rm DW}$ discrete minima separated by a vacuum energy difference $V_{\rm bias}/(N_{\rm DW}-1)$,  the averaged vacuum energy fraction at annihilation is
\begin{equation}
        \alpha^{\rm V}_{\rm ann} \simeq \frac{(N_{\rm DW}-1)V_{\rm bias}/2}{\rho_{\rm rad}} \Big|_{t=t_{\rm ann}}\simeq \frac{N_{\rm DW}-1}{2}C_d \alpha_{\rm ann}^{\rm DW} \; .
\end{equation}
For simplicity, we assume $N_{\rm DW}\simeq 2$. DW annihilation produce GWs~\cite{Saikawa:2017hiv} and provide a valid interpretation of the PTA signal~\cite{Ferreira:2022zzo,Kitajima:2023cek,Gouttenoire:2023ftk,Servant:2023mwt}. We reproduce the Bayesian analysis of the DW interpretation of the PTA signal in \cite{Gouttenoire:2023ftk}, and we report the $1\sigma$ and $2\sigma$ posteriors in blue in Fig.~\ref{fig:DW}.

The stochasticity of the spatial distribution of the false vacuum patches on super-horizon scales source curvature perturbations~\cite{Turner:1990uw,Goetz:1990qz,Goetz:1990pj,Lola:1993qu,Lazanu:2015fua,Takahashi:2020tqv,Kitajima:2022jzz,Ramberg:2022irf,Zeng:2023jut,Lu:2024dzj}. We introduce the curvature perturbation on uniform density hypersurfaces, $\zeta \equiv -\Phi - H \delta \rho / \dot{\rho}$, where $\Phi$ is the Newtonian potential and $\delta \rho$ the energy density fluctuation. On a spatially-flat hypersurface $ \Phi = 0 $, at annihilation, we have
\begin{equation}
\label{eq:zeta_DW}
\zeta(\mathbf{x}) = \frac{\delta \rho(\mathbf{x})}{4 \rho_{\rm rad}} = \alpha^{\rm V}_{\rm ann} \frac{\phi(\mathbf{x})}{4}.
\end{equation}
where $ \phi(\mathbf{x}) = \pm 1$ is the scalar field in the false and true vacuum. 
Identifying $dr/L$ as the probability to cross a wall when moving by an infinitesimal distance $dr$, we can calculate the two-point correlation function of the scalar field~\cite{Takahashi:2020tqv,Kitajima:2022jzz}
\begin{equation}
\label{eq:phi_phi_DW}
\left< \phi(\mathbf{x}) \phi(\mathbf{y}) \right> = e^{-2r/L},
\end{equation}
where $ r = |\mathbf{x} - \mathbf{y}| $. 
From Eq.~\eqref{eq:zeta_DW} and Eq.~\eqref{eq:phi_phi_DW}, we get the dimensionless curvature power spectrum:
\begin{equation}
\label{eq:DW_curvature}
\mathcal{P}_{\zeta}(k) \equiv \int \frac{d\mathbf{x}^3 \, e^{i\mathbf{k}\cdot\mathbf{x}}\left<\zeta(0)\zeta(\mathbf{x})\right>}{2\pi^2/k^3} = \frac{\alpha_{\rm V}^2}{2\pi} \frac{(kL)^3}{\left( 4 + (kL)^2 \right)^2},
\end{equation}
to which we add a sharp cut-off $\Theta(L^{-1}-k)$ to get rid of sub-horizon modes whose description goes beyond the regime of validity of Eq.~\eqref{eq:zeta_DW}, see also~\cite{Jinno:2024nwb}. We assume that the approximation $\mathcal{R}\simeq \mathcal{\zeta}$~\cite{Baumann:2022mni} remains valid on superhorizon scales in spite of the non-adiabatic fluctuations generated by the DWs.
Plugging Eq.~\eqref{eq:DW_curvature} into Eq.~\eqref{eq:sigma_j_moments}, we deduce the UCMHs and -- assuming DM is made of WIMPs -- their associated photon and neutrino fluxes in Eq.~\eqref{eq:flux_photon_DM}. Fig.~\ref{fig:DW} shows that DW interpretation of the PTA signal is incompatible with the WIMPs hypothesis.
For large $\alpha_{\rm ann}^{\rm DW}$, \textit{late-annihilating DWs} produce large curvature perturbation leading to PBHs formation~\cite{Ferrer:2018uiu,Gelmini:2023ngs,Gelmini:2022nim,Gouttenoire:2023gbn,Gouttenoire:2023ftk,Ferreira:2024eru,Lu:2024ngi,Gouttenoire:2025ofv} and wormholes to baby-universe~\cite{Gouttenoire:2023ftk,Gouttenoire:2025ofv}. We report in brown the PBH overproduction constraints and in dashed purple, the region producing at least one baby-universe in our past light-cone $f_{\rm baby}>1$, both derived in~\cite{Gouttenoire:2023ftk}.

\textit{\textbf{Conclusion.}}
We have demonstrated that interpreting the nano-Hertz GWs background observed by pulsar timing arrays as arising from early-universe phenomena -- whether via curvature peaks, first-order phase transitions, or domain wall networks -- inevitably produces enhanced small-scale density fluctuations. These fluctuations trigger the formation of UCMHs which, if dark matter is composed of WIMPs, would yield gamma-ray and neutrino fluxes that exceed current observational limits. This conclusion remains robust over a wide range of dark matter masses, annihilation channels, and kinetic decoupling temperatures~\cite{inprep}. 

Previous studies have highlighted an incompatibility between WIMPs and solar-mass PBHs, due to the correlated formation of UCMHs and PBHs~\cite{Lacki:2010zf,Adamek:2019gns}. In addition, correlations between nanohertz gravitational waves and PBHs have been established in Refs.~\cite{Nakama:2016gzw,Chen:2019xse,Inomata:2020xad,Vaskonen:2020lbd,DeLuca:2020agl,Franciolini:2023pbf,Gouttenoire:2023bqy,Ellis:2023oxs,Lewicki:2024ghw,Ferreira:2022zzo,Kitajima:2023cek,Gouttenoire:2023ftk}. By linking nano-Hertz GWs with UCMHs, the present work completes this triangle of correlations -- see also~\cite{Liu:2023tmv} for an initial investigation.

Another PTA interpretation enhancing structure formation is supermassive primordial black holes~\cite{Depta:2023qst,Gouttenoire:2023nzr}. Looking ahead, future work should explore UCMH formation from other PTA signal interpretations -- including cosmic strings~\cite{Ellis:2020ena,Blasi:2020mfx,Buchmuller:2020lbh,Buchmuller:2023aus,Ellis:2023tsl,Kume:2024adn}, inflationary tensor modes~\cite{Vagnozzi:2023lwo}, axion–gauge field dynamics~\cite{Ratzinger:2020koh,Madge:2023dxc,Geller:2023shn,Murai:2023gkv,Unal:2023srk}, or curvature peaks with large non-Gaussianities~\cite{Nakama:2016kfq,Hooper:2023nnl}. Future studies should also investigate alternative probes of UCMHs arising from primordial interpretations of the PTA, that do not rely on the WIMP hypothesis, such as stellar dynamics~\cite{Buschmann:2017ams}, CMB observations~\cite{Kawasaki:2021yek}, gravitational lensing~\cite{Delos:2023fpm}, or direct gravitational-wave detection with pulsar timing arrays~\cite{Lee:2020wfn}.

{\bf Acknowledgements.}---%

The author thanks Sten Delos, Heling Deng, Gabriele Franciolini, Ryusuke Jinno, Sokratis Trifinopoulos and Miguel Vanvlasselaer for very useful discussions.
The author acknowledge support by the Cluster of Excellence “Precision Physics, Fundamental In- teractions, and Structure of Matter” (PRISMA+ EXC 2118/1) funded by the Deutsche Forschungsgemeinschaft (DFG, German Research Foundation) within the German Excel- lence Strategy (Project No. 390831469).
The author has been partly funded by the Azrieli Foundation with the award of an Azrieli Fellowship.
\bibliography{biblio}

\begin{thebibliography}{163}%
\makeatletter
\providecommand \@ifxundefined [1]{%
 \@ifx{#1\undefined}
}%
\providecommand \@ifnum [1]{%
 \ifnum #1\expandafter \@firstoftwo
 \else \expandafter \@secondoftwo
 \fi
}%
\providecommand \@ifx [1]{%
 \ifx #1\expandafter \@firstoftwo
 \else \expandafter \@secondoftwo
 \fi
}%
\providecommand \natexlab [1]{#1}%
\providecommand \enquote  [1]{``#1''}%
\providecommand \bibnamefont  [1]{#1}%
\providecommand \bibfnamefont [1]{#1}%
\providecommand \citenamefont [1]{#1}%
\providecommand \href@noop [0]{\@secondoftwo}%
\providecommand \href [0]{\begingroup \@sanitize@url \@href}%
\providecommand \@href[1]{\@@startlink{#1}\@@href}%
\providecommand \@@href[1]{\endgroup#1\@@endlink}%
\providecommand \@sanitize@url [0]{\catcode `\\12\catcode `\$12\catcode
  `\&12\catcode `\#12\catcode `\^12\catcode `\_12\catcode `\%12\relax}%
\providecommand \@@startlink[1]{}%
\providecommand \@@endlink[0]{}%
\providecommand \url  [0]{\begingroup\@sanitize@url \@url }%
\providecommand \@url [1]{\endgroup\@href {#1}{\urlprefix }}%
\providecommand \urlprefix  [0]{URL }%
\providecommand \Eprint [0]{\href }%
\providecommand \doibase [0]{http://dx.doi.org/}%
\providecommand \selectlanguage [0]{\@gobble}%
\providecommand \bibinfo  [0]{\@secondoftwo}%
\providecommand \bibfield  [0]{\@secondoftwo}%
\providecommand \translation [1]{[#1]}%
\providecommand \BibitemOpen [0]{}%
\providecommand \bibitemStop [0]{}%
\providecommand \bibitemNoStop [0]{.\EOS\space}%
\providecommand \EOS [0]{\spacefactor3000\relax}%
\providecommand \BibitemShut  [1]{\csname bibitem#1\endcsname}%
\let\auto@bib@innerbib\@empty
\bibitem [{\citenamefont {Agazie}\ \emph {et~al.}(2023)\citenamefont {Agazie}
  \emph {et~al.}}]{NANOGrav:2023gor}%
  \BibitemOpen
  \bibfield  {author} {\bibinfo {author} {\bibfnamefont {G.}~\bibnamefont
  {Agazie}} \emph {et~al.} (\bibinfo {collaboration} {NANOGrav}),\ }\href
  {\doibase 10.3847/2041-8213/acdac6} {\bibfield  {journal} {\bibinfo
  {journal} {Astrophys. J. Lett.}\ }\textbf {\bibinfo {volume} {951}},\
  \bibinfo {pages} {L8} (\bibinfo {year} {2023})},\ \Eprint
  {http://arxiv.org/abs/2306.16213} {arXiv:2306.16213 [astro-ph.HE]}
  \BibitemShut {NoStop}%
\bibitem [{\citenamefont {Antoniadis}\ \emph {et~al.}(2023)\citenamefont
  {Antoniadis} \emph {et~al.}}]{Antoniadis:2023rey}%
  \BibitemOpen
  \bibfield  {author} {\bibinfo {author} {\bibfnamefont {J.}~\bibnamefont
  {Antoniadis}} \emph {et~al.} (\bibinfo {collaboration} {EPTA, InPTA:}),\
  }\href {\doibase 10.1051/0004-6361/202346844} {\bibfield  {journal} {\bibinfo
   {journal} {Astron. Astrophys.}\ }\textbf {\bibinfo {volume} {678}},\
  \bibinfo {pages} {A50} (\bibinfo {year} {2023})},\ \Eprint
  {http://arxiv.org/abs/2306.16214} {arXiv:2306.16214 [astro-ph.HE]}
  \BibitemShut {NoStop}%
\bibitem [{\citenamefont {Reardon}\ \emph {et~al.}(2023)\citenamefont {Reardon}
  \emph {et~al.}}]{Reardon:2023gzh}%
  \BibitemOpen
  \bibfield  {author} {\bibinfo {author} {\bibfnamefont {D.~J.}\ \bibnamefont
  {Reardon}} \emph {et~al.},\ }\href {\doibase 10.3847/2041-8213/acdd02}
  {\bibfield  {journal} {\bibinfo  {journal} {Astrophys. J. Lett.}\ }\textbf
  {\bibinfo {volume} {951}},\ \bibinfo {pages} {L6} (\bibinfo {year} {2023})},\
  \Eprint {http://arxiv.org/abs/2306.16215} {arXiv:2306.16215 [astro-ph.HE]}
  \BibitemShut {NoStop}%
\bibitem [{\citenamefont {Xu}\ \emph {et~al.}(2023)\citenamefont {Xu} \emph
  {et~al.}}]{Xu:2023wog}%
  \BibitemOpen
  \bibfield  {author} {\bibinfo {author} {\bibfnamefont {H.}~\bibnamefont {Xu}}
  \emph {et~al.},\ }\href {\doibase 10.1088/1674-4527/acdfa5} {\bibfield
  {journal} {\bibinfo  {journal} {Res. Astron. Astrophys.}\ }\textbf {\bibinfo
  {volume} {23}},\ \bibinfo {pages} {075024} (\bibinfo {year} {2023})},\
  \Eprint {http://arxiv.org/abs/2306.16216} {arXiv:2306.16216 [astro-ph.HE]}
  \BibitemShut {NoStop}%
\bibitem [{\citenamefont {Agazie}\ \emph {et~al.}(2024)\citenamefont {Agazie}
  \emph {et~al.}}]{InternationalPulsarTimingArray:2023mzf}%
  \BibitemOpen
  \bibfield  {author} {\bibinfo {author} {\bibfnamefont {G.}~\bibnamefont
  {Agazie}} \emph {et~al.} (\bibinfo {collaboration} {International Pulsar
  Timing Array}),\ }\href {\doibase 10.3847/1538-4357/ad36be} {\bibfield
  {journal} {\bibinfo  {journal} {Astrophys. J.}\ }\textbf {\bibinfo {volume}
  {966}},\ \bibinfo {pages} {105} (\bibinfo {year} {2024})},\ \Eprint
  {http://arxiv.org/abs/2309.00693} {arXiv:2309.00693 [astro-ph.HE]}
  \BibitemShut {NoStop}%
\bibitem [{\citenamefont {Giblin}\ and\ \citenamefont
  {Thrane}(2014)}]{Giblin:2014gra}%
  \BibitemOpen
  \bibfield  {author} {\bibinfo {author} {\bibfnamefont {J.~T.}\ \bibnamefont
  {Giblin}}\ and\ \bibinfo {author} {\bibfnamefont {E.}~\bibnamefont
  {Thrane}},\ }\href {\doibase 10.1103/PhysRevD.90.107502} {\bibfield
  {journal} {\bibinfo  {journal} {Phys. Rev. D}\ }\textbf {\bibinfo {volume}
  {90}},\ \bibinfo {pages} {107502} (\bibinfo {year} {2014})},\ \Eprint
  {http://arxiv.org/abs/1410.4779} {arXiv:1410.4779 [gr-qc]} \BibitemShut
  {NoStop}%
\bibitem [{\citenamefont {Ratzinger}\ and\ \citenamefont
  {Schwaller}(2021)}]{Ratzinger:2020koh}%
  \BibitemOpen
  \bibfield  {author} {\bibinfo {author} {\bibfnamefont {W.}~\bibnamefont
  {Ratzinger}}\ and\ \bibinfo {author} {\bibfnamefont {P.}~\bibnamefont
  {Schwaller}},\ }\href {\doibase 10.21468/SciPostPhys.10.2.047} {\bibfield
  {journal} {\bibinfo  {journal} {SciPost Phys.}\ }\textbf {\bibinfo {volume}
  {10}},\ \bibinfo {pages} {047} (\bibinfo {year} {2021})},\ \Eprint
  {http://arxiv.org/abs/2009.11875} {arXiv:2009.11875 [astro-ph.CO]}
  \BibitemShut {NoStop}%
\bibitem [{\citenamefont {Madge}\ \emph {et~al.}(2023)\citenamefont {Madge},
  \citenamefont {Morgante}, \citenamefont {Puchades-Ib\'a\~nez}, \citenamefont
  {Ramberg}, \citenamefont {Ratzinger}, \citenamefont {Schenk},\ and\
  \citenamefont {Schwaller}}]{Madge:2023dxc}%
  \BibitemOpen
  \bibfield  {author} {\bibinfo {author} {\bibfnamefont {E.}~\bibnamefont
  {Madge}}, \bibinfo {author} {\bibfnamefont {E.}~\bibnamefont {Morgante}},
  \bibinfo {author} {\bibfnamefont {C.}~\bibnamefont {Puchades-Ib\'a\~nez}},
  \bibinfo {author} {\bibfnamefont {N.}~\bibnamefont {Ramberg}}, \bibinfo
  {author} {\bibfnamefont {W.}~\bibnamefont {Ratzinger}}, \bibinfo {author}
  {\bibfnamefont {S.}~\bibnamefont {Schenk}}, \ and\ \bibinfo {author}
  {\bibfnamefont {P.}~\bibnamefont {Schwaller}},\ }\href {\doibase
  10.1007/JHEP10(2023)171} {\bibfield  {journal} {\bibinfo  {journal} {JHEP}\
  }\textbf {\bibinfo {volume} {10}},\ \bibinfo {pages} {171} (\bibinfo {year}
  {2023})},\ \Eprint {http://arxiv.org/abs/2306.14856} {arXiv:2306.14856
  [hep-ph]} \BibitemShut {NoStop}%
\bibitem [{\citenamefont {Afzal}\ \emph {et~al.}(2023)\citenamefont {Afzal}
  \emph {et~al.}}]{NANOGrav:2023hvm}%
  \BibitemOpen
  \bibfield  {author} {\bibinfo {author} {\bibfnamefont {A.}~\bibnamefont
  {Afzal}} \emph {et~al.} (\bibinfo {collaboration} {NANOGrav}),\ }\href
  {\doibase 10.3847/2041-8213/acdc91} {\bibfield  {journal} {\bibinfo
  {journal} {Astrophys. J. Lett.}\ }\textbf {\bibinfo {volume} {951}},\
  \bibinfo {pages} {L11} (\bibinfo {year} {2023})},\ \bibinfo {note} {[Erratum:
  Astrophys.J.Lett. 971, L27 (2024), Erratum: Astrophys.J. 971, L27 (2024)]},\
  \Eprint {http://arxiv.org/abs/2306.16219} {arXiv:2306.16219 [astro-ph.HE]}
  \BibitemShut {NoStop}%
\bibitem [{\citenamefont {Antoniadis}\ \emph {et~al.}(2024)\citenamefont
  {Antoniadis} \emph {et~al.}}]{EPTA:2023xxk}%
  \BibitemOpen
  \bibfield  {author} {\bibinfo {author} {\bibfnamefont {J.}~\bibnamefont
  {Antoniadis}} \emph {et~al.} (\bibinfo {collaboration} {EPTA, InPTA}),\
  }\href {\doibase 10.1051/0004-6361/202347433} {\bibfield  {journal} {\bibinfo
   {journal} {Astron. Astrophys.}\ }\textbf {\bibinfo {volume} {685}},\
  \bibinfo {pages} {A94} (\bibinfo {year} {2024})},\ \Eprint
  {http://arxiv.org/abs/2306.16227} {arXiv:2306.16227 [astro-ph.CO]}
  \BibitemShut {NoStop}%
\bibitem [{\citenamefont {Bian}\ \emph {et~al.}(2024)\citenamefont {Bian},
  \citenamefont {Ge}, \citenamefont {Shu}, \citenamefont {Wang}, \citenamefont
  {Yang},\ and\ \citenamefont {Zong}}]{Bian:2023dnv}%
  \BibitemOpen
  \bibfield  {author} {\bibinfo {author} {\bibfnamefont {L.}~\bibnamefont
  {Bian}}, \bibinfo {author} {\bibfnamefont {S.}~\bibnamefont {Ge}}, \bibinfo
  {author} {\bibfnamefont {J.}~\bibnamefont {Shu}}, \bibinfo {author}
  {\bibfnamefont {B.}~\bibnamefont {Wang}}, \bibinfo {author} {\bibfnamefont
  {X.-Y.}\ \bibnamefont {Yang}}, \ and\ \bibinfo {author} {\bibfnamefont
  {J.}~\bibnamefont {Zong}},\ }\href {\doibase 10.1103/PhysRevD.109.L101301}
  {\bibfield  {journal} {\bibinfo  {journal} {Phys. Rev. D}\ }\textbf {\bibinfo
  {volume} {109}},\ \bibinfo {pages} {L101301} (\bibinfo {year} {2024})},\
  \Eprint {http://arxiv.org/abs/2307.02376} {arXiv:2307.02376 [astro-ph.HE]}
  \BibitemShut {NoStop}%
\bibitem [{\citenamefont {Figueroa}\ \emph {et~al.}(2024)\citenamefont
  {Figueroa}, \citenamefont {Pieroni}, \citenamefont {Ricciardone},\ and\
  \citenamefont {Simakachorn}}]{Figueroa:2023zhu}%
  \BibitemOpen
  \bibfield  {author} {\bibinfo {author} {\bibfnamefont {D.~G.}\ \bibnamefont
  {Figueroa}}, \bibinfo {author} {\bibfnamefont {M.}~\bibnamefont {Pieroni}},
  \bibinfo {author} {\bibfnamefont {A.}~\bibnamefont {Ricciardone}}, \ and\
  \bibinfo {author} {\bibfnamefont {P.}~\bibnamefont {Simakachorn}},\ }\href
  {\doibase 10.1103/PhysRevLett.132.171002} {\bibfield  {journal} {\bibinfo
  {journal} {Phys. Rev. Lett.}\ }\textbf {\bibinfo {volume} {132}},\ \bibinfo
  {pages} {171002} (\bibinfo {year} {2024})},\ \Eprint
  {http://arxiv.org/abs/2307.02399} {arXiv:2307.02399 [astro-ph.CO]}
  \BibitemShut {NoStop}%
\bibitem [{\citenamefont {Ellis}\ \emph {et~al.}(2024)\citenamefont {Ellis},
  \citenamefont {Fairbairn}, \citenamefont {Franciolini}, \citenamefont
  {H\"utsi}, \citenamefont {Iovino}, \citenamefont {Lewicki}, \citenamefont
  {Raidal}, \citenamefont {Urrutia}, \citenamefont {Vaskonen},\ and\
  \citenamefont {Veerm\"ae}}]{Ellis:2023oxs}%
  \BibitemOpen
  \bibfield  {author} {\bibinfo {author} {\bibfnamefont {J.}~\bibnamefont
  {Ellis}}, \bibinfo {author} {\bibfnamefont {M.}~\bibnamefont {Fairbairn}},
  \bibinfo {author} {\bibfnamefont {G.}~\bibnamefont {Franciolini}}, \bibinfo
  {author} {\bibfnamefont {G.}~\bibnamefont {H\"utsi}}, \bibinfo {author}
  {\bibfnamefont {A.}~\bibnamefont {Iovino}}, \bibinfo {author} {\bibfnamefont
  {M.}~\bibnamefont {Lewicki}}, \bibinfo {author} {\bibfnamefont
  {M.}~\bibnamefont {Raidal}}, \bibinfo {author} {\bibfnamefont
  {J.}~\bibnamefont {Urrutia}}, \bibinfo {author} {\bibfnamefont
  {V.}~\bibnamefont {Vaskonen}}, \ and\ \bibinfo {author} {\bibfnamefont
  {H.}~\bibnamefont {Veerm\"ae}},\ }\href {\doibase
  10.1103/PhysRevD.109.023522} {\bibfield  {journal} {\bibinfo  {journal}
  {Phys. Rev. D}\ }\textbf {\bibinfo {volume} {109}},\ \bibinfo {pages}
  {023522} (\bibinfo {year} {2024})},\ \Eprint
  {http://arxiv.org/abs/2308.08546} {arXiv:2308.08546 [astro-ph.CO]}
  \BibitemShut {NoStop}%
\bibitem [{\citenamefont {Nakama}\ \emph {et~al.}(2017)\citenamefont {Nakama},
  \citenamefont {Silk},\ and\ \citenamefont {Kamionkowski}}]{Nakama:2016gzw}%
  \BibitemOpen
  \bibfield  {author} {\bibinfo {author} {\bibfnamefont {T.}~\bibnamefont
  {Nakama}}, \bibinfo {author} {\bibfnamefont {J.}~\bibnamefont {Silk}}, \ and\
  \bibinfo {author} {\bibfnamefont {M.}~\bibnamefont {Kamionkowski}},\ }\href
  {\doibase 10.1103/PhysRevD.95.043511} {\bibfield  {journal} {\bibinfo
  {journal} {Phys. Rev. D}\ }\textbf {\bibinfo {volume} {95}},\ \bibinfo
  {pages} {043511} (\bibinfo {year} {2017})},\ \Eprint
  {http://arxiv.org/abs/1612.06264} {arXiv:1612.06264 [astro-ph.CO]}
  \BibitemShut {NoStop}%
\bibitem [{\citenamefont {Chen}\ \emph {et~al.}(2020)\citenamefont {Chen},
  \citenamefont {Yuan},\ and\ \citenamefont {Huang}}]{Chen:2019xse}%
  \BibitemOpen
  \bibfield  {author} {\bibinfo {author} {\bibfnamefont {Z.-C.}\ \bibnamefont
  {Chen}}, \bibinfo {author} {\bibfnamefont {C.}~\bibnamefont {Yuan}}, \ and\
  \bibinfo {author} {\bibfnamefont {Q.-G.}\ \bibnamefont {Huang}},\ }\href
  {\doibase 10.1103/PhysRevLett.124.251101} {\bibfield  {journal} {\bibinfo
  {journal} {Phys. Rev. Lett.}\ }\textbf {\bibinfo {volume} {124}},\ \bibinfo
  {pages} {251101} (\bibinfo {year} {2020})},\ \Eprint
  {http://arxiv.org/abs/1910.12239} {arXiv:1910.12239 [astro-ph.CO]}
  \BibitemShut {NoStop}%
\bibitem [{\citenamefont {Inomata}\ \emph {et~al.}(2021)\citenamefont
  {Inomata}, \citenamefont {Kawasaki}, \citenamefont {Mukaida},\ and\
  \citenamefont {Yanagida}}]{Inomata:2020xad}%
  \BibitemOpen
  \bibfield  {author} {\bibinfo {author} {\bibfnamefont {K.}~\bibnamefont
  {Inomata}}, \bibinfo {author} {\bibfnamefont {M.}~\bibnamefont {Kawasaki}},
  \bibinfo {author} {\bibfnamefont {K.}~\bibnamefont {Mukaida}}, \ and\
  \bibinfo {author} {\bibfnamefont {T.~T.}\ \bibnamefont {Yanagida}},\ }\href
  {\doibase 10.1103/PhysRevLett.126.131301} {\bibfield  {journal} {\bibinfo
  {journal} {Phys. Rev. Lett.}\ }\textbf {\bibinfo {volume} {126}},\ \bibinfo
  {pages} {131301} (\bibinfo {year} {2021})},\ \Eprint
  {http://arxiv.org/abs/2011.01270} {arXiv:2011.01270 [astro-ph.CO]}
  \BibitemShut {NoStop}%
\bibitem [{\citenamefont {Vaskonen}\ and\ \citenamefont
  {Veerm\"ae}(2021)}]{Vaskonen:2020lbd}%
  \BibitemOpen
  \bibfield  {author} {\bibinfo {author} {\bibfnamefont {V.}~\bibnamefont
  {Vaskonen}}\ and\ \bibinfo {author} {\bibfnamefont {H.}~\bibnamefont
  {Veerm\"ae}},\ }\href {\doibase 10.1103/PhysRevLett.126.051303} {\bibfield
  {journal} {\bibinfo  {journal} {Phys. Rev. Lett.}\ }\textbf {\bibinfo
  {volume} {126}},\ \bibinfo {pages} {051303} (\bibinfo {year} {2021})},\
  \Eprint {http://arxiv.org/abs/2009.07832} {arXiv:2009.07832 [astro-ph.CO]}
  \BibitemShut {NoStop}%
\bibitem [{\citenamefont {De~Luca}\ \emph {et~al.}(2021)\citenamefont
  {De~Luca}, \citenamefont {Franciolini},\ and\ \citenamefont
  {Riotto}}]{DeLuca:2020agl}%
  \BibitemOpen
  \bibfield  {author} {\bibinfo {author} {\bibfnamefont {V.}~\bibnamefont
  {De~Luca}}, \bibinfo {author} {\bibfnamefont {G.}~\bibnamefont
  {Franciolini}}, \ and\ \bibinfo {author} {\bibfnamefont {A.}~\bibnamefont
  {Riotto}},\ }\href {\doibase 10.1103/PhysRevLett.126.041303} {\bibfield
  {journal} {\bibinfo  {journal} {Phys. Rev. Lett.}\ }\textbf {\bibinfo
  {volume} {126}},\ \bibinfo {pages} {041303} (\bibinfo {year} {2021})},\
  \Eprint {http://arxiv.org/abs/2009.08268} {arXiv:2009.08268 [astro-ph.CO]}
  \BibitemShut {NoStop}%
\bibitem [{\citenamefont {Franciolini}\ \emph {et~al.}(2023)\citenamefont
  {Franciolini}, \citenamefont {Iovino}, \citenamefont {Vaskonen},\ and\
  \citenamefont {Veermae}}]{Franciolini:2023pbf}%
  \BibitemOpen
  \bibfield  {author} {\bibinfo {author} {\bibfnamefont {G.}~\bibnamefont
  {Franciolini}}, \bibinfo {author} {\bibfnamefont {A.}~\bibnamefont {Iovino},
  \bibfnamefont {Junior.}}, \bibinfo {author} {\bibfnamefont {V.}~\bibnamefont
  {Vaskonen}}, \ and\ \bibinfo {author} {\bibfnamefont {H.}~\bibnamefont
  {Veermae}},\ }\href {\doibase 10.1103/PhysRevLett.131.201401} {\bibfield
  {journal} {\bibinfo  {journal} {Phys. Rev. Lett.}\ }\textbf {\bibinfo
  {volume} {131}},\ \bibinfo {pages} {201401} (\bibinfo {year} {2023})},\
  \Eprint {http://arxiv.org/abs/2306.17149} {arXiv:2306.17149 [astro-ph.CO]}
  \BibitemShut {NoStop}%
\bibitem [{\citenamefont {Gouttenoire}(2023)}]{Gouttenoire:2023bqy}%
  \BibitemOpen
  \bibfield  {author} {\bibinfo {author} {\bibfnamefont {Y.}~\bibnamefont
  {Gouttenoire}},\ }\href {\doibase 10.1103/PhysRevLett.131.171404} {\bibfield
  {journal} {\bibinfo  {journal} {Phys. Rev. Lett.}\ }\textbf {\bibinfo
  {volume} {131}},\ \bibinfo {pages} {171404} (\bibinfo {year} {2023})},\
  \Eprint {http://arxiv.org/abs/2307.04239} {arXiv:2307.04239 [hep-ph]}
  \BibitemShut {NoStop}%
\bibitem [{\citenamefont {Lewicki}\ \emph
  {et~al.}(2024{\natexlab{a}})\citenamefont {Lewicki}, \citenamefont {Toczek},\
  and\ \citenamefont {Vaskonen}}]{Lewicki:2024ghw}%
  \BibitemOpen
  \bibfield  {author} {\bibinfo {author} {\bibfnamefont {M.}~\bibnamefont
  {Lewicki}}, \bibinfo {author} {\bibfnamefont {P.}~\bibnamefont {Toczek}}, \
  and\ \bibinfo {author} {\bibfnamefont {V.}~\bibnamefont {Vaskonen}},\ }\href
  {\doibase 10.1103/PhysRevLett.133.221003} {\bibfield  {journal} {\bibinfo
  {journal} {Phys. Rev. Lett.}\ }\textbf {\bibinfo {volume} {133}},\ \bibinfo
  {pages} {221003} (\bibinfo {year} {2024}{\natexlab{a}})},\ \Eprint
  {http://arxiv.org/abs/2402.04158} {arXiv:2402.04158 [astro-ph.CO]}
  \BibitemShut {NoStop}%
\bibitem [{\citenamefont {Ferreira}\ \emph {et~al.}(2023)\citenamefont
  {Ferreira}, \citenamefont {Notari}, \citenamefont {Pujolas},\ and\
  \citenamefont {Rompineve}}]{Ferreira:2022zzo}%
  \BibitemOpen
  \bibfield  {author} {\bibinfo {author} {\bibfnamefont {R.~Z.}\ \bibnamefont
  {Ferreira}}, \bibinfo {author} {\bibfnamefont {A.}~\bibnamefont {Notari}},
  \bibinfo {author} {\bibfnamefont {O.}~\bibnamefont {Pujolas}}, \ and\
  \bibinfo {author} {\bibfnamefont {F.}~\bibnamefont {Rompineve}},\ }\href
  {\doibase 10.1088/1475-7516/2023/02/001} {\bibfield  {journal} {\bibinfo
  {journal} {JCAP}\ }\textbf {\bibinfo {volume} {02}},\ \bibinfo {pages} {001}
  (\bibinfo {year} {2023})},\ \Eprint {http://arxiv.org/abs/2204.04228}
  {arXiv:2204.04228 [astro-ph.CO]} \BibitemShut {NoStop}%
\bibitem [{\citenamefont {Kitajima}\ \emph {et~al.}(2024)\citenamefont
  {Kitajima}, \citenamefont {Lee}, \citenamefont {Murai}, \citenamefont
  {Takahashi},\ and\ \citenamefont {Yin}}]{Kitajima:2023cek}%
  \BibitemOpen
  \bibfield  {author} {\bibinfo {author} {\bibfnamefont {N.}~\bibnamefont
  {Kitajima}}, \bibinfo {author} {\bibfnamefont {J.}~\bibnamefont {Lee}},
  \bibinfo {author} {\bibfnamefont {K.}~\bibnamefont {Murai}}, \bibinfo
  {author} {\bibfnamefont {F.}~\bibnamefont {Takahashi}}, \ and\ \bibinfo
  {author} {\bibfnamefont {W.}~\bibnamefont {Yin}},\ }\href {\doibase
  10.1016/j.physletb.2024.138586} {\bibfield  {journal} {\bibinfo  {journal}
  {Phys. Lett. B}\ }\textbf {\bibinfo {volume} {851}},\ \bibinfo {pages}
  {138586} (\bibinfo {year} {2024})},\ \Eprint
  {http://arxiv.org/abs/2306.17146} {arXiv:2306.17146 [hep-ph]} \BibitemShut
  {NoStop}%
\bibitem [{\citenamefont {Gouttenoire}\ and\ \citenamefont
  {Vitagliano}(2024{\natexlab{a}})}]{Gouttenoire:2023ftk}%
  \BibitemOpen
  \bibfield  {author} {\bibinfo {author} {\bibfnamefont {Y.}~\bibnamefont
  {Gouttenoire}}\ and\ \bibinfo {author} {\bibfnamefont {E.}~\bibnamefont
  {Vitagliano}},\ }\href {\doibase 10.1103/PhysRevD.110.L061306} {\bibfield
  {journal} {\bibinfo  {journal} {Phys. Rev. D}\ }\textbf {\bibinfo {volume}
  {110}},\ \bibinfo {pages} {L061306} (\bibinfo {year} {2024}{\natexlab{a}})},\
  \Eprint {http://arxiv.org/abs/2306.17841} {arXiv:2306.17841 [gr-qc]}
  \BibitemShut {NoStop}%
\bibitem [{\citenamefont {Ricotti}\ and\ \citenamefont
  {Gould}(2009)}]{Ricotti:2009bs}%
  \BibitemOpen
  \bibfield  {author} {\bibinfo {author} {\bibfnamefont {M.}~\bibnamefont
  {Ricotti}}\ and\ \bibinfo {author} {\bibfnamefont {A.}~\bibnamefont
  {Gould}},\ }\href {\doibase 10.1088/0004-637X/707/2/979} {\bibfield
  {journal} {\bibinfo  {journal} {Astrophys. J.}\ }\textbf {\bibinfo {volume}
  {707}},\ \bibinfo {pages} {979} (\bibinfo {year} {2009})},\ \Eprint
  {http://arxiv.org/abs/0908.0735} {arXiv:0908.0735 [astro-ph.CO]} \BibitemShut
  {NoStop}%
\bibitem [{\citenamefont {Diemand}\ \emph {et~al.}(2005)\citenamefont
  {Diemand}, \citenamefont {Moore},\ and\ \citenamefont
  {Stadel}}]{Diemand:2005vz}%
  \BibitemOpen
  \bibfield  {author} {\bibinfo {author} {\bibfnamefont {J.}~\bibnamefont
  {Diemand}}, \bibinfo {author} {\bibfnamefont {B.}~\bibnamefont {Moore}}, \
  and\ \bibinfo {author} {\bibfnamefont {J.}~\bibnamefont {Stadel}},\ }\href
  {\doibase 10.1038/nature03270} {\bibfield  {journal} {\bibinfo  {journal}
  {Nature}\ }\textbf {\bibinfo {volume} {433}},\ \bibinfo {pages} {389}
  (\bibinfo {year} {2005})},\ \Eprint {http://arxiv.org/abs/astro-ph/0501589}
  {arXiv:astro-ph/0501589} \BibitemShut {NoStop}%
\bibitem [{\citenamefont {Zhao}\ \emph {et~al.}(2005)\citenamefont {Zhao},
  \citenamefont {Taylor}, \citenamefont {Silk},\ and\ \citenamefont
  {Hooper}}]{Zhao:2005py}%
  \BibitemOpen
  \bibfield  {author} {\bibinfo {author} {\bibfnamefont {H.-S.}\ \bibnamefont
  {Zhao}}, \bibinfo {author} {\bibfnamefont {J.}~\bibnamefont {Taylor}},
  \bibinfo {author} {\bibfnamefont {J.}~\bibnamefont {Silk}}, \ and\ \bibinfo
  {author} {\bibfnamefont {D.}~\bibnamefont {Hooper}},\ }\href@noop {} {\
  (\bibinfo {year} {2005})},\ \Eprint {http://arxiv.org/abs/astro-ph/0502049}
  {arXiv:astro-ph/0502049} \BibitemShut {NoStop}%
\bibitem [{\citenamefont {Moore}\ \emph {et~al.}(2005)\citenamefont {Moore},
  \citenamefont {Diemand}, \citenamefont {Stadel},\ and\ \citenamefont
  {Quinn}}]{Moore:2005uu}%
  \BibitemOpen
  \bibfield  {author} {\bibinfo {author} {\bibfnamefont {B.}~\bibnamefont
  {Moore}}, \bibinfo {author} {\bibfnamefont {J.}~\bibnamefont {Diemand}},
  \bibinfo {author} {\bibfnamefont {J.}~\bibnamefont {Stadel}}, \ and\ \bibinfo
  {author} {\bibfnamefont {T.~R.}\ \bibnamefont {Quinn}},\ }\href@noop {} {\
  (\bibinfo {year} {2005})},\ \Eprint {http://arxiv.org/abs/astro-ph/0502213}
  {arXiv:astro-ph/0502213} \BibitemShut {NoStop}%
\bibitem [{\citenamefont {Green}\ \emph {et~al.}(2005)\citenamefont {Green},
  \citenamefont {Hofmann},\ and\ \citenamefont {Schwarz}}]{Green:2005fa}%
  \BibitemOpen
  \bibfield  {author} {\bibinfo {author} {\bibfnamefont {A.~M.}\ \bibnamefont
  {Green}}, \bibinfo {author} {\bibfnamefont {S.}~\bibnamefont {Hofmann}}, \
  and\ \bibinfo {author} {\bibfnamefont {D.~J.}\ \bibnamefont {Schwarz}},\
  }\href {\doibase 10.1088/1475-7516/2005/08/003} {\bibfield  {journal}
  {\bibinfo  {journal} {JCAP}\ }\textbf {\bibinfo {volume} {08}},\ \bibinfo
  {pages} {003} (\bibinfo {year} {2005})},\ \Eprint
  {http://arxiv.org/abs/astro-ph/0503387} {arXiv:astro-ph/0503387} \BibitemShut
  {NoStop}%
\bibitem [{\citenamefont {Loeb}\ and\ \citenamefont
  {Zaldarriaga}(2005)}]{Loeb:2005pm}%
  \BibitemOpen
  \bibfield  {author} {\bibinfo {author} {\bibfnamefont {A.}~\bibnamefont
  {Loeb}}\ and\ \bibinfo {author} {\bibfnamefont {M.}~\bibnamefont
  {Zaldarriaga}},\ }\href {\doibase 10.1103/PhysRevD.71.103520} {\bibfield
  {journal} {\bibinfo  {journal} {Phys. Rev. D}\ }\textbf {\bibinfo {volume}
  {71}},\ \bibinfo {pages} {103520} (\bibinfo {year} {2005})},\ \Eprint
  {http://arxiv.org/abs/astro-ph/0504112} {arXiv:astro-ph/0504112} \BibitemShut
  {NoStop}%
\bibitem [{\citenamefont {Delos}\ and\ \citenamefont
  {White}(2022)}]{Delos:2022yhn}%
  \BibitemOpen
  \bibfield  {author} {\bibinfo {author} {\bibfnamefont {M.~S.}\ \bibnamefont
  {Delos}}\ and\ \bibinfo {author} {\bibfnamefont {S.~D.~M.}\ \bibnamefont
  {White}},\ }\href {\doibase 10.1093/mnras/stac3373} {\bibfield  {journal}
  {\bibinfo  {journal} {Mon. Not. Roy. Astron. Soc.}\ }\textbf {\bibinfo
  {volume} {518}},\ \bibinfo {pages} {3509} (\bibinfo {year} {2022})},\ \Eprint
  {http://arxiv.org/abs/2207.05082} {arXiv:2207.05082 [astro-ph.CO]}
  \BibitemShut {NoStop}%
\bibitem [{\citenamefont {Cirelli}\ \emph {et~al.}(2024)\citenamefont
  {Cirelli}, \citenamefont {Strumia},\ and\ \citenamefont
  {Zupan}}]{Cirelli:2024ssz}%
  \BibitemOpen
  \bibfield  {author} {\bibinfo {author} {\bibfnamefont {M.}~\bibnamefont
  {Cirelli}}, \bibinfo {author} {\bibfnamefont {A.}~\bibnamefont {Strumia}}, \
  and\ \bibinfo {author} {\bibfnamefont {J.}~\bibnamefont {Zupan}},\
  }\href@noop {} {\  (\bibinfo {year} {2024})},\ \Eprint
  {http://arxiv.org/abs/2406.01705} {arXiv:2406.01705 [hep-ph]} \BibitemShut
  {NoStop}%
\bibitem [{\citenamefont {Scott}\ and\ \citenamefont
  {Sivertsson}(2009)}]{Scott:2009tu}%
  \BibitemOpen
  \bibfield  {author} {\bibinfo {author} {\bibfnamefont {P.}~\bibnamefont
  {Scott}}\ and\ \bibinfo {author} {\bibfnamefont {S.}~\bibnamefont
  {Sivertsson}},\ }\href {\doibase 10.1103/PhysRevLett.103.211301} {\bibfield
  {journal} {\bibinfo  {journal} {Phys. Rev. Lett.}\ }\textbf {\bibinfo
  {volume} {103}},\ \bibinfo {pages} {211301} (\bibinfo {year} {2009})},\
  \bibinfo {note} {[Erratum: Phys.Rev.Lett. 105, 119902 (2010)]},\ \Eprint
  {http://arxiv.org/abs/0908.4082} {arXiv:0908.4082 [astro-ph.CO]} \BibitemShut
  {NoStop}%
\bibitem [{\citenamefont {Josan}\ and\ \citenamefont
  {Green}(2010)}]{Josan:2010vn}%
  \BibitemOpen
  \bibfield  {author} {\bibinfo {author} {\bibfnamefont {A.~S.}\ \bibnamefont
  {Josan}}\ and\ \bibinfo {author} {\bibfnamefont {A.~M.}\ \bibnamefont
  {Green}},\ }\href {\doibase 10.1103/PhysRevD.82.083527} {\bibfield  {journal}
  {\bibinfo  {journal} {Phys. Rev. D}\ }\textbf {\bibinfo {volume} {82}},\
  \bibinfo {pages} {083527} (\bibinfo {year} {2010})},\ \Eprint
  {http://arxiv.org/abs/1006.4970} {arXiv:1006.4970 [astro-ph.CO]} \BibitemShut
  {NoStop}%
\bibitem [{\citenamefont {Bringmann}\ \emph {et~al.}(2012)\citenamefont
  {Bringmann}, \citenamefont {Scott},\ and\ \citenamefont
  {Akrami}}]{Bringmann:2011ut}%
  \BibitemOpen
  \bibfield  {author} {\bibinfo {author} {\bibfnamefont {T.}~\bibnamefont
  {Bringmann}}, \bibinfo {author} {\bibfnamefont {P.}~\bibnamefont {Scott}}, \
  and\ \bibinfo {author} {\bibfnamefont {Y.}~\bibnamefont {Akrami}},\ }\href
  {\doibase 10.1103/PhysRevD.85.125027} {\bibfield  {journal} {\bibinfo
  {journal} {Phys. Rev. D}\ }\textbf {\bibinfo {volume} {85}},\ \bibinfo
  {pages} {125027} (\bibinfo {year} {2012})},\ \Eprint
  {http://arxiv.org/abs/1110.2484} {arXiv:1110.2484 [astro-ph.CO]} \BibitemShut
  {NoStop}%
\bibitem [{\citenamefont {Nakama}\ \emph {et~al.}(2018)\citenamefont {Nakama},
  \citenamefont {Suyama}, \citenamefont {Kohri},\ and\ \citenamefont
  {Hiroshima}}]{Nakama:2017qac}%
  \BibitemOpen
  \bibfield  {author} {\bibinfo {author} {\bibfnamefont {T.}~\bibnamefont
  {Nakama}}, \bibinfo {author} {\bibfnamefont {T.}~\bibnamefont {Suyama}},
  \bibinfo {author} {\bibfnamefont {K.}~\bibnamefont {Kohri}}, \ and\ \bibinfo
  {author} {\bibfnamefont {N.}~\bibnamefont {Hiroshima}},\ }\href {\doibase
  10.1103/PhysRevD.97.023539} {\bibfield  {journal} {\bibinfo  {journal} {Phys.
  Rev. D}\ }\textbf {\bibinfo {volume} {97}},\ \bibinfo {pages} {023539}
  (\bibinfo {year} {2018})},\ \Eprint {http://arxiv.org/abs/1712.08820}
  {arXiv:1712.08820 [astro-ph.CO]} \BibitemShut {NoStop}%
\bibitem [{\citenamefont {Delos}\ \emph
  {et~al.}(2018{\natexlab{a}})\citenamefont {Delos}, \citenamefont {Erickcek},
  \citenamefont {Bailey},\ and\ \citenamefont {Alvarez}}]{Delos:2018ueo}%
  \BibitemOpen
  \bibfield  {author} {\bibinfo {author} {\bibfnamefont {M.~S.}\ \bibnamefont
  {Delos}}, \bibinfo {author} {\bibfnamefont {A.~L.}\ \bibnamefont {Erickcek}},
  \bibinfo {author} {\bibfnamefont {A.~P.}\ \bibnamefont {Bailey}}, \ and\
  \bibinfo {author} {\bibfnamefont {M.~A.}\ \bibnamefont {Alvarez}},\ }\href
  {\doibase 10.1103/PhysRevD.98.063527} {\bibfield  {journal} {\bibinfo
  {journal} {Phys. Rev. D}\ }\textbf {\bibinfo {volume} {98}},\ \bibinfo
  {pages} {063527} (\bibinfo {year} {2018}{\natexlab{a}})},\ \Eprint
  {http://arxiv.org/abs/1806.07389} {arXiv:1806.07389 [astro-ph.CO]}
  \BibitemShut {NoStop}%
\bibitem [{\citenamefont {Blanco}\ \emph {et~al.}(2019)\citenamefont {Blanco},
  \citenamefont {Delos}, \citenamefont {Erickcek},\ and\ \citenamefont
  {Hooper}}]{Blanco:2019eij}%
  \BibitemOpen
  \bibfield  {author} {\bibinfo {author} {\bibfnamefont {C.}~\bibnamefont
  {Blanco}}, \bibinfo {author} {\bibfnamefont {M.~S.}\ \bibnamefont {Delos}},
  \bibinfo {author} {\bibfnamefont {A.~L.}\ \bibnamefont {Erickcek}}, \ and\
  \bibinfo {author} {\bibfnamefont {D.}~\bibnamefont {Hooper}},\ }\href
  {\doibase 10.1103/PhysRevD.100.103010} {\bibfield  {journal} {\bibinfo
  {journal} {Phys. Rev. D}\ }\textbf {\bibinfo {volume} {100}},\ \bibinfo
  {pages} {103010} (\bibinfo {year} {2019})},\ \Eprint
  {http://arxiv.org/abs/1906.00010} {arXiv:1906.00010 [astro-ph.CO]}
  \BibitemShut {NoStop}%
\bibitem [{\citenamefont {Hu}\ and\ \citenamefont
  {Sugiyama}(1996)}]{Hu:1995en}%
  \BibitemOpen
  \bibfield  {author} {\bibinfo {author} {\bibfnamefont {W.}~\bibnamefont
  {Hu}}\ and\ \bibinfo {author} {\bibfnamefont {N.}~\bibnamefont {Sugiyama}},\
  }\href {\doibase 10.1086/177989} {\bibfield  {journal} {\bibinfo  {journal}
  {Astrophys. J.}\ }\textbf {\bibinfo {volume} {471}},\ \bibinfo {pages} {542}
  (\bibinfo {year} {1996})},\ \Eprint {http://arxiv.org/abs/astro-ph/9510117}
  {arXiv:astro-ph/9510117} \BibitemShut {NoStop}%
\bibitem [{\citenamefont {Zyla}\ \emph {et~al.}(2020)\citenamefont {Zyla} \emph
  {et~al.}}]{ParticleDataGroup:2020ssz}%
  \BibitemOpen
  \bibfield  {author} {\bibinfo {author} {\bibfnamefont {P.~A.}\ \bibnamefont
  {Zyla}} \emph {et~al.} (\bibinfo {collaboration} {Particle Data Group}),\
  }\href {\doibase 10.1093/ptep/ptaa104} {\bibfield  {journal} {\bibinfo
  {journal} {PTEP}\ }\textbf {\bibinfo {volume} {2020}},\ \bibinfo {pages}
  {083C01} (\bibinfo {year} {2020})}\BibitemShut {NoStop}%
\bibitem [{\citenamefont {{Peebles}}(1967)}]{1967ApJ...147..859P}%
  \BibitemOpen
  \bibfield  {author} {\bibinfo {author} {\bibfnamefont {P.~J.~E.}\
  \bibnamefont {{Peebles}}},\ }\href {\doibase 10.1086/149077} {\bibfield
  {journal} {\bibinfo  {journal} {\apj}\ }\textbf {\bibinfo {volume} {147}},\
  \bibinfo {pages} {859} (\bibinfo {year} {1967})}\BibitemShut {NoStop}%
\bibitem [{\citenamefont {Dodelson}(2003)}]{Dodelson:2003ft}%
  \BibitemOpen
  \bibfield  {author} {\bibinfo {author} {\bibfnamefont {S.}~\bibnamefont
  {Dodelson}},\ }\href@noop {} {\emph {\bibinfo {title} {{Modern Cosmology}}}}\
  (\bibinfo  {publisher} {Academic Press},\ \bibinfo {address} {Amsterdam},\
  \bibinfo {year} {2003})\BibitemShut {NoStop}%
\bibitem [{\citenamefont {Sten~Delos}\ and\ \citenamefont
  {Silk}(2023)}]{StenDelos:2022jld}%
  \BibitemOpen
  \bibfield  {author} {\bibinfo {author} {\bibfnamefont {M.}~\bibnamefont
  {Sten~Delos}}\ and\ \bibinfo {author} {\bibfnamefont {J.}~\bibnamefont
  {Silk}},\ }\href {\doibase 10.1093/mnras/stad356} {\bibfield  {journal}
  {\bibinfo  {journal} {Mon. Not. Roy. Astron. Soc.}\ }\textbf {\bibinfo
  {volume} {520}},\ \bibinfo {pages} {4370} (\bibinfo {year} {2023})},\ \Eprint
  {http://arxiv.org/abs/2210.04904} {arXiv:2210.04904 [astro-ph.CO]}
  \BibitemShut {NoStop}%
\bibitem [{\citenamefont {Fillmore}\ and\ \citenamefont
  {Goldreich}(1984)}]{Fillmore:1984wk}%
  \BibitemOpen
  \bibfield  {author} {\bibinfo {author} {\bibfnamefont {J.~A.}\ \bibnamefont
  {Fillmore}}\ and\ \bibinfo {author} {\bibfnamefont {P.}~\bibnamefont
  {Goldreich}},\ }\href {\doibase 10.1086/162070} {\bibfield  {journal}
  {\bibinfo  {journal} {Astrophys. J.}\ }\textbf {\bibinfo {volume} {281}},\
  \bibinfo {pages} {1} (\bibinfo {year} {1984})}\BibitemShut {NoStop}%
\bibitem [{\citenamefont {Bertschinger}(1985)}]{Bertschinger:1985pd}%
  \BibitemOpen
  \bibfield  {author} {\bibinfo {author} {\bibfnamefont {E.}~\bibnamefont
  {Bertschinger}},\ }\href {\doibase 10.1086/191028} {\bibfield  {journal}
  {\bibinfo  {journal} {Astrophys. J. Suppl.}\ }\textbf {\bibinfo {volume}
  {58}},\ \bibinfo {pages} {39} (\bibinfo {year} {1985})}\BibitemShut {NoStop}%
\bibitem [{\citenamefont {Ishiyama}\ \emph {et~al.}(2010)\citenamefont
  {Ishiyama}, \citenamefont {Makino},\ and\ \citenamefont
  {Ebisuzaki}}]{Ishiyama:2010es}%
  \BibitemOpen
  \bibfield  {author} {\bibinfo {author} {\bibfnamefont {T.}~\bibnamefont
  {Ishiyama}}, \bibinfo {author} {\bibfnamefont {J.}~\bibnamefont {Makino}}, \
  and\ \bibinfo {author} {\bibfnamefont {T.}~\bibnamefont {Ebisuzaki}},\ }\href
  {\doibase 10.1088/2041-8205/723/2/L195} {\bibfield  {journal} {\bibinfo
  {journal} {Astrophys. J. Lett.}\ }\textbf {\bibinfo {volume} {723}},\
  \bibinfo {pages} {L195} (\bibinfo {year} {2010})},\ \Eprint
  {http://arxiv.org/abs/1006.3392} {arXiv:1006.3392 [astro-ph.CO]} \BibitemShut
  {NoStop}%
\bibitem [{\citenamefont {Ishiyama}(2014)}]{Ishiyama:2014uoa}%
  \BibitemOpen
  \bibfield  {author} {\bibinfo {author} {\bibfnamefont {T.}~\bibnamefont
  {Ishiyama}},\ }\href {\doibase 10.1088/0004-637X/788/1/27} {\bibfield
  {journal} {\bibinfo  {journal} {Astrophys. J.}\ }\textbf {\bibinfo {volume}
  {788}},\ \bibinfo {pages} {27} (\bibinfo {year} {2014})},\ \Eprint
  {http://arxiv.org/abs/1404.1650} {arXiv:1404.1650 [astro-ph.CO]} \BibitemShut
  {NoStop}%
\bibitem [{\citenamefont {Anderhalden}\ and\ \citenamefont
  {Diemand}(2013)}]{Anderhalden:2013wd}%
  \BibitemOpen
  \bibfield  {author} {\bibinfo {author} {\bibfnamefont {D.}~\bibnamefont
  {Anderhalden}}\ and\ \bibinfo {author} {\bibfnamefont {J.}~\bibnamefont
  {Diemand}},\ }\href {\doibase 10.1088/1475-7516/2013/04/009} {\bibfield
  {journal} {\bibinfo  {journal} {JCAP}\ }\textbf {\bibinfo {volume} {04}},\
  \bibinfo {pages} {009} (\bibinfo {year} {2013})},\ \bibinfo {note} {[Erratum:
  JCAP 08, E02 (2013)]},\ \Eprint {http://arxiv.org/abs/1302.0003}
  {arXiv:1302.0003 [astro-ph.CO]} \BibitemShut {NoStop}%
\bibitem [{\citenamefont {Angulo}\ \emph {et~al.}(2017)\citenamefont {Angulo},
  \citenamefont {Hahn}, \citenamefont {Ludlow},\ and\ \citenamefont
  {Bonoli}}]{Angulo:2016qof}%
  \BibitemOpen
  \bibfield  {author} {\bibinfo {author} {\bibfnamefont {R.~E.}\ \bibnamefont
  {Angulo}}, \bibinfo {author} {\bibfnamefont {O.}~\bibnamefont {Hahn}},
  \bibinfo {author} {\bibfnamefont {A.}~\bibnamefont {Ludlow}}, \ and\ \bibinfo
  {author} {\bibfnamefont {S.}~\bibnamefont {Bonoli}},\ }\href {\doibase
  10.1093/mnras/stx1658} {\bibfield  {journal} {\bibinfo  {journal} {Mon. Not.
  Roy. Astron. Soc.}\ }\textbf {\bibinfo {volume} {471}},\ \bibinfo {pages}
  {4687} (\bibinfo {year} {2017})},\ \Eprint {http://arxiv.org/abs/1604.03131}
  {arXiv:1604.03131 [astro-ph.CO]} \BibitemShut {NoStop}%
\bibitem [{\citenamefont {Ogiya}\ and\ \citenamefont
  {Hahn}(2018)}]{Ogiya:2017hbr}%
  \BibitemOpen
  \bibfield  {author} {\bibinfo {author} {\bibfnamefont {G.}~\bibnamefont
  {Ogiya}}\ and\ \bibinfo {author} {\bibfnamefont {O.}~\bibnamefont {Hahn}},\
  }\href {\doibase 10.1093/mnras/stx2639} {\bibfield  {journal} {\bibinfo
  {journal} {Mon. Not. Roy. Astron. Soc.}\ }\textbf {\bibinfo {volume} {473}},\
  \bibinfo {pages} {4339} (\bibinfo {year} {2018})},\ \Eprint
  {http://arxiv.org/abs/1707.07693} {arXiv:1707.07693 [astro-ph.CO]}
  \BibitemShut {NoStop}%
\bibitem [{\citenamefont {Delos}\ \emph
  {et~al.}(2018{\natexlab{b}})\citenamefont {Delos}, \citenamefont {Erickcek},
  \citenamefont {Bailey},\ and\ \citenamefont {Alvarez}}]{Delos:2017thv}%
  \BibitemOpen
  \bibfield  {author} {\bibinfo {author} {\bibfnamefont {M.~S.}\ \bibnamefont
  {Delos}}, \bibinfo {author} {\bibfnamefont {A.~L.}\ \bibnamefont {Erickcek}},
  \bibinfo {author} {\bibfnamefont {A.~P.}\ \bibnamefont {Bailey}}, \ and\
  \bibinfo {author} {\bibfnamefont {M.~A.}\ \bibnamefont {Alvarez}},\ }\href
  {\doibase 10.1103/PhysRevD.97.041303} {\bibfield  {journal} {\bibinfo
  {journal} {Phys. Rev. D}\ }\textbf {\bibinfo {volume} {97}},\ \bibinfo
  {pages} {041303} (\bibinfo {year} {2018}{\natexlab{b}})},\ \Eprint
  {http://arxiv.org/abs/1712.05421} {arXiv:1712.05421 [astro-ph.CO]}
  \BibitemShut {NoStop}%
\bibitem [{\citenamefont {Delos}\ \emph {et~al.}(2019)\citenamefont {Delos},
  \citenamefont {Bruff},\ and\ \citenamefont {Erickcek}}]{Delos:2019mxl}%
  \BibitemOpen
  \bibfield  {author} {\bibinfo {author} {\bibfnamefont {M.~S.}\ \bibnamefont
  {Delos}}, \bibinfo {author} {\bibfnamefont {M.}~\bibnamefont {Bruff}}, \ and\
  \bibinfo {author} {\bibfnamefont {A.~L.}\ \bibnamefont {Erickcek}},\ }\href
  {\doibase 10.1103/PhysRevD.100.023523} {\bibfield  {journal} {\bibinfo
  {journal} {Phys. Rev. D}\ }\textbf {\bibinfo {volume} {100}},\ \bibinfo
  {pages} {023523} (\bibinfo {year} {2019})},\ \Eprint
  {http://arxiv.org/abs/1905.05766} {arXiv:1905.05766 [astro-ph.CO]}
  \BibitemShut {NoStop}%
\bibitem [{\citenamefont {Ogiya}\ \emph {et~al.}(2016)\citenamefont {Ogiya},
  \citenamefont {Nagai},\ and\ \citenamefont {Ishiyama}}]{Ogiya:2016hyo}%
  \BibitemOpen
  \bibfield  {author} {\bibinfo {author} {\bibfnamefont {G.}~\bibnamefont
  {Ogiya}}, \bibinfo {author} {\bibfnamefont {D.}~\bibnamefont {Nagai}}, \ and\
  \bibinfo {author} {\bibfnamefont {T.}~\bibnamefont {Ishiyama}},\ }\href
  {\doibase 10.1093/mnras/stw1551} {\bibfield  {journal} {\bibinfo  {journal}
  {Mon. Not. Roy. Astron. Soc.}\ }\textbf {\bibinfo {volume} {461}},\ \bibinfo
  {pages} {3385} (\bibinfo {year} {2016})},\ \Eprint
  {http://arxiv.org/abs/1604.02866} {arXiv:1604.02866 [astro-ph.CO]}
  \BibitemShut {NoStop}%
\bibitem [{\citenamefont {Delos}\ and\ \citenamefont
  {Franciolini}(2023)}]{Delos:2023fpm}%
  \BibitemOpen
  \bibfield  {author} {\bibinfo {author} {\bibfnamefont {M.~S.}\ \bibnamefont
  {Delos}}\ and\ \bibinfo {author} {\bibfnamefont {G.}~\bibnamefont
  {Franciolini}},\ }\href {\doibase 10.1103/PhysRevD.107.083505} {\bibfield
  {journal} {\bibinfo  {journal} {Phys. Rev. D}\ }\textbf {\bibinfo {volume}
  {107}},\ \bibinfo {pages} {083505} (\bibinfo {year} {2023})},\ \Eprint
  {http://arxiv.org/abs/2301.13171} {arXiv:2301.13171 [astro-ph.CO]}
  \BibitemShut {NoStop}%
\bibitem [{\citenamefont {Bardeen}\ \emph {et~al.}(1986)\citenamefont
  {Bardeen}, \citenamefont {Bond}, \citenamefont {Kaiser},\ and\ \citenamefont
  {Szalay}}]{Bardeen:1985tr}%
  \BibitemOpen
  \bibfield  {author} {\bibinfo {author} {\bibfnamefont {J.~M.}\ \bibnamefont
  {Bardeen}}, \bibinfo {author} {\bibfnamefont {J.~R.}\ \bibnamefont {Bond}},
  \bibinfo {author} {\bibfnamefont {N.}~\bibnamefont {Kaiser}}, \ and\ \bibinfo
  {author} {\bibfnamefont {A.~S.}\ \bibnamefont {Szalay}},\ }\href {\doibase
  10.1086/164143} {\bibfield  {journal} {\bibinfo  {journal} {Astrophys. J.}\
  }\textbf {\bibinfo {volume} {304}},\ \bibinfo {pages} {15} (\bibinfo {year}
  {1986})}\BibitemShut {NoStop}%
\bibitem [{\citenamefont {Ferrante}\ \emph {et~al.}(2023)\citenamefont
  {Ferrante}, \citenamefont {Franciolini}, \citenamefont {Iovino},\ and\
  \citenamefont {Urbano}}]{Ferrante:2022mui}%
  \BibitemOpen
  \bibfield  {author} {\bibinfo {author} {\bibfnamefont {G.}~\bibnamefont
  {Ferrante}}, \bibinfo {author} {\bibfnamefont {G.}~\bibnamefont
  {Franciolini}}, \bibinfo {author} {\bibfnamefont {A.}~\bibnamefont {Iovino},
  \bibfnamefont {Junior.}}, \ and\ \bibinfo {author} {\bibfnamefont
  {A.}~\bibnamefont {Urbano}},\ }\href {\doibase 10.1103/PhysRevD.107.043520}
  {\bibfield  {journal} {\bibinfo  {journal} {Phys. Rev. D}\ }\textbf {\bibinfo
  {volume} {107}},\ \bibinfo {pages} {043520} (\bibinfo {year} {2023})},\
  \Eprint {http://arxiv.org/abs/2211.01728} {arXiv:2211.01728 [astro-ph.CO]}
  \BibitemShut {NoStop}%
\bibitem [{\citenamefont {Young}(2019)}]{Young:2019osy}%
  \BibitemOpen
  \bibfield  {author} {\bibinfo {author} {\bibfnamefont {S.}~\bibnamefont
  {Young}},\ }\href {\doibase 10.1142/S0218271820300025} {\bibfield  {journal}
  {\bibinfo  {journal} {Int. J. Mod. Phys. D}\ }\textbf {\bibinfo {volume}
  {29}},\ \bibinfo {pages} {2030002} (\bibinfo {year} {2019})},\ \Eprint
  {http://arxiv.org/abs/1905.01230} {arXiv:1905.01230 [astro-ph.CO]}
  \BibitemShut {NoStop}%
\bibitem [{\citenamefont {Josan}\ \emph {et~al.}(2009)\citenamefont {Josan},
  \citenamefont {Green},\ and\ \citenamefont {Malik}}]{Josan:2009qn}%
  \BibitemOpen
  \bibfield  {author} {\bibinfo {author} {\bibfnamefont {A.~S.}\ \bibnamefont
  {Josan}}, \bibinfo {author} {\bibfnamefont {A.~M.}\ \bibnamefont {Green}}, \
  and\ \bibinfo {author} {\bibfnamefont {K.~A.}\ \bibnamefont {Malik}},\ }\href
  {\doibase 10.1103/PhysRevD.79.103520} {\bibfield  {journal} {\bibinfo
  {journal} {Phys. Rev. D}\ }\textbf {\bibinfo {volume} {79}},\ \bibinfo
  {pages} {103520} (\bibinfo {year} {2009})},\ \Eprint
  {http://arxiv.org/abs/0903.3184} {arXiv:0903.3184 [astro-ph.CO]} \BibitemShut
  {NoStop}%
\bibitem [{\citenamefont {Hofmann}\ \emph {et~al.}(2001)\citenamefont
  {Hofmann}, \citenamefont {Schwarz},\ and\ \citenamefont
  {Stoecker}}]{Hofmann:2001bi}%
  \BibitemOpen
  \bibfield  {author} {\bibinfo {author} {\bibfnamefont {S.}~\bibnamefont
  {Hofmann}}, \bibinfo {author} {\bibfnamefont {D.~J.}\ \bibnamefont
  {Schwarz}}, \ and\ \bibinfo {author} {\bibfnamefont {H.}~\bibnamefont
  {Stoecker}},\ }\href {\doibase 10.1103/PhysRevD.64.083507} {\bibfield
  {journal} {\bibinfo  {journal} {Phys. Rev. D}\ }\textbf {\bibinfo {volume}
  {64}},\ \bibinfo {pages} {083507} (\bibinfo {year} {2001})},\ \Eprint
  {http://arxiv.org/abs/astro-ph/0104173} {arXiv:astro-ph/0104173} \BibitemShut
  {NoStop}%
\bibitem [{\citenamefont {Profumo}\ \emph {et~al.}(2006)\citenamefont
  {Profumo}, \citenamefont {Sigurdson},\ and\ \citenamefont
  {Kamionkowski}}]{Profumo:2006bv}%
  \BibitemOpen
  \bibfield  {author} {\bibinfo {author} {\bibfnamefont {S.}~\bibnamefont
  {Profumo}}, \bibinfo {author} {\bibfnamefont {K.}~\bibnamefont {Sigurdson}},
  \ and\ \bibinfo {author} {\bibfnamefont {M.}~\bibnamefont {Kamionkowski}},\
  }\href {\doibase 10.1103/PhysRevLett.97.031301} {\bibfield  {journal}
  {\bibinfo  {journal} {Phys. Rev. Lett.}\ }\textbf {\bibinfo {volume} {97}},\
  \bibinfo {pages} {031301} (\bibinfo {year} {2006})},\ \Eprint
  {http://arxiv.org/abs/astro-ph/0603373} {arXiv:astro-ph/0603373} \BibitemShut
  {NoStop}%
\bibitem [{\citenamefont {Bringmann}(2009)}]{Bringmann:2009vf}%
  \BibitemOpen
  \bibfield  {author} {\bibinfo {author} {\bibfnamefont {T.}~\bibnamefont
  {Bringmann}},\ }\href {\doibase 10.1088/1367-2630/11/10/105027} {\bibfield
  {journal} {\bibinfo  {journal} {New J. Phys.}\ }\textbf {\bibinfo {volume}
  {11}},\ \bibinfo {pages} {105027} (\bibinfo {year} {2009})},\ \Eprint
  {http://arxiv.org/abs/0903.0189} {arXiv:0903.0189 [astro-ph.CO]} \BibitemShut
  {NoStop}%
\bibitem [{\citenamefont {Cornell}\ and\ \citenamefont
  {Profumo}(2012)}]{Cornell:2012tb}%
  \BibitemOpen
  \bibfield  {author} {\bibinfo {author} {\bibfnamefont {J.~M.}\ \bibnamefont
  {Cornell}}\ and\ \bibinfo {author} {\bibfnamefont {S.}~\bibnamefont
  {Profumo}},\ }\href {\doibase 10.1088/1475-7516/2012/06/011} {\bibfield
  {journal} {\bibinfo  {journal} {JCAP}\ }\textbf {\bibinfo {volume} {06}},\
  \bibinfo {pages} {011} (\bibinfo {year} {2012})},\ \Eprint
  {http://arxiv.org/abs/1203.1100} {arXiv:1203.1100 [hep-ph]} \BibitemShut
  {NoStop}%
\bibitem [{\citenamefont {Ando}\ \emph {et~al.}(2019)\citenamefont {Ando},
  \citenamefont {Kamada}, \citenamefont {Sekiguchi},\ and\ \citenamefont
  {Takahashi}}]{Ando:2019rgx}%
  \BibitemOpen
  \bibfield  {author} {\bibinfo {author} {\bibfnamefont {S.}~\bibnamefont
  {Ando}}, \bibinfo {author} {\bibfnamefont {A.}~\bibnamefont {Kamada}},
  \bibinfo {author} {\bibfnamefont {T.}~\bibnamefont {Sekiguchi}}, \ and\
  \bibinfo {author} {\bibfnamefont {T.}~\bibnamefont {Takahashi}},\ }\href
  {\doibase 10.1103/PhysRevD.100.123519} {\bibfield  {journal} {\bibinfo
  {journal} {Phys. Rev. D}\ }\textbf {\bibinfo {volume} {100}},\ \bibinfo
  {pages} {123519} (\bibinfo {year} {2019})},\ \Eprint
  {http://arxiv.org/abs/1901.09992} {arXiv:1901.09992 [hep-ph]} \BibitemShut
  {NoStop}%
\bibitem [{\citenamefont {Cirelli}\ \emph {et~al.}(2011)\citenamefont
  {Cirelli}, \citenamefont {Corcella}, \citenamefont {Hektor}, \citenamefont
  {Hutsi}, \citenamefont {Kadastik}, \citenamefont {Panci}, \citenamefont
  {Raidal}, \citenamefont {Sala},\ and\ \citenamefont
  {Strumia}}]{Cirelli:2010xx}%
  \BibitemOpen
  \bibfield  {author} {\bibinfo {author} {\bibfnamefont {M.}~\bibnamefont
  {Cirelli}}, \bibinfo {author} {\bibfnamefont {G.}~\bibnamefont {Corcella}},
  \bibinfo {author} {\bibfnamefont {A.}~\bibnamefont {Hektor}}, \bibinfo
  {author} {\bibfnamefont {G.}~\bibnamefont {Hutsi}}, \bibinfo {author}
  {\bibfnamefont {M.}~\bibnamefont {Kadastik}}, \bibinfo {author}
  {\bibfnamefont {P.}~\bibnamefont {Panci}}, \bibinfo {author} {\bibfnamefont
  {M.}~\bibnamefont {Raidal}}, \bibinfo {author} {\bibfnamefont
  {F.}~\bibnamefont {Sala}}, \ and\ \bibinfo {author} {\bibfnamefont
  {A.}~\bibnamefont {Strumia}},\ }\href {\doibase
  10.1088/1475-7516/2012/10/E01} {\bibfield  {journal} {\bibinfo  {journal}
  {JCAP}\ }\textbf {\bibinfo {volume} {03}},\ \bibinfo {pages} {051} (\bibinfo
  {year} {2011})},\ \bibinfo {note} {[Erratum: JCAP 10, E01 (2012)]},\ \Eprint
  {http://arxiv.org/abs/1012.4515} {arXiv:1012.4515 [hep-ph]} \BibitemShut
  {NoStop}%
\bibitem [{\citenamefont {Nesti}\ and\ \citenamefont
  {Salucci}(2013)}]{Nesti:2013uwa}%
  \BibitemOpen
  \bibfield  {author} {\bibinfo {author} {\bibfnamefont {F.}~\bibnamefont
  {Nesti}}\ and\ \bibinfo {author} {\bibfnamefont {P.}~\bibnamefont
  {Salucci}},\ }\href {\doibase 10.1088/1475-7516/2013/07/016} {\bibfield
  {journal} {\bibinfo  {journal} {JCAP}\ }\textbf {\bibinfo {volume} {07}},\
  \bibinfo {pages} {016} (\bibinfo {year} {2013})},\ \Eprint
  {http://arxiv.org/abs/1304.5127} {arXiv:1304.5127 [astro-ph.GA]} \BibitemShut
  {NoStop}%
\bibitem [{\citenamefont {Ackermann}\ \emph {et~al.}(2012)\citenamefont
  {Ackermann} \emph {et~al.}}]{Fermi-LAT:2012edv}%
  \BibitemOpen
  \bibfield  {author} {\bibinfo {author} {\bibfnamefont {M.}~\bibnamefont
  {Ackermann}} \emph {et~al.} (\bibinfo {collaboration} {Fermi-LAT}),\ }\href
  {\doibase 10.1088/0004-637X/750/1/3} {\bibfield  {journal} {\bibinfo
  {journal} {Astrophys. J.}\ }\textbf {\bibinfo {volume} {750}},\ \bibinfo
  {pages} {3} (\bibinfo {year} {2012})},\ \Eprint
  {http://arxiv.org/abs/1202.4039} {arXiv:1202.4039 [astro-ph.HE]} \BibitemShut
  {NoStop}%
\bibitem [{\citenamefont {Daum}\ \emph {et~al.}(1995)\citenamefont {Daum} \emph
  {et~al.}}]{Frejus:1994brq}%
  \BibitemOpen
  \bibfield  {author} {\bibinfo {author} {\bibfnamefont {K.}~\bibnamefont
  {Daum}} \emph {et~al.} (\bibinfo {collaboration} {Frejus}),\ }\href {\doibase
  10.1007/BF01556368} {\bibfield  {journal} {\bibinfo  {journal} {Z. Phys. C}\
  }\textbf {\bibinfo {volume} {66}},\ \bibinfo {pages} {417} (\bibinfo {year}
  {1995})}\BibitemShut {NoStop}%
\bibitem [{\citenamefont {Abbasi}\ \emph {et~al.}(2010)\citenamefont {Abbasi}
  \emph {et~al.}}]{IceCube:2010jrf}%
  \BibitemOpen
  \bibfield  {author} {\bibinfo {author} {\bibfnamefont {R.}~\bibnamefont
  {Abbasi}} \emph {et~al.} (\bibinfo {collaboration} {IceCube}),\ }\href
  {\doibase 10.1016/j.astropartphys.2010.05.001} {\bibfield  {journal}
  {\bibinfo  {journal} {Astropart. Phys.}\ }\textbf {\bibinfo {volume} {34}},\
  \bibinfo {pages} {48} (\bibinfo {year} {2010})},\ \Eprint
  {http://arxiv.org/abs/1004.2357} {arXiv:1004.2357 [astro-ph.HE]} \BibitemShut
  {NoStop}%
\bibitem [{\citenamefont {Adrian-Martinez}\ \emph {et~al.}(2013)\citenamefont
  {Adrian-Martinez} \emph {et~al.}}]{ANTARES:2013iuz}%
  \BibitemOpen
  \bibfield  {author} {\bibinfo {author} {\bibfnamefont {S.}~\bibnamefont
  {Adrian-Martinez}} \emph {et~al.} (\bibinfo {collaboration} {ANTARES}),\
  }\href {\doibase 10.1140/epjc/s10052-013-2606-4} {\bibfield  {journal}
  {\bibinfo  {journal} {Eur. Phys. J. C}\ }\textbf {\bibinfo {volume} {73}},\
  \bibinfo {pages} {2606} (\bibinfo {year} {2013})},\ \Eprint
  {http://arxiv.org/abs/1308.1599} {arXiv:1308.1599 [astro-ph.HE]} \BibitemShut
  {NoStop}%
\bibitem [{\citenamefont {Richard}\ \emph {et~al.}(2016)\citenamefont {Richard}
  \emph {et~al.}}]{Super-Kamiokande:2015qek}%
  \BibitemOpen
  \bibfield  {author} {\bibinfo {author} {\bibfnamefont {E.}~\bibnamefont
  {Richard}} \emph {et~al.} (\bibinfo {collaboration} {Super-Kamiokande}),\
  }\href {\doibase 10.1103/PhysRevD.94.052001} {\bibfield  {journal} {\bibinfo
  {journal} {Phys. Rev. D}\ }\textbf {\bibinfo {volume} {94}},\ \bibinfo
  {pages} {052001} (\bibinfo {year} {2016})},\ \Eprint
  {http://arxiv.org/abs/1510.08127} {arXiv:1510.08127 [hep-ex]} \BibitemShut
  {NoStop}%
\bibitem [{\citenamefont {Aartsen}\ \emph {et~al.}(2016)\citenamefont {Aartsen}
  \emph {et~al.}}]{IceCube:2016umi}%
  \BibitemOpen
  \bibfield  {author} {\bibinfo {author} {\bibfnamefont {M.~G.}\ \bibnamefont
  {Aartsen}} \emph {et~al.} (\bibinfo {collaboration} {IceCube}),\ }\href
  {\doibase 10.3847/0004-637X/833/1/3} {\bibfield  {journal} {\bibinfo
  {journal} {Astrophys. J.}\ }\textbf {\bibinfo {volume} {833}},\ \bibinfo
  {pages} {3} (\bibinfo {year} {2016})},\ \Eprint
  {http://arxiv.org/abs/1607.08006} {arXiv:1607.08006 [astro-ph.HE]}
  \BibitemShut {NoStop}%
\bibitem [{\citenamefont {Fuerst}\ \emph {et~al.}(2023)\citenamefont {Fuerst}
  \emph {et~al.}}]{IceCube:2023hou}%
  \BibitemOpen
  \bibfield  {author} {\bibinfo {author} {\bibfnamefont {P.~M.}\ \bibnamefont
  {Fuerst}} \emph {et~al.} (\bibinfo {collaboration} {IceCube}),\ }\href
  {\doibase 10.22323/1.444.1046} {\bibfield  {journal} {\bibinfo  {journal}
  {PoS}\ }\textbf {\bibinfo {volume} {ICRC2023}},\ \bibinfo {pages} {1046}
  (\bibinfo {year} {2023})},\ \Eprint {http://arxiv.org/abs/2308.08233}
  {arXiv:2308.08233 [astro-ph.HE]} \BibitemShut {NoStop}%
\bibitem [{\citenamefont {Dom\`enech}(2021)}]{Domenech:2021ztg}%
  \BibitemOpen
  \bibfield  {author} {\bibinfo {author} {\bibfnamefont {G.}~\bibnamefont
  {Dom\`enech}},\ }\href {\doibase 10.3390/universe7110398} {\bibfield
  {journal} {\bibinfo  {journal} {Universe}\ }\textbf {\bibinfo {volume} {7}},\
  \bibinfo {pages} {398} (\bibinfo {year} {2021})},\ \Eprint
  {http://arxiv.org/abs/2109.01398} {arXiv:2109.01398 [gr-qc]} \BibitemShut
  {NoStop}%
\bibitem [{\citenamefont {Espinosa}\ \emph {et~al.}(2018)\citenamefont
  {Espinosa}, \citenamefont {Racco},\ and\ \citenamefont
  {Riotto}}]{Espinosa:2018eve}%
  \BibitemOpen
  \bibfield  {author} {\bibinfo {author} {\bibfnamefont {J.~R.}\ \bibnamefont
  {Espinosa}}, \bibinfo {author} {\bibfnamefont {D.}~\bibnamefont {Racco}}, \
  and\ \bibinfo {author} {\bibfnamefont {A.}~\bibnamefont {Riotto}},\ }\href
  {\doibase 10.1088/1475-7516/2018/09/012} {\bibfield  {journal} {\bibinfo
  {journal} {JCAP}\ }\textbf {\bibinfo {volume} {09}},\ \bibinfo {pages} {012}
  (\bibinfo {year} {2018})},\ \Eprint {http://arxiv.org/abs/1804.07732}
  {arXiv:1804.07732 [hep-ph]} \BibitemShut {NoStop}%
\bibitem [{\citenamefont {Kohri}\ and\ \citenamefont
  {Terada}(2018)}]{Kohri:2018awv}%
  \BibitemOpen
  \bibfield  {author} {\bibinfo {author} {\bibfnamefont {K.}~\bibnamefont
  {Kohri}}\ and\ \bibinfo {author} {\bibfnamefont {T.}~\bibnamefont {Terada}},\
  }\href {\doibase 10.1103/PhysRevD.97.123532} {\bibfield  {journal} {\bibinfo
  {journal} {Phys. Rev. D}\ }\textbf {\bibinfo {volume} {97}},\ \bibinfo
  {pages} {123532} (\bibinfo {year} {2018})},\ \Eprint
  {http://arxiv.org/abs/1804.08577} {arXiv:1804.08577 [gr-qc]} \BibitemShut
  {NoStop}%
\bibitem [{\citenamefont {Byrnes}\ \emph {et~al.}(2007)\citenamefont {Byrnes},
  \citenamefont {Koyama}, \citenamefont {Sasaki},\ and\ \citenamefont
  {Wands}}]{Byrnes:2007tm}%
  \BibitemOpen
  \bibfield  {author} {\bibinfo {author} {\bibfnamefont {C.~T.}\ \bibnamefont
  {Byrnes}}, \bibinfo {author} {\bibfnamefont {K.}~\bibnamefont {Koyama}},
  \bibinfo {author} {\bibfnamefont {M.}~\bibnamefont {Sasaki}}, \ and\ \bibinfo
  {author} {\bibfnamefont {D.}~\bibnamefont {Wands}},\ }\href {\doibase
  10.1088/1475-7516/2007/11/027} {\bibfield  {journal} {\bibinfo  {journal}
  {JCAP}\ }\textbf {\bibinfo {volume} {11}},\ \bibinfo {pages} {027} (\bibinfo
  {year} {2007})},\ \Eprint {http://arxiv.org/abs/0705.4096} {arXiv:0705.4096
  [hep-th]} \BibitemShut {NoStop}%
\bibitem [{\citenamefont {Unal}(2019)}]{Unal:2018yaa}%
  \BibitemOpen
  \bibfield  {author} {\bibinfo {author} {\bibfnamefont {C.}~\bibnamefont
  {Unal}},\ }\href {\doibase 10.1103/PhysRevD.99.041301} {\bibfield  {journal}
  {\bibinfo  {journal} {Phys. Rev. D}\ }\textbf {\bibinfo {volume} {99}},\
  \bibinfo {pages} {041301} (\bibinfo {year} {2019})},\ \Eprint
  {http://arxiv.org/abs/1811.09151} {arXiv:1811.09151 [astro-ph.CO]}
  \BibitemShut {NoStop}%
\bibitem [{\citenamefont {Cai}\ \emph {et~al.}(2019)\citenamefont {Cai},
  \citenamefont {Pi},\ and\ \citenamefont {Sasaki}}]{Cai:2018dig}%
  \BibitemOpen
  \bibfield  {author} {\bibinfo {author} {\bibfnamefont {R.-g.}\ \bibnamefont
  {Cai}}, \bibinfo {author} {\bibfnamefont {S.}~\bibnamefont {Pi}}, \ and\
  \bibinfo {author} {\bibfnamefont {M.}~\bibnamefont {Sasaki}},\ }\href
  {\doibase 10.1103/PhysRevLett.122.201101} {\bibfield  {journal} {\bibinfo
  {journal} {Phys. Rev. Lett.}\ }\textbf {\bibinfo {volume} {122}},\ \bibinfo
  {pages} {201101} (\bibinfo {year} {2019})},\ \Eprint
  {http://arxiv.org/abs/1810.11000} {arXiv:1810.11000 [astro-ph.CO]}
  \BibitemShut {NoStop}%
\bibitem [{\citenamefont {Adshead}\ \emph {et~al.}(2021)\citenamefont
  {Adshead}, \citenamefont {Lozanov},\ and\ \citenamefont
  {Weiner}}]{Adshead:2021hnm}%
  \BibitemOpen
  \bibfield  {author} {\bibinfo {author} {\bibfnamefont {P.}~\bibnamefont
  {Adshead}}, \bibinfo {author} {\bibfnamefont {K.~D.}\ \bibnamefont
  {Lozanov}}, \ and\ \bibinfo {author} {\bibfnamefont {Z.~J.}\ \bibnamefont
  {Weiner}},\ }\href {\doibase 10.1088/1475-7516/2021/10/080} {\bibfield
  {journal} {\bibinfo  {journal} {JCAP}\ }\textbf {\bibinfo {volume} {10}},\
  \bibinfo {pages} {080} (\bibinfo {year} {2021})},\ \Eprint
  {http://arxiv.org/abs/2105.01659} {arXiv:2105.01659 [astro-ph.CO]}
  \BibitemShut {NoStop}%
\bibitem [{\citenamefont {Mitridate}\ \emph {et~al.}(2023)\citenamefont
  {Mitridate}, \citenamefont {Wright}, \citenamefont {von Eckardstein},
  \citenamefont {Schr\"oder}, \citenamefont {Nay}, \citenamefont {Olum},
  \citenamefont {Schmitz},\ and\ \citenamefont {Trickle}}]{Mitridate:2023oar}%
  \BibitemOpen
  \bibfield  {author} {\bibinfo {author} {\bibfnamefont {A.}~\bibnamefont
  {Mitridate}}, \bibinfo {author} {\bibfnamefont {D.}~\bibnamefont {Wright}},
  \bibinfo {author} {\bibfnamefont {R.}~\bibnamefont {von Eckardstein}},
  \bibinfo {author} {\bibfnamefont {T.}~\bibnamefont {Schr\"oder}}, \bibinfo
  {author} {\bibfnamefont {J.}~\bibnamefont {Nay}}, \bibinfo {author}
  {\bibfnamefont {K.}~\bibnamefont {Olum}}, \bibinfo {author} {\bibfnamefont
  {K.}~\bibnamefont {Schmitz}}, \ and\ \bibinfo {author} {\bibfnamefont
  {T.}~\bibnamefont {Trickle}},\ }\href@noop {} {\  (\bibinfo {year} {2023})},\
  \Eprint {http://arxiv.org/abs/2306.16377} {arXiv:2306.16377 [hep-ph]}
  \BibitemShut {NoStop}%
\bibitem [{\citenamefont {Musco}\ \emph {et~al.}(2021)\citenamefont {Musco},
  \citenamefont {De~Luca}, \citenamefont {Franciolini},\ and\ \citenamefont
  {Riotto}}]{Musco:2020jjb}%
  \BibitemOpen
  \bibfield  {author} {\bibinfo {author} {\bibfnamefont {I.}~\bibnamefont
  {Musco}}, \bibinfo {author} {\bibfnamefont {V.}~\bibnamefont {De~Luca}},
  \bibinfo {author} {\bibfnamefont {G.}~\bibnamefont {Franciolini}}, \ and\
  \bibinfo {author} {\bibfnamefont {A.}~\bibnamefont {Riotto}},\ }\href
  {\doibase 10.1103/PhysRevD.103.063538} {\bibfield  {journal} {\bibinfo
  {journal} {Phys. Rev. D}\ }\textbf {\bibinfo {volume} {103}},\ \bibinfo
  {pages} {063538} (\bibinfo {year} {2021})},\ \Eprint
  {http://arxiv.org/abs/2011.03014} {arXiv:2011.03014 [astro-ph.CO]}
  \BibitemShut {NoStop}%
\bibitem [{\citenamefont {Escriv\`a}\ \emph {et~al.}(2020)\citenamefont
  {Escriv\`a}, \citenamefont {Germani},\ and\ \citenamefont
  {Sheth}}]{Escriva:2019phb}%
  \BibitemOpen
  \bibfield  {author} {\bibinfo {author} {\bibfnamefont {A.}~\bibnamefont
  {Escriv\`a}}, \bibinfo {author} {\bibfnamefont {C.}~\bibnamefont {Germani}},
  \ and\ \bibinfo {author} {\bibfnamefont {R.~K.}\ \bibnamefont {Sheth}},\
  }\href {\doibase 10.1103/PhysRevD.101.044022} {\bibfield  {journal} {\bibinfo
   {journal} {Phys. Rev. D}\ }\textbf {\bibinfo {volume} {101}},\ \bibinfo
  {pages} {044022} (\bibinfo {year} {2020})},\ \Eprint
  {http://arxiv.org/abs/1907.13311} {arXiv:1907.13311 [gr-qc]} \BibitemShut
  {NoStop}%
\bibitem [{\citenamefont {Frosina}\ and\ \citenamefont
  {Urbano}(2023)}]{Frosina:2023nxu}%
  \BibitemOpen
  \bibfield  {author} {\bibinfo {author} {\bibfnamefont {L.}~\bibnamefont
  {Frosina}}\ and\ \bibinfo {author} {\bibfnamefont {A.}~\bibnamefont
  {Urbano}},\ }\href {\doibase 10.1103/PhysRevD.108.103544} {\bibfield
  {journal} {\bibinfo  {journal} {Phys. Rev. D}\ }\textbf {\bibinfo {volume}
  {108}},\ \bibinfo {pages} {103544} (\bibinfo {year} {2023})},\ \Eprint
  {http://arxiv.org/abs/2308.06915} {arXiv:2308.06915 [astro-ph.CO]}
  \BibitemShut {NoStop}%
\bibitem [{\citenamefont {Iovino}\ \emph {et~al.}(2024)\citenamefont {Iovino},
  \citenamefont {Perna}, \citenamefont {Riotto},\ and\ \citenamefont
  {Veerm\"ae}}]{Iovino:2024tyg}%
  \BibitemOpen
  \bibfield  {author} {\bibinfo {author} {\bibfnamefont {A.~J.}\ \bibnamefont
  {Iovino}}, \bibinfo {author} {\bibfnamefont {G.}~\bibnamefont {Perna}},
  \bibinfo {author} {\bibfnamefont {A.}~\bibnamefont {Riotto}}, \ and\ \bibinfo
  {author} {\bibfnamefont {H.}~\bibnamefont {Veerm\"ae}},\ }\href {\doibase
  10.1088/1475-7516/2024/10/050} {\bibfield  {journal} {\bibinfo  {journal}
  {JCAP}\ }\textbf {\bibinfo {volume} {10}},\ \bibinfo {pages} {050} (\bibinfo
  {year} {2024})},\ \Eprint {http://arxiv.org/abs/2406.20089} {arXiv:2406.20089
  [astro-ph.CO]} \BibitemShut {NoStop}%
\bibitem [{\citenamefont {Caprini}\ \emph {et~al.}(2016)\citenamefont {Caprini}
  \emph {et~al.}}]{Caprini:2015zlo}%
  \BibitemOpen
  \bibfield  {author} {\bibinfo {author} {\bibfnamefont {C.}~\bibnamefont
  {Caprini}} \emph {et~al.},\ }\href {\doibase 10.1088/1475-7516/2016/04/001}
  {\bibfield  {journal} {\bibinfo  {journal} {JCAP}\ }\textbf {\bibinfo
  {volume} {04}},\ \bibinfo {pages} {001} (\bibinfo {year} {2016})},\ \Eprint
  {http://arxiv.org/abs/1512.06239} {arXiv:1512.06239 [astro-ph.CO]}
  \BibitemShut {NoStop}%
\bibitem [{\citenamefont {Caprini}\ \emph {et~al.}(2020)\citenamefont {Caprini}
  \emph {et~al.}}]{Caprini:2019egz}%
  \BibitemOpen
  \bibfield  {author} {\bibinfo {author} {\bibfnamefont {C.}~\bibnamefont
  {Caprini}} \emph {et~al.},\ }\href {\doibase 10.1088/1475-7516/2020/03/024}
  {\bibfield  {journal} {\bibinfo  {journal} {JCAP}\ }\textbf {\bibinfo
  {volume} {03}},\ \bibinfo {pages} {024} (\bibinfo {year} {2020})},\ \Eprint
  {http://arxiv.org/abs/1910.13125} {arXiv:1910.13125 [astro-ph.CO]}
  \BibitemShut {NoStop}%
\bibitem [{\citenamefont {Gouttenoire}(2022)}]{Gouttenoire:2022gwi}%
  \BibitemOpen
  \bibfield  {author} {\bibinfo {author} {\bibfnamefont {Y.}~\bibnamefont
  {Gouttenoire}},\ }\href {\doibase 10.1007/978-3-031-11862-3} {\emph {\bibinfo
  {title} {{Beyond the Standard Model Cocktail}}}},\ Springer Theses\ (\bibinfo
   {publisher} {Springer},\ \bibinfo {address} {Cham},\ \bibinfo {year}
  {2022})\ \Eprint {http://arxiv.org/abs/2207.01633} {arXiv:2207.01633
  [hep-ph]} \BibitemShut {NoStop}%
\bibitem [{\citenamefont {Arzoumanian}\ \emph {et~al.}(2021)\citenamefont
  {Arzoumanian} \emph {et~al.}}]{NANOGrav:2021flc}%
  \BibitemOpen
  \bibfield  {author} {\bibinfo {author} {\bibfnamefont {Z.}~\bibnamefont
  {Arzoumanian}} \emph {et~al.} (\bibinfo {collaboration} {NANOGrav}),\ }\href
  {\doibase 10.1103/PhysRevLett.127.251302} {\bibfield  {journal} {\bibinfo
  {journal} {Phys. Rev. Lett.}\ }\textbf {\bibinfo {volume} {127}},\ \bibinfo
  {pages} {251302} (\bibinfo {year} {2021})},\ \Eprint
  {http://arxiv.org/abs/2104.13930} {arXiv:2104.13930 [astro-ph.CO]}
  \BibitemShut {NoStop}%
\bibitem [{\citenamefont {Xue}\ \emph {et~al.}(2021)\citenamefont {Xue} \emph
  {et~al.}}]{Xue:2021gyq}%
  \BibitemOpen
  \bibfield  {author} {\bibinfo {author} {\bibfnamefont {X.}~\bibnamefont
  {Xue}} \emph {et~al.},\ }\href {\doibase 10.1103/PhysRevLett.127.251303}
  {\bibfield  {journal} {\bibinfo  {journal} {Phys. Rev. Lett.}\ }\textbf
  {\bibinfo {volume} {127}},\ \bibinfo {pages} {251303} (\bibinfo {year}
  {2021})},\ \Eprint {http://arxiv.org/abs/2110.03096} {arXiv:2110.03096
  [astro-ph.CO]} \BibitemShut {NoStop}%
\bibitem [{\citenamefont {Roper~Pol}\ \emph {et~al.}(2022)\citenamefont
  {Roper~Pol}, \citenamefont {Caprini}, \citenamefont {Neronov},\ and\
  \citenamefont {Semikoz}}]{RoperPol:2022iel}%
  \BibitemOpen
  \bibfield  {author} {\bibinfo {author} {\bibfnamefont {A.}~\bibnamefont
  {Roper~Pol}}, \bibinfo {author} {\bibfnamefont {C.}~\bibnamefont {Caprini}},
  \bibinfo {author} {\bibfnamefont {A.}~\bibnamefont {Neronov}}, \ and\
  \bibinfo {author} {\bibfnamefont {D.}~\bibnamefont {Semikoz}},\ }\href
  {\doibase 10.1103/PhysRevD.105.123502} {\bibfield  {journal} {\bibinfo
  {journal} {Phys. Rev. D}\ }\textbf {\bibinfo {volume} {105}},\ \bibinfo
  {pages} {123502} (\bibinfo {year} {2022})},\ \Eprint
  {http://arxiv.org/abs/2201.05630} {arXiv:2201.05630 [astro-ph.CO]}
  \BibitemShut {NoStop}%
\bibitem [{\citenamefont {Bringmann}\ \emph {et~al.}(2023)\citenamefont
  {Bringmann}, \citenamefont {Depta}, \citenamefont {Konstandin}, \citenamefont
  {Schmidt-Hoberg},\ and\ \citenamefont {Tasillo}}]{Bringmann:2023opz}%
  \BibitemOpen
  \bibfield  {author} {\bibinfo {author} {\bibfnamefont {T.}~\bibnamefont
  {Bringmann}}, \bibinfo {author} {\bibfnamefont {P.~F.}\ \bibnamefont
  {Depta}}, \bibinfo {author} {\bibfnamefont {T.}~\bibnamefont {Konstandin}},
  \bibinfo {author} {\bibfnamefont {K.}~\bibnamefont {Schmidt-Hoberg}}, \ and\
  \bibinfo {author} {\bibfnamefont {C.}~\bibnamefont {Tasillo}},\ }\href
  {\doibase 10.1088/1475-7516/2023/11/053} {\bibfield  {journal} {\bibinfo
  {journal} {JCAP}\ }\textbf {\bibinfo {volume} {11}},\ \bibinfo {pages} {053}
  (\bibinfo {year} {2023})},\ \Eprint {http://arxiv.org/abs/2306.09411}
  {arXiv:2306.09411 [astro-ph.CO]} \BibitemShut {NoStop}%
\bibitem [{\citenamefont {Baldes}\ and\ \citenamefont
  {Garcia-Cely}(2019)}]{Baldes:2018emh}%
  \BibitemOpen
  \bibfield  {author} {\bibinfo {author} {\bibfnamefont {I.}~\bibnamefont
  {Baldes}}\ and\ \bibinfo {author} {\bibfnamefont {C.}~\bibnamefont
  {Garcia-Cely}},\ }\href {\doibase 10.1007/JHEP05(2019)190} {\bibfield
  {journal} {\bibinfo  {journal} {JHEP}\ }\textbf {\bibinfo {volume} {05}},\
  \bibinfo {pages} {190} (\bibinfo {year} {2019})},\ \Eprint
  {http://arxiv.org/abs/1809.01198} {arXiv:1809.01198 [hep-ph]} \BibitemShut
  {NoStop}%
\bibitem [{\citenamefont {Lewicki}\ \emph
  {et~al.}(2024{\natexlab{b}})\citenamefont {Lewicki}, \citenamefont {Toczek},\
  and\ \citenamefont {Vaskonen}}]{Lewicki:2024sfw}%
  \BibitemOpen
  \bibfield  {author} {\bibinfo {author} {\bibfnamefont {M.}~\bibnamefont
  {Lewicki}}, \bibinfo {author} {\bibfnamefont {P.}~\bibnamefont {Toczek}}, \
  and\ \bibinfo {author} {\bibfnamefont {V.}~\bibnamefont {Vaskonen}},\
  }\href@noop {} {\  (\bibinfo {year} {2024}{\natexlab{b}})},\ \Eprint
  {http://arxiv.org/abs/2412.10366} {arXiv:2412.10366 [astro-ph.CO]}
  \BibitemShut {NoStop}%
\bibitem [{\citenamefont {Gon\c{c}alves}\ \emph {et~al.}(2025)\citenamefont
  {Gon\c{c}alves}, \citenamefont {Marfatia}, \citenamefont {Morais},\ and\
  \citenamefont {Pasechnik}}]{Goncalves:2025uwh}%
  \BibitemOpen
  \bibfield  {author} {\bibinfo {author} {\bibfnamefont {J.~a.}\ \bibnamefont
  {Gon\c{c}alves}}, \bibinfo {author} {\bibfnamefont {D.}~\bibnamefont
  {Marfatia}}, \bibinfo {author} {\bibfnamefont {A.~P.}\ \bibnamefont
  {Morais}}, \ and\ \bibinfo {author} {\bibfnamefont {R.}~\bibnamefont
  {Pasechnik}},\ }\href@noop {} {\  (\bibinfo {year} {2025})},\ \Eprint
  {http://arxiv.org/abs/2501.11619} {arXiv:2501.11619 [hep-ph]} \BibitemShut
  {NoStop}%
\bibitem [{\citenamefont {Balan}\ \emph {et~al.}(2025)\citenamefont {Balan},
  \citenamefont {Bringmann}, \citenamefont {Kahlhoefer}, \citenamefont
  {Matuszak},\ and\ \citenamefont {Tasillo}}]{Balan:2025uke}%
  \BibitemOpen
  \bibfield  {author} {\bibinfo {author} {\bibfnamefont {S.}~\bibnamefont
  {Balan}}, \bibinfo {author} {\bibfnamefont {T.}~\bibnamefont {Bringmann}},
  \bibinfo {author} {\bibfnamefont {F.}~\bibnamefont {Kahlhoefer}}, \bibinfo
  {author} {\bibfnamefont {J.}~\bibnamefont {Matuszak}}, \ and\ \bibinfo
  {author} {\bibfnamefont {C.}~\bibnamefont {Tasillo}},\ }\href@noop {} {\
  (\bibinfo {year} {2025})},\ \Eprint {http://arxiv.org/abs/2502.19478}
  {arXiv:2502.19478 [hep-ph]} \BibitemShut {NoStop}%
\bibitem [{\citenamefont {Costa}\ \emph {et~al.}(2025)\citenamefont {Costa},
  \citenamefont {Hoefken~Zink}, \citenamefont {Lucente}, \citenamefont
  {Pascoli},\ and\ \citenamefont {Rosauro-Alcaraz}}]{Costa:2025csj}%
  \BibitemOpen
  \bibfield  {author} {\bibinfo {author} {\bibfnamefont {F.}~\bibnamefont
  {Costa}}, \bibinfo {author} {\bibfnamefont {J.}~\bibnamefont {Hoefken~Zink}},
  \bibinfo {author} {\bibfnamefont {M.}~\bibnamefont {Lucente}}, \bibinfo
  {author} {\bibfnamefont {S.}~\bibnamefont {Pascoli}}, \ and\ \bibinfo
  {author} {\bibfnamefont {S.}~\bibnamefont {Rosauro-Alcaraz}},\ }\href@noop {}
  {\  (\bibinfo {year} {2025})},\ \Eprint {http://arxiv.org/abs/2501.15649}
  {arXiv:2501.15649 [hep-ph]} \BibitemShut {NoStop}%
\bibitem [{\citenamefont {Sasaki}\ \emph {et~al.}(1982)\citenamefont {Sasaki},
  \citenamefont {Kodama},\ and\ \citenamefont {Sato}}]{Sasaki:1982fi}%
  \BibitemOpen
  \bibfield  {author} {\bibinfo {author} {\bibfnamefont {M.}~\bibnamefont
  {Sasaki}}, \bibinfo {author} {\bibfnamefont {H.}~\bibnamefont {Kodama}}, \
  and\ \bibinfo {author} {\bibfnamefont {K.}~\bibnamefont {Sato}},\ }\href
  {\doibase 10.1143/PTP.68.1561} {\bibfield  {journal} {\bibinfo  {journal}
  {Prog. Theor. Phys.}\ }\textbf {\bibinfo {volume} {68}},\ \bibinfo {pages}
  {1561} (\bibinfo {year} {1982})}\BibitemShut {NoStop}%
\bibitem [{\citenamefont {Liu}\ \emph {et~al.}(2023)\citenamefont {Liu},
  \citenamefont {Bian}, \citenamefont {Cai}, \citenamefont {Guo},\ and\
  \citenamefont {Wang}}]{Liu:2022lvz}%
  \BibitemOpen
  \bibfield  {author} {\bibinfo {author} {\bibfnamefont {J.}~\bibnamefont
  {Liu}}, \bibinfo {author} {\bibfnamefont {L.}~\bibnamefont {Bian}}, \bibinfo
  {author} {\bibfnamefont {R.-G.}\ \bibnamefont {Cai}}, \bibinfo {author}
  {\bibfnamefont {Z.-K.}\ \bibnamefont {Guo}}, \ and\ \bibinfo {author}
  {\bibfnamefont {S.-J.}\ \bibnamefont {Wang}},\ }\href {\doibase
  10.1103/PhysRevLett.130.051001} {\bibfield  {journal} {\bibinfo  {journal}
  {Phys. Rev. Lett.}\ }\textbf {\bibinfo {volume} {130}},\ \bibinfo {pages}
  {051001} (\bibinfo {year} {2023})},\ \Eprint
  {http://arxiv.org/abs/2208.14086} {arXiv:2208.14086 [astro-ph.CO]}
  \BibitemShut {NoStop}%
\bibitem [{\citenamefont {Giombi}\ and\ \citenamefont
  {Hindmarsh}(2024)}]{Giombi:2023jqq}%
  \BibitemOpen
  \bibfield  {author} {\bibinfo {author} {\bibfnamefont {L.}~\bibnamefont
  {Giombi}}\ and\ \bibinfo {author} {\bibfnamefont {M.}~\bibnamefont
  {Hindmarsh}},\ }\href {\doibase 10.1088/1475-7516/2024/03/059} {\bibfield
  {journal} {\bibinfo  {journal} {JCAP}\ }\textbf {\bibinfo {volume} {03}},\
  \bibinfo {pages} {059} (\bibinfo {year} {2024})},\ \Eprint
  {http://arxiv.org/abs/2307.12080} {arXiv:2307.12080 [astro-ph.CO]}
  \BibitemShut {NoStop}%
\bibitem [{\citenamefont {Elor}\ \emph {et~al.}(2024)\citenamefont {Elor},
  \citenamefont {Jinno}, \citenamefont {Kumar}, \citenamefont {McGehee},\ and\
  \citenamefont {Tsai}}]{Elor:2023xbz}%
  \BibitemOpen
  \bibfield  {author} {\bibinfo {author} {\bibfnamefont {G.}~\bibnamefont
  {Elor}}, \bibinfo {author} {\bibfnamefont {R.}~\bibnamefont {Jinno}},
  \bibinfo {author} {\bibfnamefont {S.}~\bibnamefont {Kumar}}, \bibinfo
  {author} {\bibfnamefont {R.}~\bibnamefont {McGehee}}, \ and\ \bibinfo
  {author} {\bibfnamefont {Y.}~\bibnamefont {Tsai}},\ }\href {\doibase
  10.1103/PhysRevLett.133.211003} {\bibfield  {journal} {\bibinfo  {journal}
  {Phys. Rev. Lett.}\ }\textbf {\bibinfo {volume} {133}},\ \bibinfo {pages}
  {211003} (\bibinfo {year} {2024})},\ \Eprint
  {http://arxiv.org/abs/2311.16222} {arXiv:2311.16222 [hep-ph]} \BibitemShut
  {NoStop}%
\bibitem [{\citenamefont {Buckley}\ \emph {et~al.}(2024)\citenamefont
  {Buckley}, \citenamefont {Du}, \citenamefont {Fernandez},\ and\ \citenamefont
  {Weikert}}]{Buckley:2024nen}%
  \BibitemOpen
  \bibfield  {author} {\bibinfo {author} {\bibfnamefont {M.~R.}\ \bibnamefont
  {Buckley}}, \bibinfo {author} {\bibfnamefont {P.}~\bibnamefont {Du}},
  \bibinfo {author} {\bibfnamefont {N.}~\bibnamefont {Fernandez}}, \ and\
  \bibinfo {author} {\bibfnamefont {M.~J.}\ \bibnamefont {Weikert}},\ }\href
  {\doibase 10.1088/1475-7516/2024/07/031} {\bibfield  {journal} {\bibinfo
  {journal} {JCAP}\ }\textbf {\bibinfo {volume} {07}},\ \bibinfo {pages} {031}
  (\bibinfo {year} {2024})},\ \Eprint {http://arxiv.org/abs/2402.13309}
  {arXiv:2402.13309 [hep-ph]} \BibitemShut {NoStop}%
\bibitem [{\citenamefont {Cai}\ \emph {et~al.}(2024)\citenamefont {Cai},
  \citenamefont {Hao},\ and\ \citenamefont {Wang}}]{Cai:2024nln}%
  \BibitemOpen
  \bibfield  {author} {\bibinfo {author} {\bibfnamefont {R.-G.}\ \bibnamefont
  {Cai}}, \bibinfo {author} {\bibfnamefont {Y.-S.}\ \bibnamefont {Hao}}, \ and\
  \bibinfo {author} {\bibfnamefont {S.-J.}\ \bibnamefont {Wang}},\ }\href
  {\doibase 10.1007/s11433-024-2416-3} {\bibfield  {journal} {\bibinfo
  {journal} {Sci. China Phys. Mech. Astron.}\ }\textbf {\bibinfo {volume}
  {67}},\ \bibinfo {pages} {290411} (\bibinfo {year} {2024})},\ \Eprint
  {http://arxiv.org/abs/2404.06506} {arXiv:2404.06506 [astro-ph.CO]}
  \BibitemShut {NoStop}%
\bibitem [{\citenamefont {Jinno}\ and\ \citenamefont
  {Kume}(2025)}]{Jinno:2024nwb}%
  \BibitemOpen
  \bibfield  {author} {\bibinfo {author} {\bibfnamefont {R.}~\bibnamefont
  {Jinno}}\ and\ \bibinfo {author} {\bibfnamefont {J.}~\bibnamefont {Kume}},\
  }\href {\doibase 10.1088/1475-7516/2025/02/057} {\bibfield  {journal}
  {\bibinfo  {journal} {JCAP}\ }\textbf {\bibinfo {volume} {02}},\ \bibinfo
  {pages} {057} (\bibinfo {year} {2025})},\ \Eprint
  {http://arxiv.org/abs/2408.10770} {arXiv:2408.10770 [gr-qc]} \BibitemShut
  {NoStop}%
\bibitem [{\citenamefont {Franciolini}\ \emph {et~al.}(2025)\citenamefont
  {Franciolini}, \citenamefont {Gouttenoire},\ and\ \citenamefont
  {Jinno}}]{Franciolini:2025ztf}%
  \BibitemOpen
  \bibfield  {author} {\bibinfo {author} {\bibfnamefont {G.}~\bibnamefont
  {Franciolini}}, \bibinfo {author} {\bibfnamefont {Y.}~\bibnamefont
  {Gouttenoire}}, \ and\ \bibinfo {author} {\bibfnamefont {R.}~\bibnamefont
  {Jinno}},\ }\href@noop {} {\  (\bibinfo {year} {2025})},\ \Eprint
  {http://arxiv.org/abs/2503.01962} {arXiv:2503.01962 [hep-ph]} \BibitemShut
  {NoStop}%
\bibitem [{\citenamefont {Kodama}\ \emph {et~al.}(1982)\citenamefont {Kodama},
  \citenamefont {Sasaki},\ and\ \citenamefont {Sato}}]{Kodama:1982sf}%
  \BibitemOpen
  \bibfield  {author} {\bibinfo {author} {\bibfnamefont {H.}~\bibnamefont
  {Kodama}}, \bibinfo {author} {\bibfnamefont {M.}~\bibnamefont {Sasaki}}, \
  and\ \bibinfo {author} {\bibfnamefont {K.}~\bibnamefont {Sato}},\ }\href
  {\doibase 10.1143/PTP.68.1979} {\bibfield  {journal} {\bibinfo  {journal}
  {Prog. Theor. Phys.}\ }\textbf {\bibinfo {volume} {68}},\ \bibinfo {pages}
  {1979} (\bibinfo {year} {1982})}\BibitemShut {NoStop}%
\bibitem [{\citenamefont {Liu}\ \emph {et~al.}(2022)\citenamefont {Liu},
  \citenamefont {Bian}, \citenamefont {Cai}, \citenamefont {Guo},\ and\
  \citenamefont {Wang}}]{Liu:2021svg}%
  \BibitemOpen
  \bibfield  {author} {\bibinfo {author} {\bibfnamefont {J.}~\bibnamefont
  {Liu}}, \bibinfo {author} {\bibfnamefont {L.}~\bibnamefont {Bian}}, \bibinfo
  {author} {\bibfnamefont {R.-G.}\ \bibnamefont {Cai}}, \bibinfo {author}
  {\bibfnamefont {Z.-K.}\ \bibnamefont {Guo}}, \ and\ \bibinfo {author}
  {\bibfnamefont {S.-J.}\ \bibnamefont {Wang}},\ }\href {\doibase
  10.1103/PhysRevD.105.L021303} {\bibfield  {journal} {\bibinfo  {journal}
  {Phys. Rev. D}\ }\textbf {\bibinfo {volume} {105}},\ \bibinfo {pages}
  {L021303} (\bibinfo {year} {2022})},\ \Eprint
  {http://arxiv.org/abs/2106.05637} {arXiv:2106.05637 [astro-ph.CO]}
  \BibitemShut {NoStop}%
\bibitem [{\citenamefont {Hashino}\ \emph {et~al.}(2022)\citenamefont
  {Hashino}, \citenamefont {Kanemura},\ and\ \citenamefont
  {Takahashi}}]{Hashino:2021qoq}%
  \BibitemOpen
  \bibfield  {author} {\bibinfo {author} {\bibfnamefont {K.}~\bibnamefont
  {Hashino}}, \bibinfo {author} {\bibfnamefont {S.}~\bibnamefont {Kanemura}}, \
  and\ \bibinfo {author} {\bibfnamefont {T.}~\bibnamefont {Takahashi}},\ }\href
  {\doibase 10.1016/j.physletb.2022.137261} {\bibfield  {journal} {\bibinfo
  {journal} {Phys. Lett. B}\ }\textbf {\bibinfo {volume} {833}},\ \bibinfo
  {pages} {137261} (\bibinfo {year} {2022})},\ \Eprint
  {http://arxiv.org/abs/2111.13099} {arXiv:2111.13099 [hep-ph]} \BibitemShut
  {NoStop}%
\bibitem [{\citenamefont {Kawana}\ \emph {et~al.}(2023)\citenamefont {Kawana},
  \citenamefont {Kim},\ and\ \citenamefont {Lu}}]{Kawana:2022olo}%
  \BibitemOpen
  \bibfield  {author} {\bibinfo {author} {\bibfnamefont {K.}~\bibnamefont
  {Kawana}}, \bibinfo {author} {\bibfnamefont {T.}~\bibnamefont {Kim}}, \ and\
  \bibinfo {author} {\bibfnamefont {P.}~\bibnamefont {Lu}},\ }\href {\doibase
  10.1103/PhysRevD.108.103531} {\bibfield  {journal} {\bibinfo  {journal}
  {Phys. Rev. D}\ }\textbf {\bibinfo {volume} {108}},\ \bibinfo {pages}
  {103531} (\bibinfo {year} {2023})},\ \Eprint
  {http://arxiv.org/abs/2212.14037} {arXiv:2212.14037 [astro-ph.CO]}
  \BibitemShut {NoStop}%
\bibitem [{\citenamefont {Lewicki}\ \emph {et~al.}(2023)\citenamefont
  {Lewicki}, \citenamefont {Toczek},\ and\ \citenamefont
  {Vaskonen}}]{Lewicki:2023ioy}%
  \BibitemOpen
  \bibfield  {author} {\bibinfo {author} {\bibfnamefont {M.}~\bibnamefont
  {Lewicki}}, \bibinfo {author} {\bibfnamefont {P.}~\bibnamefont {Toczek}}, \
  and\ \bibinfo {author} {\bibfnamefont {V.}~\bibnamefont {Vaskonen}},\ }\href
  {\doibase 10.1007/JHEP09(2023)092} {\bibfield  {journal} {\bibinfo  {journal}
  {JHEP}\ }\textbf {\bibinfo {volume} {09}},\ \bibinfo {pages} {092} (\bibinfo
  {year} {2023})},\ \Eprint {http://arxiv.org/abs/2305.04924} {arXiv:2305.04924
  [astro-ph.CO]} \BibitemShut {NoStop}%
\bibitem [{\citenamefont {Gouttenoire}\ and\ \citenamefont
  {Volansky}(2024)}]{Gouttenoire:2023naa}%
  \BibitemOpen
  \bibfield  {author} {\bibinfo {author} {\bibfnamefont {Y.}~\bibnamefont
  {Gouttenoire}}\ and\ \bibinfo {author} {\bibfnamefont {T.}~\bibnamefont
  {Volansky}},\ }\href {\doibase 10.1103/PhysRevD.110.043514} {\bibfield
  {journal} {\bibinfo  {journal} {Phys. Rev. D}\ }\textbf {\bibinfo {volume}
  {110}},\ \bibinfo {pages} {043514} (\bibinfo {year} {2024})},\ \Eprint
  {http://arxiv.org/abs/2305.04942} {arXiv:2305.04942 [hep-ph]} \BibitemShut
  {NoStop}%
\bibitem [{\citenamefont {Baldes}\ and\ \citenamefont
  {Olea-Romacho}(2024)}]{Baldes:2023rqv}%
  \BibitemOpen
  \bibfield  {author} {\bibinfo {author} {\bibfnamefont {I.}~\bibnamefont
  {Baldes}}\ and\ \bibinfo {author} {\bibfnamefont {M.~O.}\ \bibnamefont
  {Olea-Romacho}},\ }\href {\doibase 10.1007/JHEP01(2024)133} {\bibfield
  {journal} {\bibinfo  {journal} {JHEP}\ }\textbf {\bibinfo {volume} {01}},\
  \bibinfo {pages} {133} (\bibinfo {year} {2024})},\ \Eprint
  {http://arxiv.org/abs/2307.11639} {arXiv:2307.11639 [hep-ph]} \BibitemShut
  {NoStop}%
\bibitem [{\citenamefont {Gouttenoire}(2024)}]{Gouttenoire:2023pxh}%
  \BibitemOpen
  \bibfield  {author} {\bibinfo {author} {\bibfnamefont {Y.}~\bibnamefont
  {Gouttenoire}},\ }\href {\doibase 10.1016/j.physletb.2024.138800} {\bibfield
  {journal} {\bibinfo  {journal} {Phys. Lett. B}\ }\textbf {\bibinfo {volume}
  {855}},\ \bibinfo {pages} {138800} (\bibinfo {year} {2024})},\ \Eprint
  {http://arxiv.org/abs/2311.13640} {arXiv:2311.13640 [hep-ph]} \BibitemShut
  {NoStop}%
\bibitem [{\citenamefont {Jinno}\ \emph {et~al.}(2024)\citenamefont {Jinno},
  \citenamefont {Kume},\ and\ \citenamefont {Yamada}}]{Jinno:2023vnr}%
  \BibitemOpen
  \bibfield  {author} {\bibinfo {author} {\bibfnamefont {R.}~\bibnamefont
  {Jinno}}, \bibinfo {author} {\bibfnamefont {J.}~\bibnamefont {Kume}}, \ and\
  \bibinfo {author} {\bibfnamefont {M.}~\bibnamefont {Yamada}},\ }\href
  {\doibase 10.1016/j.physletb.2024.138465} {\bibfield  {journal} {\bibinfo
  {journal} {Phys. Lett. B}\ }\textbf {\bibinfo {volume} {849}},\ \bibinfo
  {pages} {138465} (\bibinfo {year} {2024})},\ \Eprint
  {http://arxiv.org/abs/2310.06901} {arXiv:2310.06901 [hep-ph]} \BibitemShut
  {NoStop}%
\bibitem [{\citenamefont {Flores}\ \emph {et~al.}(2024)\citenamefont {Flores},
  \citenamefont {Kusenko},\ and\ \citenamefont {Sasaki}}]{Flores:2024lng}%
  \BibitemOpen
  \bibfield  {author} {\bibinfo {author} {\bibfnamefont {M.~M.}\ \bibnamefont
  {Flores}}, \bibinfo {author} {\bibfnamefont {A.}~\bibnamefont {Kusenko}}, \
  and\ \bibinfo {author} {\bibfnamefont {M.}~\bibnamefont {Sasaki}},\ }\href
  {\doibase 10.1103/PhysRevD.110.015005} {\bibfield  {journal} {\bibinfo
  {journal} {Phys. Rev. D}\ }\textbf {\bibinfo {volume} {110}},\ \bibinfo
  {pages} {015005} (\bibinfo {year} {2024})},\ \Eprint
  {http://arxiv.org/abs/2402.13341} {arXiv:2402.13341 [hep-ph]} \BibitemShut
  {NoStop}%
\bibitem [{\citenamefont {Ai}\ \emph {et~al.}(2024)\citenamefont {Ai},
  \citenamefont {Heurtier},\ and\ \citenamefont {Jung}}]{Ai:2024cka}%
  \BibitemOpen
  \bibfield  {author} {\bibinfo {author} {\bibfnamefont {W.-Y.}\ \bibnamefont
  {Ai}}, \bibinfo {author} {\bibfnamefont {L.}~\bibnamefont {Heurtier}}, \ and\
  \bibinfo {author} {\bibfnamefont {T.~H.}\ \bibnamefont {Jung}},\ }\href@noop
  {} {\  (\bibinfo {year} {2024})},\ \Eprint {http://arxiv.org/abs/2409.02175}
  {arXiv:2409.02175 [astro-ph.CO]} \BibitemShut {NoStop}%
\bibitem [{\citenamefont {Hashino}\ \emph {et~al.}(2025)\citenamefont
  {Hashino}, \citenamefont {Kanemura}, \citenamefont {Takahashi}, \citenamefont
  {Tanaka},\ and\ \citenamefont {Yoo}}]{Hashino:2025fse}%
  \BibitemOpen
  \bibfield  {author} {\bibinfo {author} {\bibfnamefont {K.}~\bibnamefont
  {Hashino}}, \bibinfo {author} {\bibfnamefont {S.}~\bibnamefont {Kanemura}},
  \bibinfo {author} {\bibfnamefont {T.}~\bibnamefont {Takahashi}}, \bibinfo
  {author} {\bibfnamefont {M.}~\bibnamefont {Tanaka}}, \ and\ \bibinfo {author}
  {\bibfnamefont {C.-M.}\ \bibnamefont {Yoo}},\ }\href@noop {} {\  (\bibinfo
  {year} {2025})},\ \Eprint {http://arxiv.org/abs/2501.11040} {arXiv:2501.11040
  [hep-ph]} \BibitemShut {NoStop}%
\bibitem [{\citenamefont {Murai}\ \emph {et~al.}(2025)\citenamefont {Murai},
  \citenamefont {Sakurai},\ and\ \citenamefont {Takahashi}}]{Murai:2025hse}%
  \BibitemOpen
  \bibfield  {author} {\bibinfo {author} {\bibfnamefont {K.}~\bibnamefont
  {Murai}}, \bibinfo {author} {\bibfnamefont {K.}~\bibnamefont {Sakurai}}, \
  and\ \bibinfo {author} {\bibfnamefont {F.}~\bibnamefont {Takahashi}},\
  }\href@noop {} {\  (\bibinfo {year} {2025})},\ \Eprint
  {http://arxiv.org/abs/2502.02291} {arXiv:2502.02291 [astro-ph.CO]}
  \BibitemShut {NoStop}%
\bibitem [{\citenamefont {Zou}\ \emph {et~al.}(2025)\citenamefont {Zou},
  \citenamefont {Zhu}, \citenamefont {Zhao},\ and\ \citenamefont
  {Bian}}]{Zou:2025sow}%
  \BibitemOpen
  \bibfield  {author} {\bibinfo {author} {\bibfnamefont {J.}~\bibnamefont
  {Zou}}, \bibinfo {author} {\bibfnamefont {Z.}~\bibnamefont {Zhu}}, \bibinfo
  {author} {\bibfnamefont {Z.}~\bibnamefont {Zhao}}, \ and\ \bibinfo {author}
  {\bibfnamefont {L.}~\bibnamefont {Bian}},\ }\href@noop {} {\  (\bibinfo
  {year} {2025})},\ \Eprint {http://arxiv.org/abs/2502.20166} {arXiv:2502.20166
  [hep-ph]} \BibitemShut {NoStop}%
\bibitem [{\citenamefont {Gouttenoire}\ and\ \citenamefont
  {Vitagliano}(2024{\natexlab{b}})}]{Gouttenoire:2023gbn}%
  \BibitemOpen
  \bibfield  {author} {\bibinfo {author} {\bibfnamefont {Y.}~\bibnamefont
  {Gouttenoire}}\ and\ \bibinfo {author} {\bibfnamefont {E.}~\bibnamefont
  {Vitagliano}},\ }\href {\doibase 10.1103/PhysRevD.109.123507} {\bibfield
  {journal} {\bibinfo  {journal} {Phys. Rev. D}\ }\textbf {\bibinfo {volume}
  {109}},\ \bibinfo {pages} {123507} (\bibinfo {year} {2024}{\natexlab{b}})},\
  \Eprint {http://arxiv.org/abs/2311.07670} {arXiv:2311.07670 [hep-ph]}
  \BibitemShut {NoStop}%
\bibitem [{\citenamefont {Gouttenoire}\ \emph {et~al.}(2025)\citenamefont
  {Gouttenoire}, \citenamefont {King}, \citenamefont {Roshan}, \citenamefont
  {Wang}, \citenamefont {White},\ and\ \citenamefont
  {Yamazaki}}]{Gouttenoire:2025ofv}%
  \BibitemOpen
  \bibfield  {author} {\bibinfo {author} {\bibfnamefont {Y.}~\bibnamefont
  {Gouttenoire}}, \bibinfo {author} {\bibfnamefont {S.~F.}\ \bibnamefont
  {King}}, \bibinfo {author} {\bibfnamefont {R.}~\bibnamefont {Roshan}},
  \bibinfo {author} {\bibfnamefont {X.}~\bibnamefont {Wang}}, \bibinfo {author}
  {\bibfnamefont {G.}~\bibnamefont {White}}, \ and\ \bibinfo {author}
  {\bibfnamefont {M.}~\bibnamefont {Yamazaki}},\ }\href@noop {} {\  (\bibinfo
  {year} {2025})},\ \Eprint {http://arxiv.org/abs/2501.16414} {arXiv:2501.16414
  [hep-ph]} \BibitemShut {NoStop}%
\bibitem [{\citenamefont {Hiramatsu}\ \emph {et~al.}(2014)\citenamefont
  {Hiramatsu}, \citenamefont {Kawasaki},\ and\ \citenamefont
  {Saikawa}}]{Hiramatsu:2013qaa}%
  \BibitemOpen
  \bibfield  {author} {\bibinfo {author} {\bibfnamefont {T.}~\bibnamefont
  {Hiramatsu}}, \bibinfo {author} {\bibfnamefont {M.}~\bibnamefont {Kawasaki}},
  \ and\ \bibinfo {author} {\bibfnamefont {K.}~\bibnamefont {Saikawa}},\ }\href
  {\doibase 10.1088/1475-7516/2014/02/031} {\bibfield  {journal} {\bibinfo
  {journal} {JCAP}\ }\textbf {\bibinfo {volume} {02}},\ \bibinfo {pages} {031}
  (\bibinfo {year} {2014})},\ \Eprint {http://arxiv.org/abs/1309.5001}
  {arXiv:1309.5001 [astro-ph.CO]} \BibitemShut {NoStop}%
\bibitem [{\citenamefont {Ipser}\ and\ \citenamefont
  {Sikivie}(1984)}]{Ipser:1983db}%
  \BibitemOpen
  \bibfield  {author} {\bibinfo {author} {\bibfnamefont {J.}~\bibnamefont
  {Ipser}}\ and\ \bibinfo {author} {\bibfnamefont {P.}~\bibnamefont
  {Sikivie}},\ }\href {\doibase 10.1103/PhysRevD.30.712} {\bibfield  {journal}
  {\bibinfo  {journal} {Phys. Rev. D}\ }\textbf {\bibinfo {volume} {30}},\
  \bibinfo {pages} {712} (\bibinfo {year} {1984})}\BibitemShut {NoStop}%
\bibitem [{\citenamefont {Zeldovich}\ \emph {et~al.}(1974)\citenamefont
  {Zeldovich}, \citenamefont {Kobzarev},\ and\ \citenamefont
  {Okun}}]{Zeldovich:1974uw}%
  \BibitemOpen
  \bibfield  {author} {\bibinfo {author} {\bibfnamefont {Y.~B.}\ \bibnamefont
  {Zeldovich}}, \bibinfo {author} {\bibfnamefont {I.~Y.}\ \bibnamefont
  {Kobzarev}}, \ and\ \bibinfo {author} {\bibfnamefont {L.~B.}\ \bibnamefont
  {Okun}},\ }\href@noop {} {\bibfield  {journal} {\bibinfo  {journal} {Zh.
  Eksp. Teor. Fiz.}\ }\textbf {\bibinfo {volume} {67}},\ \bibinfo {pages} {3}
  (\bibinfo {year} {1974})}\BibitemShut {NoStop}%
\bibitem [{\citenamefont {Kawasaki}\ \emph {et~al.}(2015)\citenamefont
  {Kawasaki}, \citenamefont {Saikawa},\ and\ \citenamefont
  {Sekiguchi}}]{Kawasaki:2014sqa}%
  \BibitemOpen
  \bibfield  {author} {\bibinfo {author} {\bibfnamefont {M.}~\bibnamefont
  {Kawasaki}}, \bibinfo {author} {\bibfnamefont {K.}~\bibnamefont {Saikawa}}, \
  and\ \bibinfo {author} {\bibfnamefont {T.}~\bibnamefont {Sekiguchi}},\ }\href
  {\doibase 10.1103/PhysRevD.91.065014} {\bibfield  {journal} {\bibinfo
  {journal} {Phys. Rev. D}\ }\textbf {\bibinfo {volume} {91}},\ \bibinfo
  {pages} {065014} (\bibinfo {year} {2015})},\ \Eprint
  {http://arxiv.org/abs/1412.0789} {arXiv:1412.0789 [hep-ph]} \BibitemShut
  {NoStop}%
\bibitem [{\citenamefont {Saikawa}(2017)}]{Saikawa:2017hiv}%
  \BibitemOpen
  \bibfield  {author} {\bibinfo {author} {\bibfnamefont {K.}~\bibnamefont
  {Saikawa}},\ }\href {\doibase 10.3390/universe3020040} {\bibfield  {journal}
  {\bibinfo  {journal} {Universe}\ }\textbf {\bibinfo {volume} {3}},\ \bibinfo
  {pages} {40} (\bibinfo {year} {2017})},\ \Eprint
  {http://arxiv.org/abs/1703.02576} {arXiv:1703.02576 [hep-ph]} \BibitemShut
  {NoStop}%
\bibitem [{\citenamefont {Servant}\ and\ \citenamefont
  {Simakachorn}(2023)}]{Servant:2023mwt}%
  \BibitemOpen
  \bibfield  {author} {\bibinfo {author} {\bibfnamefont {G.}~\bibnamefont
  {Servant}}\ and\ \bibinfo {author} {\bibfnamefont {P.}~\bibnamefont
  {Simakachorn}},\ }\href {\doibase 10.1103/PhysRevD.108.123516} {\bibfield
  {journal} {\bibinfo  {journal} {Phys. Rev. D}\ }\textbf {\bibinfo {volume}
  {108}},\ \bibinfo {pages} {123516} (\bibinfo {year} {2023})},\ \Eprint
  {http://arxiv.org/abs/2307.03121} {arXiv:2307.03121 [hep-ph]} \BibitemShut
  {NoStop}%
\bibitem [{\citenamefont {Turner}\ \emph {et~al.}(1991)\citenamefont {Turner},
  \citenamefont {Watkins},\ and\ \citenamefont {Widrow}}]{Turner:1990uw}%
  \BibitemOpen
  \bibfield  {author} {\bibinfo {author} {\bibfnamefont {M.~S.}\ \bibnamefont
  {Turner}}, \bibinfo {author} {\bibfnamefont {R.}~\bibnamefont {Watkins}}, \
  and\ \bibinfo {author} {\bibfnamefont {L.~M.}\ \bibnamefont {Widrow}},\
  }\href {\doibase 10.1086/185928} {\bibfield  {journal} {\bibinfo  {journal}
  {Astrophys. J. Lett.}\ }\textbf {\bibinfo {volume} {367}},\ \bibinfo {pages}
  {L43} (\bibinfo {year} {1991})}\BibitemShut {NoStop}%
\bibitem [{\citenamefont {Goetz}\ and\ \citenamefont
  {Notzold}(1991)}]{Goetz:1990qz}%
  \BibitemOpen
  \bibfield  {author} {\bibinfo {author} {\bibfnamefont {G.}~\bibnamefont
  {Goetz}}\ and\ \bibinfo {author} {\bibfnamefont {D.}~\bibnamefont
  {Notzold}},\ }\href {\doibase 10.1016/S0550-3213(05)80037-9} {\bibfield
  {journal} {\bibinfo  {journal} {Nucl. Phys. B}\ }\textbf {\bibinfo {volume}
  {351}},\ \bibinfo {pages} {645} (\bibinfo {year} {1991})}\BibitemShut
  {NoStop}%
\bibitem [{\citenamefont {Goetz}\ and\ \citenamefont
  {Notzold}(1990)}]{Goetz:1990pj}%
  \BibitemOpen
  \bibfield  {author} {\bibinfo {author} {\bibfnamefont {G.}~\bibnamefont
  {Goetz}}\ and\ \bibinfo {author} {\bibfnamefont {D.}~\bibnamefont
  {Notzold}},\ }\href {\doibase 10.1103/PhysRevLett.65.2229} {\bibfield
  {journal} {\bibinfo  {journal} {Phys. Rev. Lett.}\ }\textbf {\bibinfo
  {volume} {65}},\ \bibinfo {pages} {2229} (\bibinfo {year} {1990})},\ \bibinfo
  {note} {[Erratum: Phys.Rev.Lett. 65, 3458 (1990)]}\BibitemShut {NoStop}%
\bibitem [{\citenamefont {Lola}\ and\ \citenamefont
  {Ross}(1993)}]{Lola:1993qu}%
  \BibitemOpen
  \bibfield  {author} {\bibinfo {author} {\bibfnamefont {S.}~\bibnamefont
  {Lola}}\ and\ \bibinfo {author} {\bibfnamefont {G.~G.}\ \bibnamefont
  {Ross}},\ }\href {\doibase 10.1016/0550-3213(93)90176-P} {\bibfield
  {journal} {\bibinfo  {journal} {Nucl. Phys. B}\ }\textbf {\bibinfo {volume}
  {406}},\ \bibinfo {pages} {452} (\bibinfo {year} {1993})}\BibitemShut
  {NoStop}%
\bibitem [{\citenamefont {Lazanu}\ \emph {et~al.}(2015)\citenamefont {Lazanu},
  \citenamefont {Martins},\ and\ \citenamefont {Shellard}}]{Lazanu:2015fua}%
  \BibitemOpen
  \bibfield  {author} {\bibinfo {author} {\bibfnamefont {A.}~\bibnamefont
  {Lazanu}}, \bibinfo {author} {\bibfnamefont {C.~J. A.~P.}\ \bibnamefont
  {Martins}}, \ and\ \bibinfo {author} {\bibfnamefont {E.~P.~S.}\ \bibnamefont
  {Shellard}},\ }\href {\doibase 10.1016/j.physletb.2015.06.034} {\bibfield
  {journal} {\bibinfo  {journal} {Phys. Lett. B}\ }\textbf {\bibinfo {volume}
  {747}},\ \bibinfo {pages} {426} (\bibinfo {year} {2015})},\ \Eprint
  {http://arxiv.org/abs/1505.03673} {arXiv:1505.03673 [astro-ph.CO]}
  \BibitemShut {NoStop}%
\bibitem [{\citenamefont {Takahashi}\ and\ \citenamefont
  {Yin}(2021)}]{Takahashi:2020tqv}%
  \BibitemOpen
  \bibfield  {author} {\bibinfo {author} {\bibfnamefont {F.}~\bibnamefont
  {Takahashi}}\ and\ \bibinfo {author} {\bibfnamefont {W.}~\bibnamefont
  {Yin}},\ }\href {\doibase 10.1088/1475-7516/2021/04/007} {\bibfield
  {journal} {\bibinfo  {journal} {JCAP}\ }\textbf {\bibinfo {volume} {04}},\
  \bibinfo {pages} {007} (\bibinfo {year} {2021})},\ \Eprint
  {http://arxiv.org/abs/2012.11576} {arXiv:2012.11576 [hep-ph]} \BibitemShut
  {NoStop}%
\bibitem [{\citenamefont {Kitajima}\ \emph {et~al.}(2022)\citenamefont
  {Kitajima}, \citenamefont {Kozai}, \citenamefont {Takahashi},\ and\
  \citenamefont {Yin}}]{Kitajima:2022jzz}%
  \BibitemOpen
  \bibfield  {author} {\bibinfo {author} {\bibfnamefont {N.}~\bibnamefont
  {Kitajima}}, \bibinfo {author} {\bibfnamefont {F.}~\bibnamefont {Kozai}},
  \bibinfo {author} {\bibfnamefont {F.}~\bibnamefont {Takahashi}}, \ and\
  \bibinfo {author} {\bibfnamefont {W.}~\bibnamefont {Yin}},\ }\href {\doibase
  10.1088/1475-7516/2022/10/043} {\bibfield  {journal} {\bibinfo  {journal}
  {JCAP}\ }\textbf {\bibinfo {volume} {10}},\ \bibinfo {pages} {043} (\bibinfo
  {year} {2022})},\ \Eprint {http://arxiv.org/abs/2205.05083} {arXiv:2205.05083
  [astro-ph.CO]} \BibitemShut {NoStop}%
\bibitem [{\citenamefont {Ramberg}\ \emph {et~al.}(2023)\citenamefont
  {Ramberg}, \citenamefont {Ratzinger},\ and\ \citenamefont
  {Schwaller}}]{Ramberg:2022irf}%
  \BibitemOpen
  \bibfield  {author} {\bibinfo {author} {\bibfnamefont {N.}~\bibnamefont
  {Ramberg}}, \bibinfo {author} {\bibfnamefont {W.}~\bibnamefont {Ratzinger}},
  \ and\ \bibinfo {author} {\bibfnamefont {P.}~\bibnamefont {Schwaller}},\
  }\href {\doibase 10.1088/1475-7516/2023/02/039} {\bibfield  {journal}
  {\bibinfo  {journal} {JCAP}\ }\textbf {\bibinfo {volume} {02}},\ \bibinfo
  {pages} {039} (\bibinfo {year} {2023})},\ \Eprint
  {http://arxiv.org/abs/2209.14313} {arXiv:2209.14313 [hep-ph]} \BibitemShut
  {NoStop}%
\bibitem [{\citenamefont {Zeng}\ \emph {et~al.}(2023)\citenamefont {Zeng},
  \citenamefont {Liu},\ and\ \citenamefont {Guo}}]{Zeng:2023jut}%
  \BibitemOpen
  \bibfield  {author} {\bibinfo {author} {\bibfnamefont {Z.-M.}\ \bibnamefont
  {Zeng}}, \bibinfo {author} {\bibfnamefont {J.}~\bibnamefont {Liu}}, \ and\
  \bibinfo {author} {\bibfnamefont {Z.-K.}\ \bibnamefont {Guo}},\ }\href
  {\doibase 10.1103/PhysRevD.108.063005} {\bibfield  {journal} {\bibinfo
  {journal} {Phys. Rev. D}\ }\textbf {\bibinfo {volume} {108}},\ \bibinfo
  {pages} {063005} (\bibinfo {year} {2023})},\ \Eprint
  {http://arxiv.org/abs/2301.07230} {arXiv:2301.07230 [astro-ph.CO]}
  \BibitemShut {NoStop}%
\bibitem [{\citenamefont {Lu}(2024)}]{Lu:2024dzj}%
  \BibitemOpen
  \bibfield  {author} {\bibinfo {author} {\bibfnamefont {B.-Q.}\ \bibnamefont
  {Lu}},\ }\href@noop {} {\  (\bibinfo {year} {2024})},\ \Eprint
  {http://arxiv.org/abs/2412.07677} {arXiv:2412.07677 [gr-qc]} \BibitemShut
  {NoStop}%
\bibitem [{\citenamefont {Baumann}(2022)}]{Baumann:2022mni}%
  \BibitemOpen
  \bibfield  {author} {\bibinfo {author} {\bibfnamefont {D.}~\bibnamefont
  {Baumann}},\ }\href {\doibase 10.1017/9781108937092} {\emph {\bibinfo {title}
  {{Cosmology}}}}\ (\bibinfo  {publisher} {Cambridge University Press},\
  \bibinfo {year} {2022})\BibitemShut {NoStop}%
\bibitem [{\citenamefont {Ferrer}\ \emph {et~al.}(2019)\citenamefont {Ferrer},
  \citenamefont {Masso}, \citenamefont {Panico}, \citenamefont {Pujolas},\ and\
  \citenamefont {Rompineve}}]{Ferrer:2018uiu}%
  \BibitemOpen
  \bibfield  {author} {\bibinfo {author} {\bibfnamefont {F.}~\bibnamefont
  {Ferrer}}, \bibinfo {author} {\bibfnamefont {E.}~\bibnamefont {Masso}},
  \bibinfo {author} {\bibfnamefont {G.}~\bibnamefont {Panico}}, \bibinfo
  {author} {\bibfnamefont {O.}~\bibnamefont {Pujolas}}, \ and\ \bibinfo
  {author} {\bibfnamefont {F.}~\bibnamefont {Rompineve}},\ }\href {\doibase
  10.1103/PhysRevLett.122.101301} {\bibfield  {journal} {\bibinfo  {journal}
  {Phys. Rev. Lett.}\ }\textbf {\bibinfo {volume} {122}},\ \bibinfo {pages}
  {101301} (\bibinfo {year} {2019})},\ \Eprint
  {http://arxiv.org/abs/1807.01707} {arXiv:1807.01707 [hep-ph]} \BibitemShut
  {NoStop}%
\bibitem [{\citenamefont {Gelmini}\ \emph
  {et~al.}(2023{\natexlab{a}})\citenamefont {Gelmini}, \citenamefont {Hyman},
  \citenamefont {Simpson},\ and\ \citenamefont {Vitagliano}}]{Gelmini:2023ngs}%
  \BibitemOpen
  \bibfield  {author} {\bibinfo {author} {\bibfnamefont {G.~B.}\ \bibnamefont
  {Gelmini}}, \bibinfo {author} {\bibfnamefont {J.}~\bibnamefont {Hyman}},
  \bibinfo {author} {\bibfnamefont {A.}~\bibnamefont {Simpson}}, \ and\
  \bibinfo {author} {\bibfnamefont {E.}~\bibnamefont {Vitagliano}},\ }\href
  {\doibase 10.1088/1475-7516/2023/06/055} {\bibfield  {journal} {\bibinfo
  {journal} {JCAP}\ }\textbf {\bibinfo {volume} {06}},\ \bibinfo {pages} {055}
  (\bibinfo {year} {2023}{\natexlab{a}})},\ \Eprint
  {http://arxiv.org/abs/2303.14107} {arXiv:2303.14107 [hep-ph]} \BibitemShut
  {NoStop}%
\bibitem [{\citenamefont {Gelmini}\ \emph
  {et~al.}(2023{\natexlab{b}})\citenamefont {Gelmini}, \citenamefont
  {Simpson},\ and\ \citenamefont {Vitagliano}}]{Gelmini:2022nim}%
  \BibitemOpen
  \bibfield  {author} {\bibinfo {author} {\bibfnamefont {G.~B.}\ \bibnamefont
  {Gelmini}}, \bibinfo {author} {\bibfnamefont {A.}~\bibnamefont {Simpson}}, \
  and\ \bibinfo {author} {\bibfnamefont {E.}~\bibnamefont {Vitagliano}},\
  }\href {\doibase 10.1088/1475-7516/2023/02/031} {\bibfield  {journal}
  {\bibinfo  {journal} {JCAP}\ }\textbf {\bibinfo {volume} {02}},\ \bibinfo
  {pages} {031} (\bibinfo {year} {2023}{\natexlab{b}})},\ \Eprint
  {http://arxiv.org/abs/2207.07126} {arXiv:2207.07126 [hep-ph]} \BibitemShut
  {NoStop}%
\bibitem [{\citenamefont {Ferreira}\ \emph {et~al.}(2024)\citenamefont
  {Ferreira}, \citenamefont {Notari}, \citenamefont {Pujol\`as},\ and\
  \citenamefont {Rompineve}}]{Ferreira:2024eru}%
  \BibitemOpen
  \bibfield  {author} {\bibinfo {author} {\bibfnamefont {R.~Z.}\ \bibnamefont
  {Ferreira}}, \bibinfo {author} {\bibfnamefont {A.}~\bibnamefont {Notari}},
  \bibinfo {author} {\bibfnamefont {O.}~\bibnamefont {Pujol\`as}}, \ and\
  \bibinfo {author} {\bibfnamefont {F.}~\bibnamefont {Rompineve}},\ }\href
  {\doibase 10.1088/1475-7516/2024/06/020} {\bibfield  {journal} {\bibinfo
  {journal} {JCAP}\ }\textbf {\bibinfo {volume} {06}},\ \bibinfo {pages} {020}
  (\bibinfo {year} {2024})},\ \Eprint {http://arxiv.org/abs/2401.14331}
  {arXiv:2401.14331 [astro-ph.CO]} \BibitemShut {NoStop}%
\bibitem [{\citenamefont {Lu}\ \emph {et~al.}(2024)\citenamefont {Lu},
  \citenamefont {Chiang},\ and\ \citenamefont {Li}}]{Lu:2024ngi}%
  \BibitemOpen
  \bibfield  {author} {\bibinfo {author} {\bibfnamefont {B.-Q.}\ \bibnamefont
  {Lu}}, \bibinfo {author} {\bibfnamefont {C.-W.}\ \bibnamefont {Chiang}}, \
  and\ \bibinfo {author} {\bibfnamefont {T.}~\bibnamefont {Li}},\ }\href@noop
  {} {\  (\bibinfo {year} {2024})},\ \Eprint {http://arxiv.org/abs/2409.09986}
  {arXiv:2409.09986 [astro-ph.CO]} \BibitemShut {NoStop}%
\bibitem [{\citenamefont {Gouttenoire}(2025)}]{inprep}%
  \BibitemOpen
  \bibfield  {author} {\bibinfo {author} {\bibfnamefont {Y.}~\bibnamefont
  {Gouttenoire}},\ }\href@noop {} {\  (\bibinfo {year} {2025})},\ \bibinfo
  {note} {in preparation}\BibitemShut {NoStop}%
\bibitem [{\citenamefont {Lacki}\ and\ \citenamefont
  {Beacom}(2010)}]{Lacki:2010zf}%
  \BibitemOpen
  \bibfield  {author} {\bibinfo {author} {\bibfnamefont {B.~C.}\ \bibnamefont
  {Lacki}}\ and\ \bibinfo {author} {\bibfnamefont {J.~F.}\ \bibnamefont
  {Beacom}},\ }\href {\doibase 10.1088/2041-8205/720/1/L67} {\bibfield
  {journal} {\bibinfo  {journal} {Astrophys. J. Lett.}\ }\textbf {\bibinfo
  {volume} {720}},\ \bibinfo {pages} {L67} (\bibinfo {year} {2010})},\ \Eprint
  {http://arxiv.org/abs/1003.3466} {arXiv:1003.3466 [astro-ph.CO]} \BibitemShut
  {NoStop}%
\bibitem [{\citenamefont {Adamek}\ \emph {et~al.}(2019)\citenamefont {Adamek},
  \citenamefont {Byrnes}, \citenamefont {Gosenca},\ and\ \citenamefont
  {Hotchkiss}}]{Adamek:2019gns}%
  \BibitemOpen
  \bibfield  {author} {\bibinfo {author} {\bibfnamefont {J.}~\bibnamefont
  {Adamek}}, \bibinfo {author} {\bibfnamefont {C.~T.}\ \bibnamefont {Byrnes}},
  \bibinfo {author} {\bibfnamefont {M.}~\bibnamefont {Gosenca}}, \ and\
  \bibinfo {author} {\bibfnamefont {S.}~\bibnamefont {Hotchkiss}},\ }\href
  {\doibase 10.1103/PhysRevD.100.023506} {\bibfield  {journal} {\bibinfo
  {journal} {Phys. Rev. D}\ }\textbf {\bibinfo {volume} {100}},\ \bibinfo
  {pages} {023506} (\bibinfo {year} {2019})},\ \Eprint
  {http://arxiv.org/abs/1901.08528} {arXiv:1901.08528 [astro-ph.CO]}
  \BibitemShut {NoStop}%
\bibitem [{\citenamefont {Liu}(2023)}]{Liu:2023tmv}%
  \BibitemOpen
  \bibfield  {author} {\bibinfo {author} {\bibfnamefont {J.}~\bibnamefont
  {Liu}},\ }\href {\doibase 10.1103/PhysRevD.108.123544} {\bibfield  {journal}
  {\bibinfo  {journal} {Phys. Rev. D}\ }\textbf {\bibinfo {volume} {108}},\
  \bibinfo {pages} {123544} (\bibinfo {year} {2023})}\BibitemShut {NoStop}%
\bibitem [{\citenamefont {Depta}\ \emph {et~al.}(2023)\citenamefont {Depta},
  \citenamefont {Schmidt-Hoberg}, \citenamefont {Schwaller},\ and\
  \citenamefont {Tasillo}}]{Depta:2023qst}%
  \BibitemOpen
  \bibfield  {author} {\bibinfo {author} {\bibfnamefont {P.~F.}\ \bibnamefont
  {Depta}}, \bibinfo {author} {\bibfnamefont {K.}~\bibnamefont
  {Schmidt-Hoberg}}, \bibinfo {author} {\bibfnamefont {P.}~\bibnamefont
  {Schwaller}}, \ and\ \bibinfo {author} {\bibfnamefont {C.}~\bibnamefont
  {Tasillo}},\ }\href@noop {} {\  (\bibinfo {year} {2023})},\ \Eprint
  {http://arxiv.org/abs/2306.17836} {arXiv:2306.17836 [astro-ph.CO]}
  \BibitemShut {NoStop}%
\bibitem [{\citenamefont {Gouttenoire}\ \emph {et~al.}(2024)\citenamefont
  {Gouttenoire}, \citenamefont {Trifinopoulos}, \citenamefont {Valogiannis},\
  and\ \citenamefont {Vanvlasselaer}}]{Gouttenoire:2023nzr}%
  \BibitemOpen
  \bibfield  {author} {\bibinfo {author} {\bibfnamefont {Y.}~\bibnamefont
  {Gouttenoire}}, \bibinfo {author} {\bibfnamefont {S.}~\bibnamefont
  {Trifinopoulos}}, \bibinfo {author} {\bibfnamefont {G.}~\bibnamefont
  {Valogiannis}}, \ and\ \bibinfo {author} {\bibfnamefont {M.}~\bibnamefont
  {Vanvlasselaer}},\ }\href {\doibase 10.1103/PhysRevD.109.123002} {\bibfield
  {journal} {\bibinfo  {journal} {Phys. Rev. D}\ }\textbf {\bibinfo {volume}
  {109}},\ \bibinfo {pages} {123002} (\bibinfo {year} {2024})},\ \Eprint
  {http://arxiv.org/abs/2307.01457} {arXiv:2307.01457 [astro-ph.CO]}
  \BibitemShut {NoStop}%
\bibitem [{\citenamefont {Ellis}\ and\ \citenamefont
  {Lewicki}(2021)}]{Ellis:2020ena}%
  \BibitemOpen
  \bibfield  {author} {\bibinfo {author} {\bibfnamefont {J.}~\bibnamefont
  {Ellis}}\ and\ \bibinfo {author} {\bibfnamefont {M.}~\bibnamefont
  {Lewicki}},\ }\href {\doibase 10.1103/PhysRevLett.126.041304} {\bibfield
  {journal} {\bibinfo  {journal} {Phys. Rev. Lett.}\ }\textbf {\bibinfo
  {volume} {126}},\ \bibinfo {pages} {041304} (\bibinfo {year} {2021})},\
  \Eprint {http://arxiv.org/abs/2009.06555} {arXiv:2009.06555 [astro-ph.CO]}
  \BibitemShut {NoStop}%
\bibitem [{\citenamefont {Blasi}\ \emph {et~al.}(2021)\citenamefont {Blasi},
  \citenamefont {Brdar},\ and\ \citenamefont {Schmitz}}]{Blasi:2020mfx}%
  \BibitemOpen
  \bibfield  {author} {\bibinfo {author} {\bibfnamefont {S.}~\bibnamefont
  {Blasi}}, \bibinfo {author} {\bibfnamefont {V.}~\bibnamefont {Brdar}}, \ and\
  \bibinfo {author} {\bibfnamefont {K.}~\bibnamefont {Schmitz}},\ }\href
  {\doibase 10.1103/PhysRevLett.126.041305} {\bibfield  {journal} {\bibinfo
  {journal} {Phys. Rev. Lett.}\ }\textbf {\bibinfo {volume} {126}},\ \bibinfo
  {pages} {041305} (\bibinfo {year} {2021})},\ \Eprint
  {http://arxiv.org/abs/2009.06607} {arXiv:2009.06607 [astro-ph.CO]}
  \BibitemShut {NoStop}%
\bibitem [{\citenamefont {Buchmuller}\ \emph {et~al.}(2020)\citenamefont
  {Buchmuller}, \citenamefont {Domcke},\ and\ \citenamefont
  {Schmitz}}]{Buchmuller:2020lbh}%
  \BibitemOpen
  \bibfield  {author} {\bibinfo {author} {\bibfnamefont {W.}~\bibnamefont
  {Buchmuller}}, \bibinfo {author} {\bibfnamefont {V.}~\bibnamefont {Domcke}},
  \ and\ \bibinfo {author} {\bibfnamefont {K.}~\bibnamefont {Schmitz}},\ }\href
  {\doibase 10.1016/j.physletb.2020.135914} {\bibfield  {journal} {\bibinfo
  {journal} {Phys. Lett. B}\ }\textbf {\bibinfo {volume} {811}},\ \bibinfo
  {pages} {135914} (\bibinfo {year} {2020})},\ \Eprint
  {http://arxiv.org/abs/2009.10649} {arXiv:2009.10649 [astro-ph.CO]}
  \BibitemShut {NoStop}%
\bibitem [{\citenamefont {Buchmuller}\ \emph {et~al.}(2023)\citenamefont
  {Buchmuller}, \citenamefont {Domcke},\ and\ \citenamefont
  {Schmitz}}]{Buchmuller:2023aus}%
  \BibitemOpen
  \bibfield  {author} {\bibinfo {author} {\bibfnamefont {W.}~\bibnamefont
  {Buchmuller}}, \bibinfo {author} {\bibfnamefont {V.}~\bibnamefont {Domcke}},
  \ and\ \bibinfo {author} {\bibfnamefont {K.}~\bibnamefont {Schmitz}},\ }\href
  {\doibase 10.1088/1475-7516/2023/11/020} {\bibfield  {journal} {\bibinfo
  {journal} {JCAP}\ }\textbf {\bibinfo {volume} {11}},\ \bibinfo {pages} {020}
  (\bibinfo {year} {2023})},\ \Eprint {http://arxiv.org/abs/2307.04691}
  {arXiv:2307.04691 [hep-ph]} \BibitemShut {NoStop}%
\bibitem [{\citenamefont {Ellis}\ \emph {et~al.}(2023)\citenamefont {Ellis},
  \citenamefont {Lewicki}, \citenamefont {Lin},\ and\ \citenamefont
  {Vaskonen}}]{Ellis:2023tsl}%
  \BibitemOpen
  \bibfield  {author} {\bibinfo {author} {\bibfnamefont {J.}~\bibnamefont
  {Ellis}}, \bibinfo {author} {\bibfnamefont {M.}~\bibnamefont {Lewicki}},
  \bibinfo {author} {\bibfnamefont {C.}~\bibnamefont {Lin}}, \ and\ \bibinfo
  {author} {\bibfnamefont {V.}~\bibnamefont {Vaskonen}},\ }\href {\doibase
  10.1103/PhysRevD.108.103511} {\bibfield  {journal} {\bibinfo  {journal}
  {Phys. Rev. D}\ }\textbf {\bibinfo {volume} {108}},\ \bibinfo {pages}
  {103511} (\bibinfo {year} {2023})},\ \Eprint
  {http://arxiv.org/abs/2306.17147} {arXiv:2306.17147 [astro-ph.CO]}
  \BibitemShut {NoStop}%
\bibitem [{\citenamefont {Kume}\ and\ \citenamefont
  {Hindmarsh}(2024)}]{Kume:2024adn}%
  \BibitemOpen
  \bibfield  {author} {\bibinfo {author} {\bibfnamefont {J.}~\bibnamefont
  {Kume}}\ and\ \bibinfo {author} {\bibfnamefont {M.}~\bibnamefont
  {Hindmarsh}},\ }\href {\doibase 10.1088/1475-7516/2024/12/001} {\bibfield
  {journal} {\bibinfo  {journal} {JCAP}\ }\textbf {\bibinfo {volume} {12}},\
  \bibinfo {pages} {001} (\bibinfo {year} {2024})},\ \Eprint
  {http://arxiv.org/abs/2404.02705} {arXiv:2404.02705 [astro-ph.CO]}
  \BibitemShut {NoStop}%
\bibitem [{\citenamefont {Vagnozzi}(2023)}]{Vagnozzi:2023lwo}%
  \BibitemOpen
  \bibfield  {author} {\bibinfo {author} {\bibfnamefont {S.}~\bibnamefont
  {Vagnozzi}},\ }\href {\doibase 10.1016/j.jheap.2023.07.001} {\bibfield
  {journal} {\bibinfo  {journal} {JHEAp}\ }\textbf {\bibinfo {volume} {39}},\
  \bibinfo {pages} {81} (\bibinfo {year} {2023})},\ \Eprint
  {http://arxiv.org/abs/2306.16912} {arXiv:2306.16912 [astro-ph.CO]}
  \BibitemShut {NoStop}%
\bibitem [{\citenamefont {Geller}\ \emph {et~al.}(2024)\citenamefont {Geller},
  \citenamefont {Ghosh}, \citenamefont {Lu},\ and\ \citenamefont
  {Tsai}}]{Geller:2023shn}%
  \BibitemOpen
  \bibfield  {author} {\bibinfo {author} {\bibfnamefont {M.}~\bibnamefont
  {Geller}}, \bibinfo {author} {\bibfnamefont {S.}~\bibnamefont {Ghosh}},
  \bibinfo {author} {\bibfnamefont {S.}~\bibnamefont {Lu}}, \ and\ \bibinfo
  {author} {\bibfnamefont {Y.}~\bibnamefont {Tsai}},\ }\href {\doibase
  10.1103/PhysRevD.109.063537} {\bibfield  {journal} {\bibinfo  {journal}
  {Phys. Rev. D}\ }\textbf {\bibinfo {volume} {109}},\ \bibinfo {pages}
  {063537} (\bibinfo {year} {2024})},\ \Eprint
  {http://arxiv.org/abs/2307.03724} {arXiv:2307.03724 [hep-ph]} \BibitemShut
  {NoStop}%
\bibitem [{\citenamefont {Murai}\ and\ \citenamefont
  {Yin}(2023)}]{Murai:2023gkv}%
  \BibitemOpen
  \bibfield  {author} {\bibinfo {author} {\bibfnamefont {K.}~\bibnamefont
  {Murai}}\ and\ \bibinfo {author} {\bibfnamefont {W.}~\bibnamefont {Yin}},\
  }\href {\doibase 10.1007/JHEP10(2023)062} {\bibfield  {journal} {\bibinfo
  {journal} {JHEP}\ }\textbf {\bibinfo {volume} {10}},\ \bibinfo {pages} {062}
  (\bibinfo {year} {2023})},\ \Eprint {http://arxiv.org/abs/2307.00628}
  {arXiv:2307.00628 [hep-ph]} \BibitemShut {NoStop}%
\bibitem [{\citenamefont {Unal}\ \emph {et~al.}(2024)\citenamefont {Unal},
  \citenamefont {Papageorgiou},\ and\ \citenamefont {Obata}}]{Unal:2023srk}%
  \BibitemOpen
  \bibfield  {author} {\bibinfo {author} {\bibfnamefont {C.}~\bibnamefont
  {Unal}}, \bibinfo {author} {\bibfnamefont {A.}~\bibnamefont {Papageorgiou}},
  \ and\ \bibinfo {author} {\bibfnamefont {I.}~\bibnamefont {Obata}},\ }\href
  {\doibase 10.1016/j.physletb.2024.138873} {\bibfield  {journal} {\bibinfo
  {journal} {Phys. Lett. B}\ }\textbf {\bibinfo {volume} {856}},\ \bibinfo
  {pages} {138873} (\bibinfo {year} {2024})},\ \Eprint
  {http://arxiv.org/abs/2307.02322} {arXiv:2307.02322 [astro-ph.CO]}
  \BibitemShut {NoStop}%
\bibitem [{\citenamefont {Nakama}\ \emph {et~al.}(2016)\citenamefont {Nakama},
  \citenamefont {Suyama},\ and\ \citenamefont {Yokoyama}}]{Nakama:2016kfq}%
  \BibitemOpen
  \bibfield  {author} {\bibinfo {author} {\bibfnamefont {T.}~\bibnamefont
  {Nakama}}, \bibinfo {author} {\bibfnamefont {T.}~\bibnamefont {Suyama}}, \
  and\ \bibinfo {author} {\bibfnamefont {J.}~\bibnamefont {Yokoyama}},\ }\href
  {\doibase 10.1103/PhysRevD.94.103522} {\bibfield  {journal} {\bibinfo
  {journal} {Phys. Rev. D}\ }\textbf {\bibinfo {volume} {94}},\ \bibinfo
  {pages} {103522} (\bibinfo {year} {2016})},\ \Eprint
  {http://arxiv.org/abs/1609.02245} {arXiv:1609.02245 [gr-qc]} \BibitemShut
  {NoStop}%
\bibitem [{\citenamefont {Hooper}\ \emph {et~al.}(2024)\citenamefont {Hooper},
  \citenamefont {Ireland}, \citenamefont {Krnjaic},\ and\ \citenamefont
  {Stebbins}}]{Hooper:2023nnl}%
  \BibitemOpen
  \bibfield  {author} {\bibinfo {author} {\bibfnamefont {D.}~\bibnamefont
  {Hooper}}, \bibinfo {author} {\bibfnamefont {A.}~\bibnamefont {Ireland}},
  \bibinfo {author} {\bibfnamefont {G.}~\bibnamefont {Krnjaic}}, \ and\
  \bibinfo {author} {\bibfnamefont {A.}~\bibnamefont {Stebbins}},\ }\href
  {\doibase 10.1088/1475-7516/2024/04/021} {\bibfield  {journal} {\bibinfo
  {journal} {JCAP}\ }\textbf {\bibinfo {volume} {04}},\ \bibinfo {pages} {021}
  (\bibinfo {year} {2024})},\ \Eprint {http://arxiv.org/abs/2308.00756}
  {arXiv:2308.00756 [astro-ph.CO]} \BibitemShut {NoStop}%
\bibitem [{\citenamefont {Buschmann}\ \emph {et~al.}(2018)\citenamefont
  {Buschmann}, \citenamefont {Kopp}, \citenamefont {Safdi},\ and\ \citenamefont
  {Wu}}]{Buschmann:2017ams}%
  \BibitemOpen
  \bibfield  {author} {\bibinfo {author} {\bibfnamefont {M.}~\bibnamefont
  {Buschmann}}, \bibinfo {author} {\bibfnamefont {J.}~\bibnamefont {Kopp}},
  \bibinfo {author} {\bibfnamefont {B.~R.}\ \bibnamefont {Safdi}}, \ and\
  \bibinfo {author} {\bibfnamefont {C.-L.}\ \bibnamefont {Wu}},\ }\href
  {\doibase 10.1103/PhysRevLett.120.211101} {\bibfield  {journal} {\bibinfo
  {journal} {Phys. Rev. Lett.}\ }\textbf {\bibinfo {volume} {120}},\ \bibinfo
  {pages} {211101} (\bibinfo {year} {2018})},\ \Eprint
  {http://arxiv.org/abs/1711.03554} {arXiv:1711.03554 [astro-ph.GA]}
  \BibitemShut {NoStop}%
\bibitem [{\citenamefont {Kawasaki}\ \emph {et~al.}(2022)\citenamefont
  {Kawasaki}, \citenamefont {Nakatsuka},\ and\ \citenamefont
  {Nakayama}}]{Kawasaki:2021yek}%
  \BibitemOpen
  \bibfield  {author} {\bibinfo {author} {\bibfnamefont {M.}~\bibnamefont
  {Kawasaki}}, \bibinfo {author} {\bibfnamefont {H.}~\bibnamefont {Nakatsuka}},
  \ and\ \bibinfo {author} {\bibfnamefont {K.}~\bibnamefont {Nakayama}},\
  }\href {\doibase 10.1088/1475-7516/2022/03/061} {\bibfield  {journal}
  {\bibinfo  {journal} {JCAP}\ }\textbf {\bibinfo {volume} {03}},\ \bibinfo
  {pages} {061} (\bibinfo {year} {2022})},\ \Eprint
  {http://arxiv.org/abs/2110.12620} {arXiv:2110.12620 [astro-ph.CO]}
  \BibitemShut {NoStop}%
\bibitem [{\citenamefont {Lee}\ \emph {et~al.}(2021)\citenamefont {Lee},
  \citenamefont {Mitridate}, \citenamefont {Trickle},\ and\ \citenamefont
  {Zurek}}]{Lee:2020wfn}%
  \BibitemOpen
  \bibfield  {author} {\bibinfo {author} {\bibfnamefont {V.~S.~H.}\
  \bibnamefont {Lee}}, \bibinfo {author} {\bibfnamefont {A.}~\bibnamefont
  {Mitridate}}, \bibinfo {author} {\bibfnamefont {T.}~\bibnamefont {Trickle}},
  \ and\ \bibinfo {author} {\bibfnamefont {K.~M.}\ \bibnamefont {Zurek}},\
  }\href {\doibase 10.1007/JHEP06(2021)028} {\bibfield  {journal} {\bibinfo
  {journal} {JHEP}\ }\textbf {\bibinfo {volume} {06}},\ \bibinfo {pages} {028}
  (\bibinfo {year} {2021})},\ \Eprint {http://arxiv.org/abs/2012.09857}
  {arXiv:2012.09857 [astro-ph.CO]} \BibitemShut {NoStop}%
\end{thebibliography}%

\end{document}